\documentclass[fleqn,usenatbib]{mnras}

\usepackage{newtxtext,newtxmath}

\usepackage[T1]{fontenc}

\DeclareRobustCommand{\VAN}[3]{#2}
\let\VANthebibliography\thebibliography
\def\thebibliography{\DeclareRobustCommand{\VAN}[3]{##3}\VANthebibliography}

\usepackage{graphicx}	
\usepackage{amsmath}	
\usepackage{listings}
\usepackage{color}
\definecolor{dkgreen}{rgb}{0,0.6,0}
\definecolor{gray}{rgb}{0.5,0.5,0.5}
\definecolor{mauve}{rgb}{0.58,0,0.82}
\definecolor{golden}{rgb}{0.86,0.65,0.01}
\defcitealias{Elbadry2018}{ER18}

\title[A million binaries from Gaia]{A million binaries from Gaia eDR3: sample selection and validation of Gaia parallax uncertainties} 

\author[El-Badry et al.]{
Kareem El-Badry,$^{1,2}$\thanks{E-mail: kelbadry@berkeley.edu}
Hans-Walter Rix,$^{2}$
and Tyler M. Heintz$^{3}$
\\
$^{1}$Department of Astronomy and Theoretical Astrophysics Center, University of California Berkeley, Berkeley, CA 94720, USA\\
$^{2}$Max-Planck Institute for Astronomy, K\"onigstuhl 17, D-69117 Heidelberg, Germany\\
$^{3}$Department of Astronomy, Boston University, 725 Commonwealth Ave., Boston, MA 02215, US
}

\date{Accepted to MNRAS}

\pubyear{2021}

\begin{document}
\label{firstpage}
\pagerange{\pageref{firstpage}--\pageref{lastpage}}
\maketitle

\begin{abstract}
We construct from {\it Gaia} eDR3 an extensive catalog of spatially resolved binary stars within $\approx$ 1\,kpc of the Sun, with projected separations ranging from a few au to 1 pc. We estimate the probability that each pair is a chance alignment empirically, using the {\it Gaia} catalog itself to calculate the rate of chance alignments as a function of observables. The catalog contains 1.3 (1.1) million binaries with >90\% (>99\%) probability of being bound, including 16,000 white dwarf -- main sequence (WD+MS) binaries and 1,400 WD+WD binaries. We make the full catalog publicly available, as well as the queries and code to produce it. We then use this sample to calibrate the published {\it Gaia} DR3 parallax {\it uncertainties}, making use of the binary components' near-identical parallaxes. We show that these uncertainties are generally reliable for faint stars ($G\gtrsim 18$), but are underestimated significantly for brighter stars. The underestimates are generally $\le 30\%$ for isolated sources with well-behaved astrometry, but are larger (up to $\sim$80\%) for apparently well-behaved sources with a companion within $\lesssim 4$ arcsec, and much larger for sources with poor astrometric fits. We provide an empirical fitting function to inflate published $\sigma_{\varpi}$ values for isolated sources. The public catalog offers wide ranging follow-up opportunities: from calibrating spectroscopic surveys, to precisely constraining ages of field stars, to the masses and the initial-final mass relation of white dwarfs, to dynamically probing the Galactic tidal field. 
\end{abstract}




\section{Introduction}

About half of all solar-type stars are members of binary systems, and a majority of these are so widely separated that the two components never interact \citep[e.g.][]{Moe2017}. With orbital periods ranging from $\sim$10 to $\sim 10^8$ years, most of these binaries can in some sense be viewed as clusters of two: the components formed from the same gas cloud and have orbited one another ever since. They thus have essentially the same age, initial composition, and distance, but generally different masses and occasionally different evolutionary phases. This makes wide binaries useful for calibrating stellar models as well as spectroscopic and astrometric surveys. 

At angular separations greater than about one arcsecond, wide binaries are easily resolvable as two point sources. Distinguishing physically bound binary stars from chance alignments (``optical doubles'') has been a long-standing challenge for binary star astronomy. Indeed, the first systematic binary star catalog was constructed under the assumption that all close pairs were chance alignments \citep{Herschel1782}, an assumption that was only shown to be incorrect two decades later \citep{Herschel1803}.

For bright binaries at close angular separations, chance alignments can be excluded probabilistically. However, the contamination rate from chance alignments increases at wider separations and fainter magnitudes. Inclusion of proper motion data can aid the selection of genuine binaries, in which the two stars have nearly identical proper motions  \citep[e.g.][]{Luyten_1971, Luyten_1979, Salim_2003, Chaname_2004, Dhital_2010}. Many wide binary searches have specifically targeted high-proper motion stars, which have fewer phase-space neighbors that can be mistaken for binary companions. If available, parallaxes and radial velocities are also useful for distinguishing binaries from chance alignments \citep[e.g.][]{Close_1990, Andrews_2017}. 

Prior to the {\it Gaia} mission \citep{Gaia_2016}, useful parallaxes for this purpose were only available for (relatively) small samples of nearby and bright stars. {\it Gaia} DR2 \citep{Brown_2018} dramatically expanded the sample of stars with well-measured parallaxes and proper motions, enabling the construction of unprecedentedly pure and extensive wide binary samples. \citet[][hereafter ER18]{Elbadry2018} searched {\it Gaia} DR2 for pairs of stars within 200 pc of the Sun with parallaxes and proper motions consistent with being gravitationally bound, and projected separations of up to 50,000 AU (0.24 pc). Their catalog prioritized purity over completeness, and thus imposed relatively strict cuts on astrometric and photometric quality and signal-to-noise ratio (SNR). This resulted in a catalog of $\sim 55,000$ binaries with an estimated contamination rate of $\sim 0.1$\%. 

Using the same basic strategy but less stringent cuts on astrometric SNR, \citet{Tian2020} extended the \citetalias{Elbadry2018} binary search to larger distances ($d<4$\,kpc) and wider separations ($s<1$\,pc). This produced a substantially larger sample of $\sim 800,000$ binary candidates, but with a higher contamination rate: chance-alignments dominate their catalog at $s\gtrsim 20,000$\,au, though higher-purity subsamples can be selected by imposing stricter cuts on astrometric SNR. Their catalog contained 325,000 binaries with $s < 20,000$\,au; it is expected to be reasonably pure in this separation regime. 

Another binary catalog was produced by \citet{Hartman2020}, who combined {\it Gaia} DR2 astrometry with a catalog of high-proper motion stars not contained in {\it Gaia} DR2. In order to reduce contamination from chance alignments, they limited their search to binaries with proper motions larger than $40\,\rm mas\,yr^{-1}$; this translates to a distance limit of order 200 pc for typical stars in the Galactic disk, but to a larger search volume for stars on halo-like orbits with large tangential velocities. Their primary binary catalog contains $\sim 100,000$ binary candidates, with projected separations as large as 10 pc.

In this paper, we use {\it Gaia} eDR3 data \citep{Gaia2020} to further expand the sample of known wide binaries. Compared to DR2, eDR3 astrometry is based on a $\sim$1.5\,$\times$ longer time baseline. This yields significant improvements in parallax and especially proper motion uncertainties. For example, the median uncertainties in \texttt{parallax} and \texttt{pmra} (one-dimenstional proper motion) at $G=18$ have respectively improved from 0.165 mas to 0.120 mas, and from $0.280\,\rm mas\,yr^{-1}$ to  $0.123\,\rm mas\,yr^{-1}$ \citep[][]{Lindegren_2018, Lindegren2020}. Improvements at the bright end are more significant, due to better handling of systematics; i.e., at $G=13$, the median parallax uncertainty decreased from 0.029 mas to 0.015 mas. This improved astrometric precision allows us to distinguish bound binaries from chance alignments to larger distances and wider separations than was possible with DR2. Our approach is a compromise between the strategies adopted by \citet{Tian2020} and \citetalias{Elbadry2018}. Like \citet{Tian2020}, we search out to wide separations and relatively low astrometric SNR, so that at wide separations and faint magnitudes, the full catalog is dominated by chance alignments. However, we also empirically estimate and assign each binary a probability that it is a chance alignment, making it straightforward to select pure subsets of the catalog. 

The remainder of the paper is organized as follows. In Section~\ref{sec:selection}, we describe how our binary candidate sample is selected and cleaned. Section~\ref{sec:chance_align} describes how we quantify the contamination rate from chance alignments and estimate the probability that each binary candidate is bound. Section~\ref{sec:basic_prop} details basic properties of the catalog and a cross-match with the LAMOST survey. In Section~\ref{sec:parallax_uncert}, we use the binary sample to validate the {\it Gaia} eDR3 parallax uncertainties. We summarize and discuss our results in Section~\ref{sec:discussion}. The public catalog is described in Section~\ref{sec:cat_description}. Details about the calculation of chance-alignment probabilities are provided in the Appendix.

\section{Sample selection}
\label{sec:selection}
To reduce contamination from chance alignments, we limit our sample to pairs in which both components have moderately precise astrometry. We retrieved from the {\it Gaia} archive all sources with parallaxes greater than 1 mas (corresponding to a nominal distance limit of 1 kpc, which in reality is blurred due to parallax errors), fractional parallax uncertainties less than 20\%, absolute parallax uncertainties less than 2 mas, and non-missing $G-$band magnitudes. This was achieved with the following ADQL query:
\begin{lstlisting}
select *
from gaiaedr3.gaia_source
where parallax > 1
and parallax_over_error > 5
and parallax_error < 2
and phot_g_mean_mag is not null
\end{lstlisting}
The query returns a total of $N=64,407,853$ sources, corresponding to $N(N+1)/2 \approx 2\times 10^{15}$ possible pairs. Of these, we consider as initial binary candidates all pairs that satisfy the following: 
\begin{itemize}
    \item {\it Projected separation less than 1 parsec}: the angular separation between the two stars, $\theta$, must satisfy 
    \begin{equation}
        \frac{\theta}{{\rm arcsec}}\leq206.265\times\frac{\varpi_1}{{\rm mas}},
        \label{eq:theta}
    \end{equation}
    where $\varpi_1$ is the parallax of the star with the brighter $G$ magnitude. The maximum search radius of 1 pc (corresponding to an orbital period of $\sim 10^8$ years) is chosen because a vanishingly small number of bound binaries are expected to exist at separations wider than this, where the Galactic tidal field becomes comparable to the gravitational attraction of the two stars \citep[e.g.][]{Binney2008}. The separation beyond which the Galactic tidal field dominates a binary's internal acceleration is called the Jacobi radius. In the Solar neighborhood, it is given by $r_{J}\approx 1.35\,{\rm pc}\times\left(M_{{\rm tot}}/M_{\odot}\right)$, where $M_{\rm tot}$ is the total mass of the binary \citep{Jiang_2010}. At separations slightly below $r_J$, binaries are efficiently disrupted by gravitational perturbations from objects such as other stars and molecular clouds \citep[e.g.][]{Weinberg_1987}.
    \item {\it Parallaxes consistent within 3 (or 6) sigma}: the parallaxes of the two components, $\varpi_1$ and $\varpi_2$, must satisfy 
    \begin{equation}
    \left|\varpi_{1}-\varpi_{2}\right|<b\sqrt{\sigma_{\varpi,1}^{2}+\sigma_{\varpi,2}^{2}},
    \label{eq:varpi}
    \end{equation}
    where $\sigma_{\varpi,i}$ is the parallax uncertainty of the $i$-th component, and $b=3$ for pairs with $\theta > 4$ arcsec, or $b=6$ for pairs with $\theta < 4$ arcsec. The less stringent cut at $\theta < 4$ arcsec is adopted because the chance alignment rate there is low and parallax uncertainties are significantly underestimated at close angular separations (Section~\ref{sec:parallax_uncert}).
    \item {\it Proper motions consistent with a Keplerian orbit}: The two stars in a wide binary will have proper motions that are similar, but, due to orbital motion, not identical. We require
    \begin{equation}
    \Delta\mu\leq\Delta\mu_{{\rm orbit}}+2\sigma_{\Delta\mu}, 
    \label{eq:deltamu}
    \end{equation}
    where  $\Delta \mu$ is the observed scalar proper motion difference, $\sigma_{\Delta\mu}$ its uncertainty, and $\Delta\mu_{{\rm orbit}}$ the maximum proper motion difference expected due to orbital motion. The first two quantities are calculated as 
    \begin{align}
\Delta\mu&=\left[(\mu_{\alpha,1}^{*}-\mu_{\alpha,2}^{*})^{2}+(\mu_{\delta,1}-\mu_{\delta,2})^{2}\right]^{1/2}, \label{eq:delta_mu} \\
\sigma_{\Delta\mu}&=\frac{1}{\Delta\mu}\left[\left(\sigma_{\mu_{\alpha,1}^{*}}^{2}+\sigma_{\mu_{\alpha,2}^{*}}^{2}\right)\Delta\mu_{\alpha}^{2}+\left(\sigma_{\mu_{\delta,1}}^{2}+\sigma_{\mu_{\delta,2}}^{2}\right)\Delta\mu_{\delta}^{2}\right]^{1/2},
\label{eq:sigma_delta_mu}
\end{align}
where $\Delta\mu_{\alpha}^{2}=(\mu_{\alpha,1}^{*}-\mu_{\alpha,2}^{*})^{2}$ and $\Delta\mu_{\delta}^{2}=(\mu_{\delta,1}-\mu_{\delta,2})^{2}$. Here  $\mu_{\alpha,i}^{*}\equiv \mu_{\alpha,i}\cos\delta_{i}$, $\alpha$ and $\delta$ denote right ascension and declination, and $\mu_{\alpha}$ and $\mu_{\delta}$, the proper motion in the right ascension and declination directions. Following \citetalias{Elbadry2018}, we take 
\begin{align}
\Delta \mu_{{\rm orbit}}=0.44\,{\rm mas\,yr}^{-1}\times \left(\frac{\varpi}{{\rm mas}}\right)^{3/2}\left(\frac{\theta}{{\rm arcsec}}\right)^{-1/2},
\label{eq:orbital_delta_mu}
\end{align}
which is the maximum proper motion difference expected for a circular orbit of total mass $5 M_{\odot}$; it corresponds to a projected physical velocity difference 
\begin{align}
\Delta V_{\rm orb}=2.1\,{\rm km\,s^{-1}}\times\left(\frac{s}{{\rm 1000\,au}}\right)^{-1/2},
\label{eq:deltav_orb}
\end{align}
\end{itemize}
where $s=1000\,{\rm au}\times\left(\theta/{\rm arcsec}\right)\left(\varpi/{\rm mas}\right)^{-1}$ is the projected separation. This quantity is of course not equal to the full 3D separation, or to the semimajor axis, $a$. For randomly oriented orbits with a realistic eccentricity distribution, $s$ and $a$ usually agree within a factor of 2 (see \citetalias{Elbadry2018}, their Appendix B). For our purposes, it is generally an acceptable approximation to assume $a\sim s$. We note that the cut on proper motion difference removes from the sample a significant fraction of unresolved hierarchical triples and higher-order multiples (see Section~\ref{sec:triples}).

We apply the cuts on projected separation, parallax difference, and proper motion difference to all possible pairs. The projected separation cut reduces the list of possible binaries to $10^{10}$; additionally requiring consistent parallaxes and proper motions reduces it to $10^8$ initial candidate pairs. A random sample of 1\% of these is plotted in the top panel of Figure~\ref{fig:sky_distributions}, and their projected separation distribution is shown in gray in Figure~\ref{fig:sep_dist_cleaning}. A large  majority of initial candidate pairs are chance alignments, not genuine binaries. This is evident both from the fact that many candidate pairs are in the Galactic bulge, LMC, and SMC, and from the fact that their separation distribution peaks at the widest separations (Figure~\ref{fig:sep_dist_cleaning}), where true binaries are rare. 

Stars in the bulge, LMC, and SMC would ideally be excluded by the requirement of $\varpi > 1$ and $\varpi/\sigma_{\varpi} > 5$ in the initial query, but a significant fraction of sources in crowded fields have spurious parallaxes \citep[e.g.][]{Fabricius2020}. Most of these spurious background pairs can be excluded by imposing astrometric quality cuts -- for example, we find that adding the requirement of \texttt{astrometric\_sigma5d\_max} < 1 to our initial query reduces the number of initial candidates by a factor of 10 while only removing a minority of genuine binaries\footnote{\texttt{astrometric\_sigma5d\_max} is the longest principal axis in the 5-dimensional error ellipsoid, in mas. A large value indicates that at least one of the astrometric parameters is poorly determined in the 5-parameter solution. More information on this and other {\it Gaia } flags can be found in the \href{https://gea.esac.esa.int/archive/documentation/GEDR3/Gaia_archive/chap_datamodel/sec_dm_main_tables/ssec_dm_gaia_source.html}{Gaia eDR3 data model}. }. We opted against applying such cuts because they do remove some real binaries, and we find that chance alignments with spurious parallaxes can be efficiently filtered out by the cleaning described in Section~\ref{sec:clusters}.

In addition to chance alignments of background sources with spurious parallaxes, our initial selection also efficiently selects members of star clusters and moving groups, about 100 of which can be seen in the top panel of Figure~\ref{fig:sky_distributions}. These are not spurious, in the sense that they really do contain many pairs of stars within our search volume that are close in phase space and in some cases mutually bound \citep[e.g.][]{Oh2017}. However, most of them are not binaries and will become unbound when the clusters dissolve. 

\subsection{Cleaning clusters, background pairs, and triples}
\label{sec:clusters}

 \begin{figure}
	\includegraphics[width=\columnwidth]{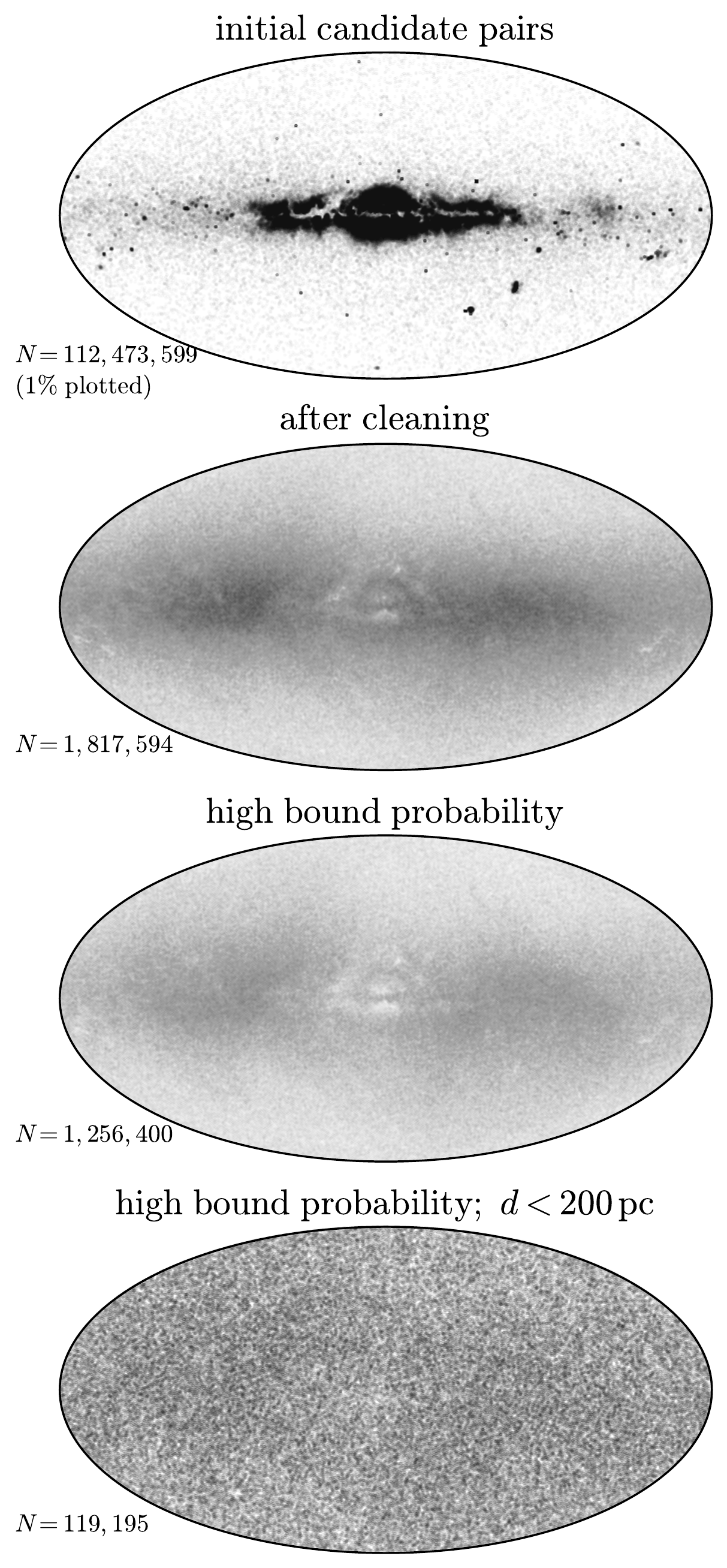}
    \caption{Sky distribution of binary candidates in Galactic coordinates. Top panel shows a random subset of all pairs with consistent parallaxes and proper motions. The vast majority of these are not genuine binaries. Many clusters are visible, as well as the Magellanic clouds and inner Galaxy. These are beyond the nominal 1 kpc search limit but enter the dataset due to spurious parallaxes. 2nd panel shows the sky distribution after clusters, moving groups, and resolved triples have been filtered (Section~\ref{sec:clusters}). This removes the majority of spurious background pairs. 3rd panel shows the pairs with $\mathcal{R} < 0.1$ (corresponding approximately to  90\% bound probability; see Section~\ref{sec:chance_alignment_prob}). Some structure remains, primarily tracing dust. Bottom panel shows sources within 200 pc, which are distributed almost uniformly on the sky.}
    \label{fig:sky_distributions}
\end{figure}

\begin{figure}
	\includegraphics[width=\columnwidth]{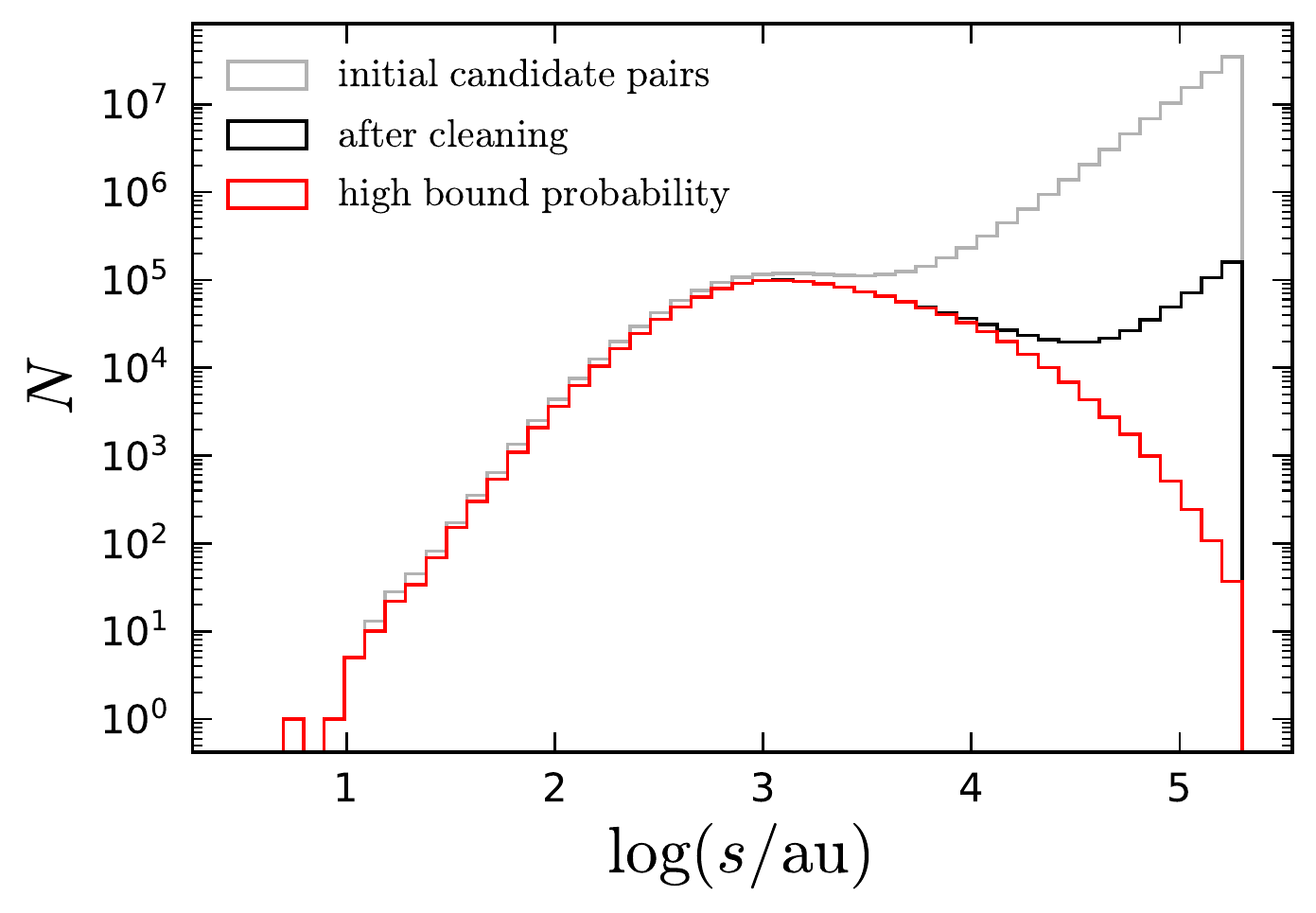}
    \caption{Separation distribution of binary candidates at three stages of the selection procedure, corresponding to the top three panels of Figure~\ref{fig:sky_distributions}. Most initial candidates (gray) are chance alignments, which dominate at $s \gtrsim 3,000\,$au. Cleaning resolved triples,  clusters, and moving groups removes a large fraction of chance alignments (black), but they still dominate at $s \gtrsim 30,000$\,au. The separation distribution of binaries with high bound probability ($\gtrsim 90\%$; red; see Section~\ref{sec:chance_alignment_prob}) falls off steeply at wide separations.}
    \label{fig:sep_dist_cleaning}
\end{figure}

We clean the list of initial binary candidates in several passes. First, beginning with all the sources returned by our initial ADQL query, we count for each source the number of phase-space neighbors that are brighter than $G=18$ and consistent with the size and velocity dispersion of a typical cluster. We define neighboring sources as those that satisfy the following:
\begin{itemize}
    \item Projected separation less than 5 pc; i.e., $\theta\leq17.19\,{\rm arcmin}\times\left(\varpi/{\rm mas}\right)$.
    \item Proper motions within 5\,km\,s$^{-1}$; this translates to a proper motion difference $\Delta\mu\leq1.05\,{\rm mas\,yr^{-1}\times}\left(\varpi/{\rm mas}\right)$, with a 2$\sigma$ tolerance.
    \item Parallaxes consistent within 2 sigma; i.e., $\Delta\varpi\leq2\sqrt[]{\sigma_{\varpi,1}^{2}+\sigma_{\varpi,2}^{2}}$.
\end{itemize}
We remove from our binary candidate list all pairs in which either component has more than 30 neighbors as defined above. Only 6.5 million of the 64 million sources in the search sample have more than 30 neighbors, and inspection reveals that a majority of these are not in the search volume at all, but rather are spurious sources in the Galactic bulge, LMC and SMC. Removing candidates containing these sources shrinks the candidate list from 112,473,599 pairs to 2,881,543 and removes most of the obvious structure seen in the top panel of Figure~\ref{fig:sky_distributions}.

Next, we remove all overlapping pairs. That is, if either component of a binary candidate is a member of another binary candidate, we remove both pairs. This removes genuine resolved triples, which are efficiently identified by our initial search \citep[e.g.][]{Perpinyaes2019} and are not rare \citep{Tokovinin2014}. It also removes some additional chance alignments. This cut shrinks the candidate sample from 2,881,543 to 1,918,362.   

Finally, we search for members of small clusters or moving groups not removed in the first pass. Adopting the phase-space coordinates of the brighter component of each pair as representing the pair, we count the number of neighboring pairs for each candidate, defined using the same three criteria used when counting neighboring sources (without any magnitude cut). We reject all candidates that have more than 1 neighboring pair, shrinking the sample from 1,918,362 to 1,817,594.

It is important to note that some real binaries will be removed during the filtering of resolved triples, clusters, and moving groups. In regions of high stellar density, a distant tertiary candidate that is really a chance alignment can cause a genuine binary to be rejected as a triple. Similarly, some bound binaries {\it do} exist within clusters, and these will all be rejected. An upper limit of $\sim 15\%$ can be set on the fraction of true binaries lost during cleaning by comparing the gray and black histograms in Figure~\ref{fig:sep_dist_cleaning} at close separations; this is an upper limit because some of the pairs removed at close separations are genuinely members of resolved triples, moving groups, and clusters.

The binary candidate sample after removal of resolved triples and suspected cluster members is shown in the 2nd panel of Figure~\ref{fig:sky_distributions} and with the black histogram in Figure~\ref{fig:sep_dist_cleaning}. This is the catalog published along with this work; it contains both high-confidence binaries and pairs that are very likely chance alignments. The contamination rate from chance alignments increases rapidly with separation (Section~\ref{sec:chance_align}), while the true binary separation distribution decreases monotonically over the range of separations to which we are sensitive (e.g., \citetalias{Elbadry2018}). The separation at which the number of binary candidates per dex of separation begins to increase ($\log(s/{\rm au})\approx 4.5$ for the black histogram in Figure~\ref{fig:sep_dist_cleaning}) thus marks the separation at which chance alignments begin to dominate the full sample. That is, when  considering the full catalog, a majority of binary candidates with $s\gtrsim 30,000$\,au are chance alignments. It is, however, possible to select subsets of the catalog that are free of chance alignments out to much wider separations; see Section~\ref{sec:chance_align}.

We define the ``primary'' and  ``secondary'' components, denoted ``1'' and ``2'', as the component with the brighter and fainter $G$ magnitude, respectively. Both components are on the main sequence in a majority of binaries; in these cases, the ``primary'' is generally also the more massive component. For binaries containing white dwarfs, the secondary will often be more massive than the primary.


\section{Chance alignments}
\label{sec:chance_align}

\begin{figure*}
	\includegraphics[width=\textwidth]{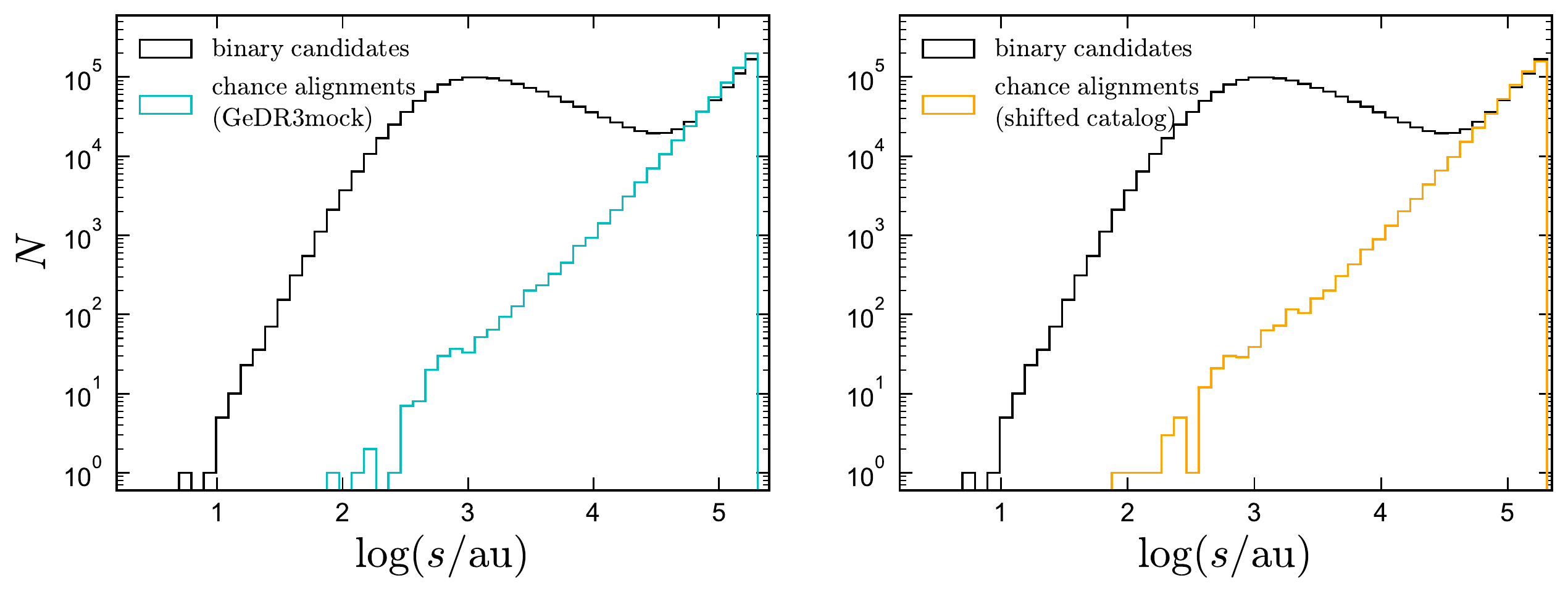}
    \caption{Two methods for estimating the contamination rate from chance alignments. Left panel compares the separation distribution of binary candidates (black) to that of candidates selected from the eDR3 mock catalog (\citealt{Rybizki2020}; cyan), which contains no true binaries.  Right panel compares the same binary candidates (black) to the separation distribution of candidates when stars are artificially shifted 0.5 degrees from their true positions when selecting candidate companions (orange; by construction, these are not true binaries).
    Both methods show that chance alignments dominate the full catalog at wide separations ($s\gtrsim 30,000$\,au, where the separation distribution begins to increase), but true binaries dominate at closer separations. }
    \label{fig:chance_align_estimate}
\end{figure*}

The contamination rate from chance alignments depends on a variety of factors, including angular separation, parallax and proper motion, their respective uncertainties, and the local source density. We use two complementary approaches to constrain the chance alignment rate for different subsets of the catalog. 

First, we repeat our binary selection procedure on the \texttt{GeDR3mock} catalog produced by \citet{Rybizki2020}. This catalog is built on a realization of the Besan\c{c}on model of the Milky Way \citep{Robin_2003} produced with \texttt{Galaxia} \citep{Sharma_2011}. It contains no binaries, so by construction all binary candidates selected from the mock catalog are chance alignments. It does, however, contain a realistic population of open clusters, a variety of stellar populations, an approximation of the {\it Gaia} eDR3 selection function, and realistic astrometric uncertainties. We repeat the full binary selection process described in Section~\ref{sec:selection} on the mock catalog, including filtering of clusters and resolved triples. We remove pairs with angular separations $\theta < 0.5$ arcsec by hand, because the {\it Gaia} eDR3 sensitivity drops precipitously at closer separations \citep{Fabricius2020}, and this is not accounted for in the mock catalog. The separation distribution of candidates selected from the mock catalog is shown in the left panel of Figure~\ref{fig:chance_align_estimate}.  

Second, we produce an empirical chance alignment sample based on the actual {\it Gaia} eDR3 catalog, following the method introduced by \citet{Lepine2007}. Prior to selecting potential binary companions to each star, we artificially shift it from its true position by $\approx 0.5$ degrees, increasing its reported RA by $0.5\sec(\delta)$ degrees. We then repeat the binary search, treating each star's shifted coordinates as its true coordinates when searching for possible companions. This process avoids selecting real binaries, since stars are shifted away from their true companions,
but preserves chance alignment statistics, because the source density within our 1 kpc search volume does not vary much on 0.5 degree scales. Copying and shifting the catalog effectively doubles the number of possible chance alignments, increasing the total number of pairs from $N(N+1)/2$ to $N(N+1)$, so we retain members of the shifted chance alignment catalog with 50\% probability. We again remove pairs with $\theta < 0.5$ arcsec. The separation distribution of chance alignments produced in this way is shown in the right panel of Figure~\ref{fig:chance_align_estimate}.

The two methods predict similar chance alignment rates.  At wide separations ($\log(s/{\rm au}) \gtrsim 4.5$ for the full catalog), the number of chance alignments is similar to the number of pairs in the binary candidates catalog. In this regime, most candidates are chance alignments. Not surprisingly, chance alignments begin to dominate at the separation where the binary candidate separation distribution begins to increase. The number of chance alignments per logarithmic separation interval increases steeply with separation: there are 100 times more chance alignments with $5 < \log(s/{\rm au}) < 5.1$ than there are with $4 < \log(s/{\rm au}) < 4.1$. This is a consequence of the larger available area for background stars to be found in at wider separations. At fixed distance, the area in which chance alignments can appear scales as $\sim 2\pi s\,{\rm d}s$. At wide separations, the chance alignment distributions thus scale as ${\rm d}N/{\rm d}\log s\sim s^{2}$. This scaling does not hold exactly, in part due to small-scale clustering and in part due to the less stringent parallax consistency required at small angular separations, but it provides a good approximation to the chance alignment rate at wide separations ($\log(s/{\rm au}) \gtrsim 4$).

We use the ``shifted'' chance alignment catalog in the rest of our analysis, because we find that it reproduces the separation distribution of binary candidates in the large-separation limit somewhat more reliably than the mock catalog.


\begin{figure*}
    \centering
    \includegraphics[width=\textwidth]{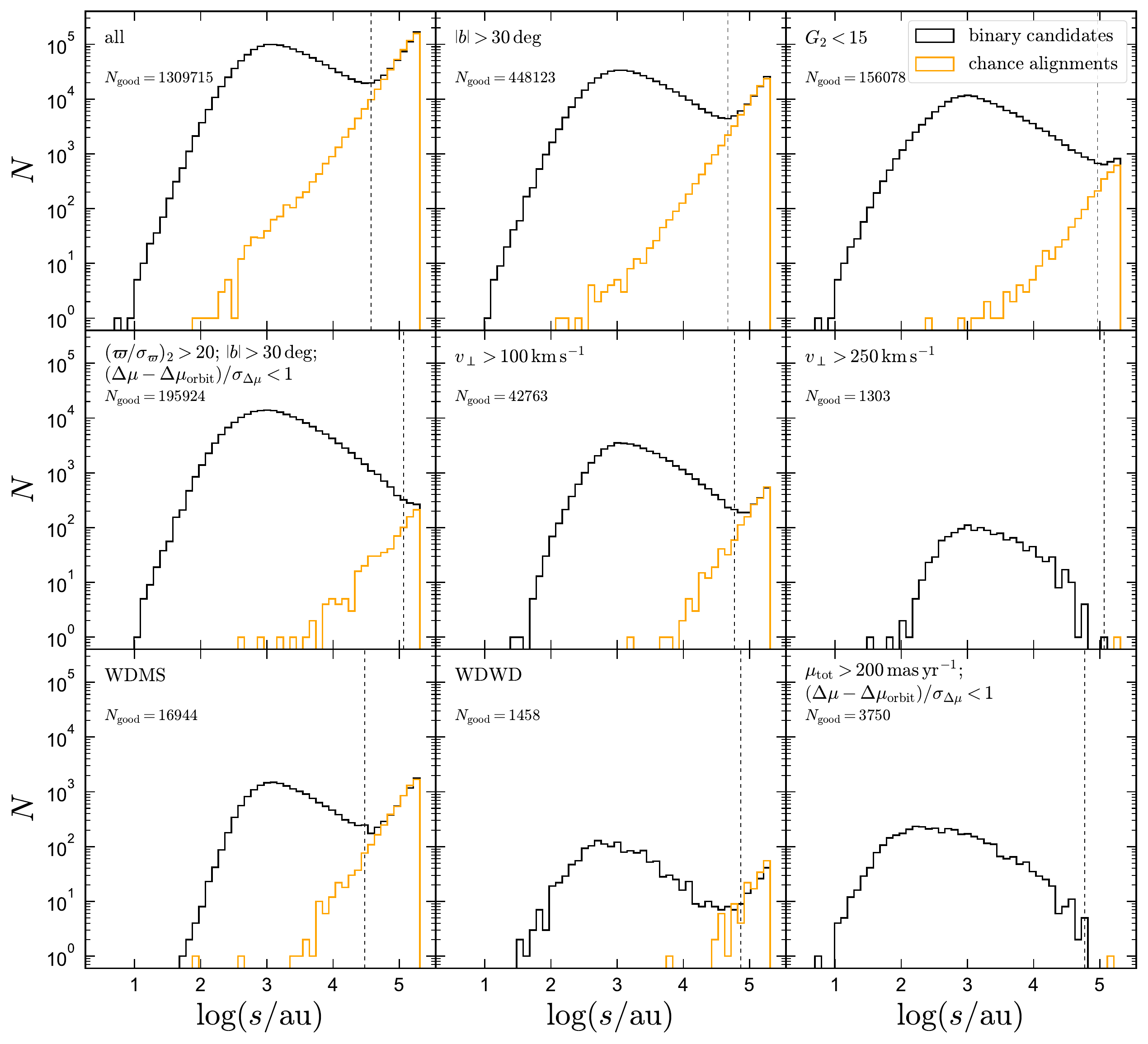}
    \caption{Contamination rate from chance alignments in various subsets of the catalog. Black histograms show binary candidates. Orange histograms show a catalog of chance alignments constructed by shifting stars 0.5 degrees when selecting candidate binaries. It contains no real binaries, but it has similar chance alignment statistics to the true binary candidates catalog. For the set of cuts illustrated in each panel, the vertical dashed line shows the widest separation at which there are more than $2\times$ the number of binary candidates as chance alignments. $N_{\rm good}$ is the number of binary candidates in that subset closer than this separation. Panels show the full catalog, high galactic latitudes, binaries with both components brighter than $G=15$, precise parallaxes at high Galactic latitudes, high tangential velocities (e.g. halo-like orbits), WD+MS and WD+WD binaries, and high-proper motion pairs.}
    \label{fig:subsets}
\end{figure*}

\subsection{Chance alignment rate for subsets of the catalog}
\label{sec:subsets}
Although chance alignments dominate the full sample at $s\gtrsim 30,000\,\rm au$, it is possible to select subsets of the catalog that are free of chance alignments out to wider separations. This is illustrated in Figure~\ref{fig:subsets}, which shows the separation distributions of various subsets of the binary candidate and the shifted chance alignment catalogs. As expected, the chance alignment rate at fixed separation is lower for samples with high galactic latitude, bright component stars, small fractional parallax errors, large space velocities, or large proper motions. Nevertheless, it is challenging to select any subset that remains pure beyond $s\sim 10^5$\,au ($\sim 0.5$\,pc) without also dramatically reducing the sample size. We make the catalog of shifted chance alignments publicly available in order to facilitate estimation of the chance alignment rate in various subsamples of the catalog.

\subsection{Estimating chance alignment probabilities}
\label{sec:chance_alignment_prob}
As illustrated in Figure~\ref{fig:subsets}, chance alignments and true binaries are found in different, but overlapping, regions of parameter space. We estimate the probability that a particular binary candidate is bound by comparing, at its location in parameter space, the local density of binary candidates and that of chance alignments from the shifted catalog. The process is described in detail in Appendix~\ref{sec:kde}. The ``densities'' are evaluated in a seven-dimensional space using a Gaussian kernel density estimate (KDE). The dimensions are (1) angular separation, (2) distance,  (3) parallax difference uncertainty, (4) local sky density, (5) tangential velocity, (6) parallax difference over error, and (7) proper motion difference over error. We re-scale these quantities to all have similar, order-unity dynamic range before fitting KDEs to both the binary candidate and the chance alignment distributions.

\begin{figure*}
    \centering
    \includegraphics[width=\textwidth]{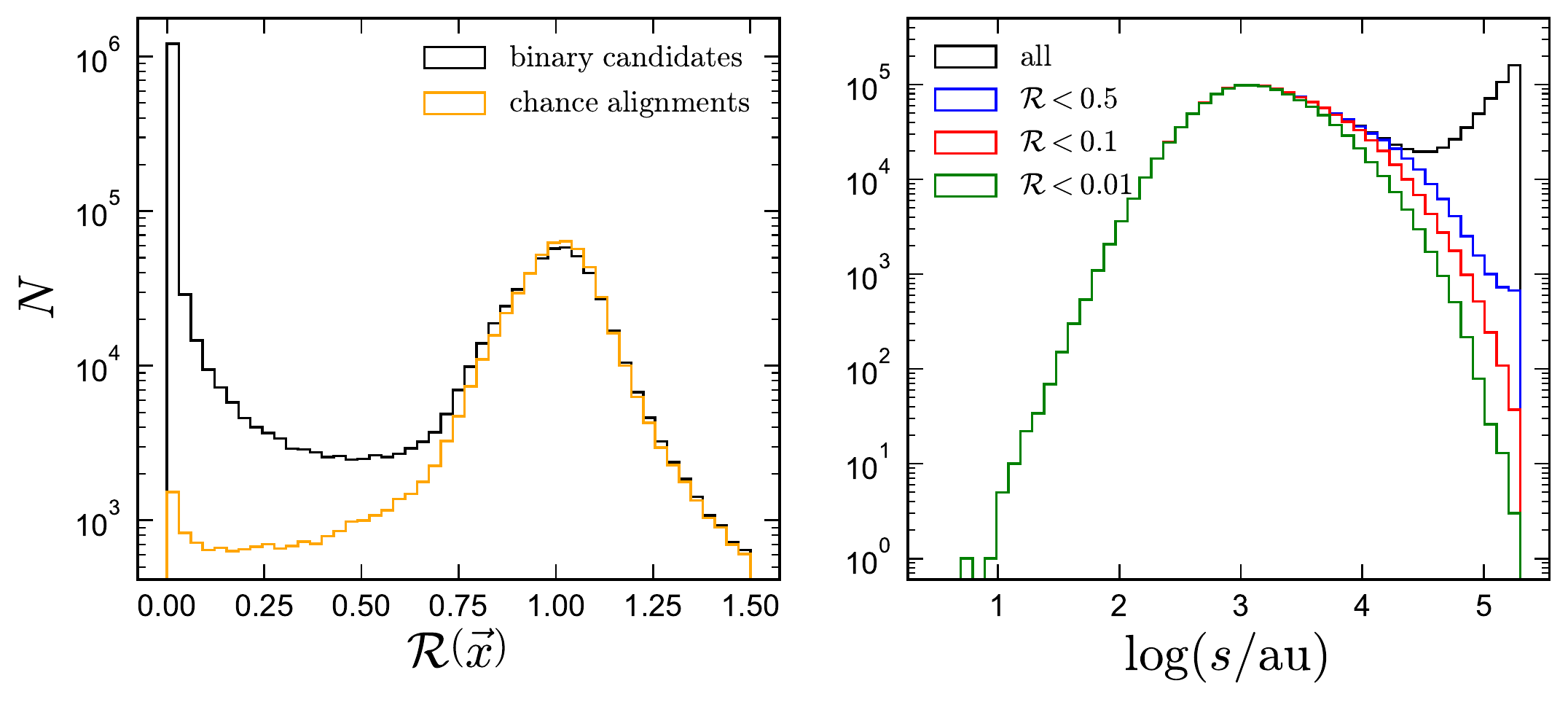}
    \caption{Left: ratio of the local density of chance alignments to binary candidates (Equation~\ref{eq:ratio}), for all pairs in the binary candidate sample (black) and the shifted chance alignment sample (orange). Likely chance alignments have $\mathcal{R} \sim 1$; objects with high probability of being bound have $\mathcal{R}\sim 0$. Right: separation distribution of binary candidates below different $\mathcal{R}$ thresholds. The chance-alignment rate increases at wide separations, so the separation distribution of low-$\mathcal{R}$ candidates falls off more steeply for progressively smaller $\mathcal{R}$ thresholds.}
    \label{fig:hist_ratios}
\end{figure*}

We denote the KDE-estimated density of chance alignments at a point $\vec{x}$ in the 7-dimensional parameter space as $\mathcal{N}_{{\rm chance\,align}}\left(\vec{x}\right)$, and that of binary candidates as $\mathcal{N}_{{\rm candidates}}\left(\vec{x}\right)$. The latter quantity is expected to be the sum of the chance-alignment and true binary densities.  We then calculate the ratio of these two quantities, 
\begin{equation}
    \mathcal{R}\left(\vec{x}\right)=\mathcal{N}_{{\rm chance\,align}}\left(\vec{x}\right)/\mathcal{N}_{{\rm candidates}}\left(\vec{x}\right).
    \label{eq:ratio}
\end{equation}
This ratio approximately represents the probability that a binary candidate at position $\vec{x}$ is a chance alignment, so selecting only candidates with small $\mathcal{R}$ is an efficient method for eliminating chance alignments. $\mathcal{R}$ is not strictly a probability -- for example, it is not strictly less than one (Figure~\ref{fig:hist_ratios}) -- but it is a serviceable approximation for one. We calculate $\mathcal{R}$ values for all members of the binary candidate and chance alignment catalogs.

Figure~\ref{fig:hist_ratios} (left) shows the distribution of $\mathcal{R}$ values for both catalogs. There is a narrow population of binary candidates with $\mathcal{R}$ near zero; these are objects that have a high probability of being bound. There is a second population of candidates with $\mathcal{R}\sim 1$; these objects are likely chance alignments. The separation distributions of binary candidates with $\mathcal{R}$ below several thresholds are shown in the right panel of Figure~\ref{fig:hist_ratios}. As expected, the vast majority of binary candidates with close separations ($\log(s/{\rm au})\lesssim 4$) have low $\mathcal{R}$ values, indicating a high bound probability. At wider separations, the separation distribution of high-probability binaries falls off precipitously. We emphasize that this drop-off is steeper than that of the separation distribution of all binaries, because for a low $\mathcal{R}$ threshold, more true binaries will be excluded from the high-confidence sample than chance alignments will be included in it.

\begin{figure}
    \centering
    \includegraphics[width=\columnwidth]{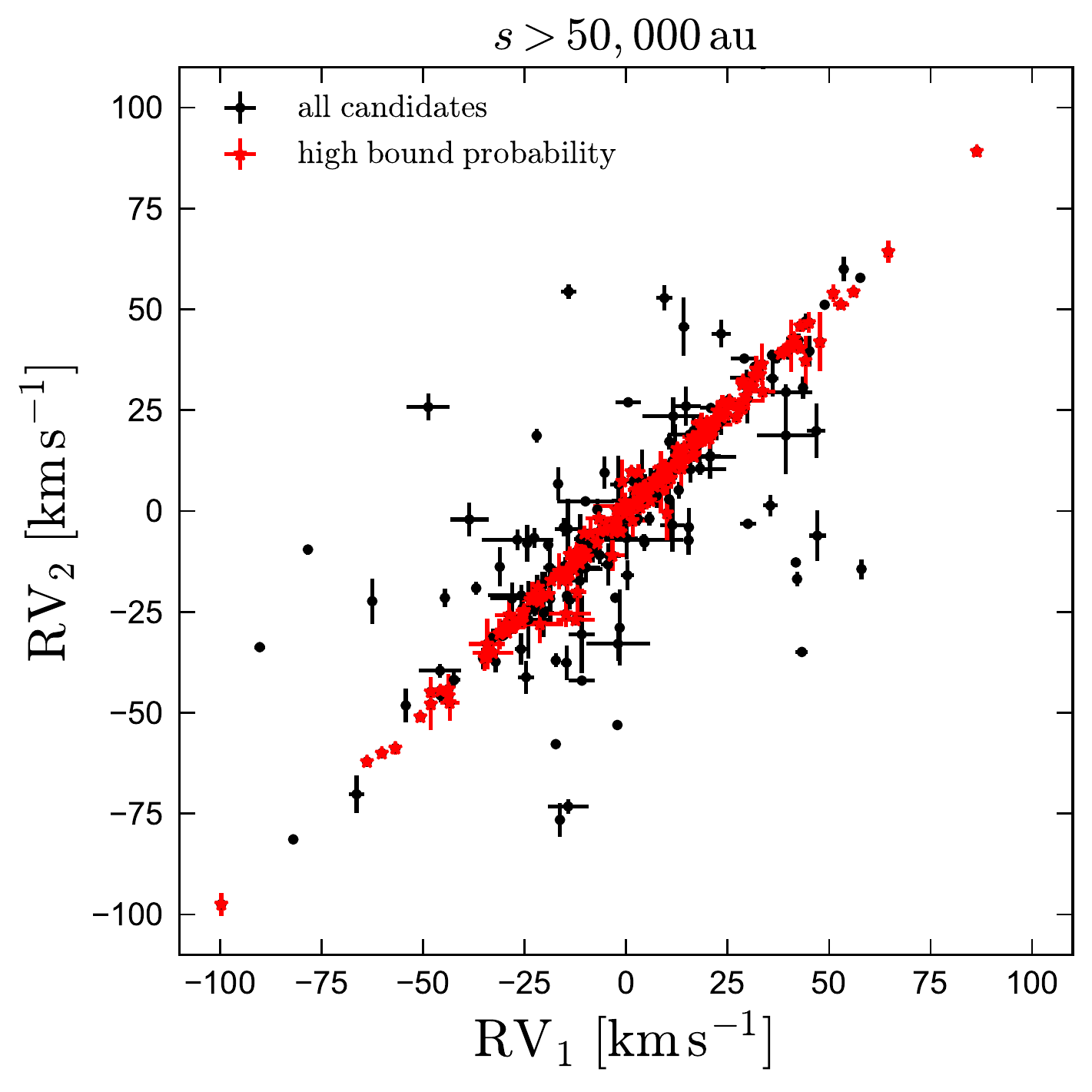}
    \caption{{\it Gaia}  RVS velocities of the components of binary candidates wider than 50,000 au. These are not used in constructing the catalog or calculating bound probabilities, but are useful for validation. Black points show all candidate binaries for which both components have $\sigma_{\rm RV} <10\,\rm km\,s^{-1}$; red points show the subset of these with $\mathcal{R}<0.1$, corresponding approximately to a 90\% bound probability. These all fall close to the one-to-one line,  suggesting that they are indeed bound. }
    \label{fig:rvs}
\end{figure}

We validate the use of $\mathcal{R}$ as a proxy for the chance-alignment probability in Figure~\ref{fig:rvs}, which compares the {\it Gaia} DR2 radial velocities \citep{Sartoretti_2018, Katz_2018} of the two components of candidates in which both stars have measured RVs.  We only plot candidates in which both components have $\sigma_{\rm RV} < 10\,\rm km\,s^{-1}$ and the separation is wider than $ 50,000$\,au, where the full catalog is dominated by chance alignments. One expects the RVs of the two components to be similar for genuine wide binaries. For chance alignments, the RVs of the two components should be drawn from a broad distribution with width comparable to the local velocity dispersion of the Galactic disk, and thus they will generally be inconsistent. For the full sample, there are indeed plenty of pairs with obviously inconsistent RVs, as chance alignments dominate at wide separations, even for bright pairs (e.g. Figure~\ref{fig:subsets}).  Red points in Figure~\ref{fig:rvs} show binaries with $\mathcal{R} < 0.1$. As expected, these pairs all have RVs close to the one-to-one line and are likely all bound. We stress that RVs are not used in creating the catalog or calculating $\mathcal{R}$ values, so this result bolsters our confidence in the chance alignment ratios calculated from the shifted catalog. 

We note that only bright stars ($G\lesssim 13.5$) had RVs published in {\it Gaia} DR2 (with no new RVs added in eDR3), so the fraction of chance alignments among binaries where both components have RVs is lower than in the full sample at the same separation. RVs for fainter stars are compared in Section~\ref{sec:lamost}, where we take RVs from the LAMOST survey.

In the rest of the paper, we define the ``high bound probability'' or ``high confidence'' subset of the catalog as the subset with $\mathcal{R} < 0.1$; this corresponds approximately to $>90\%$ probability of being bound. This does {\it not} mean that 10\% of the pairs in this subset are chance alignments: most candidates in it have $\mathcal{R}\ll 0.1$ (Figure~\ref{fig:hist_ratios}). Interpreting $\mathcal{R}$ as the probability that a given pair is a chance alignment, we estimate that 4,600 of the 1.26 million candidates with $\mathcal{R} < 0.1$ are chance alignments (0.4\%). For $\mathcal{R} < 0.01$, the same fraction is 870 out of  1.15 million (0.08\%). 
We make the full candidate catalog and $\mathcal{R}$ values available, including pairs that are likely chance alignments.  

\begin{figure*}
    \centering
    \includegraphics[width=\textwidth]{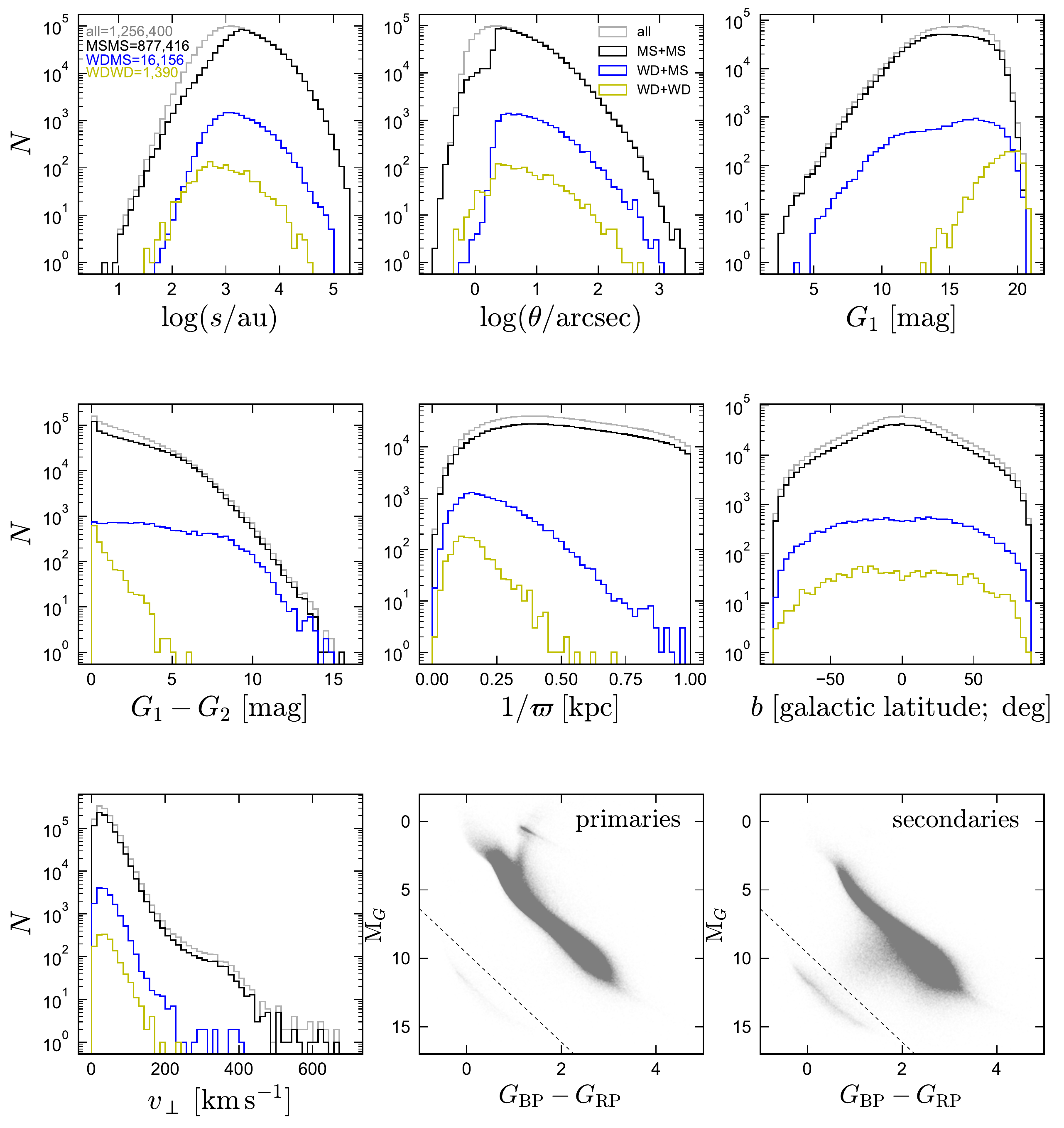}
    \caption{Basic properties of the high-bound probability binary sample: physical and angular separation, apparent magnitude of the primary, magnitude difference, distance, galactic latitude, plane-of-the-sky velocity, and color-magnitude diagrams. We designate stars as white dwarfs (WDs) or main sequence (MS; including giants) based on whether they fall below or above the dashed line in the bottom right panels. Black, blue, and yellow histograms show properties of MS+MS, WD+MS, and WD+WD binaries. We only classify binaries as MS+MS, WD+MS, or WD+WD if both components have a $G_{\rm BP}-G_{\rm RP}$ color. A majority of binaries with separation $\theta < 2$ arcsec do not have {\it Gaia} colors for a least one component and are therefore not classified; this accounts for most of the difference between the black and gray histograms (see also Table~\ref{tab:numbers}). }
    \label{fig:basic_properties}
\end{figure*}

\section{Basic properties of the catalog}
\label{sec:basic_prop}
Basic properties of the binary catalog are shown in Figure~\ref{fig:basic_properties} and listed in Table~\ref{tab:numbers}. Only the high-confidence pairs ($\mathcal{R}<0.1$) are shown. Following~\citetalias{Elbadry2018}, we classify stars as white dwarfs (WDs) or main-sequence (MS) based on their position in the {\it Gaia} color-absolute magnitude diagram (CMD): defining $M_{G}=G+5\log\left(\varpi/100\right)$, we classify as WDs objects with $M_{G}>3.25\left(G_{{\rm BP}}-G_{{\rm RP}}\right)+9.625$; all other stars with measured $G_{\rm BP}-G_{\rm RP}$ colors are classified as MS stars. Under the ansatz that the two components have the same distance, we use the (usually more precise) parallax of the primary, (the brighter star) for both components when calculating $M_{G}$.

\begin{figure*}
    \centering
    \includegraphics[width=\textwidth]{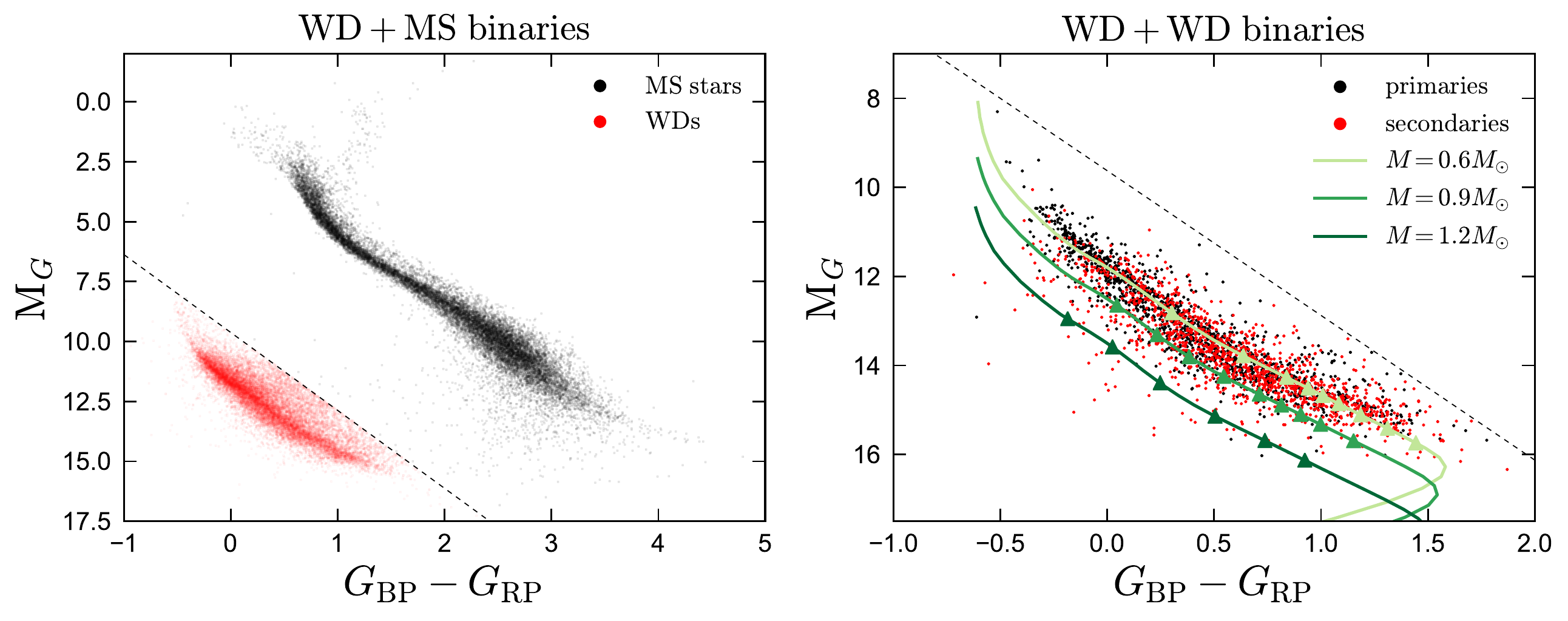}
    \caption{Color--absolute magnitude diagrams of high-confidence WD+MS and WD+WD binaries. Left: black points show MS components, which are usually the primaries; WD components are shown in red. Right: primary and secondary WDs are shown in black and red. We overplot cooling models for hydrogen-atmosphere WDs with masses between 0.6 and 1.2 $M_{\odot}$. Symbols mark 1 Gyr intervals of cooling age, with the first (leftmost) symbol at 1 Gyr.}
    \label{fig:wd_cmds}
\end{figure*}

The adopted WD/MS boundary in the CMD is shown with a dashed line in Figures~\ref{fig:basic_properties} and~\ref{fig:wd_cmds}. The boundary is not entirely unambiguous -- particularly for the WD+MS binaries, there are a few objects near the boundary that may be misclassified -- but a majority of objects do fall clearly on the WD or MS sequences. We note that the ``MS'' classification serves only to exclude WDs. The CMDs in Figure~\ref{fig:basic_properties} show that while most non-WD stars are indeed on the main sequence, the ``MS'' class also includes some giants, subgiants, pre-main sequence stars, and brown dwarfs. The number of binary candidates and high-confidence binaries in each class is summarized in Table~\ref{tab:numbers}.

\begin{table}
\begin{tabular}{lrrl}
Classification & $N_{\rm candidates}$ & $N_{\mathcal{R}<0.1}$ & Description \\
\hline
        MSMS       &    1,412,903    &   877,416    &      both MS       \\
        WDMS       &      22,563    &    16,156   &        one WD, one MS \\
        WDWD       &      1,565   &      1,390  &           both WD  \\
        MS??       &      378,877  &     360,180  &     one MS, one no colors        \\
        WD??       &     646    &    547   &        one WD, one no colors     \\
        ????       &     1,040   &    711   &         both no colors    \\
\hline 
        Total      & 1,817,594   &   1,256,400 & \\
        \hline
\end{tabular}
\caption{\label{tab:numbers} Contents of the binary candidate catalog. All stars with measured $G_{\rm BP}-G_{\rm RP}$ colors are classified as ``MS'' or ``WD'' depending whether they fall above or below the dashed lines in Figure~\ref{fig:wd_cmds}. $N_{\rm candidates}$ is the number of candidate binaries with separations up to 1 pc (black histogram in Figure~\ref{fig:sep_dist_cleaning}); $N_{\mathcal{R}<0.1}$ is the number with high bound probability (red histogram in Figure~\ref{fig:sep_dist_cleaning}).}
\end{table}

The data acquisition window for BP/RP spectra is $2.1 \times 3.5$ arcsec wide, preventing colors from being measured for most close pairs \citep{Arenou2018}. The majority of sources with a comparably bright companion within 2 arcsec thus do not have measured $G_{\rm BP} - G_{\rm RP}$ colors and cannot be classified as WD or MS stars based on {\it Gaia} data alone. Components lacking a color measurement are denoted ``??'' in Table~\ref{tab:numbers} and in the catalog. About 30\% of all high-confidence binaries have one component with unknown color; 0.05\% lack colors for both components. In 98\% of cases where only one component has a color, it is the brighter component. In many cases where no {\it Gaia} color is available, colors from other surveys (e.g. Pan-STARRS) should be sufficient to distinguish WD and MS components.

The catalog contains  $8.8\times 10^5$ high-confidence MS+MS binaries, more than 16,000 high-confidence WD+MS binaries, and 1,390 high-confidence WD+WD binaries.  Angular separations range from 0.2 arcsec to one degree. The peak of the angular separation distribution is at 1.2 arcsec. This is simply a result of the {\it Gaia} eDR3 angular resolution, since the intrinsic separation distribution falls off monotonically with increasing separation over all separations that are well-represented in the catalog \citep[e.g.][]{Duchene_2013}. There are 271 pairs with separations between 0.2 and 0.4 arcsec, including 24 below 0.3 arcsec. \citet[][their Figure 7]{Fabricius2020} found some indication that the {\it Gaia} eDR3 catalog may contain spurious duplicated sources at separations below 0.4 arcsec (that is, single sources that were erroneously classified as two sources), so it is possible that a small fraction of the closest pairs in the catalog are spurious. We do not, however, find any increase in the angular separation distribution at close separations, as might be expected to arise from a population of  duplicated sources.

The median magnitude of high-confidence primaries is $G=15.2$, and that of secondaries is $G=17.7$. Most WDs in the catalog are significantly fainter: the median magnitudes of primaries and secondaries in the WD+WD sample are 19.1 and 19.8. For WD+MS binaries, the same values are 15.5 for the primaries (in most cases, the MS star) and 19.4 for the secondaries. The median parallax of the full high-confidence sample is 2.05 mas ($1/\varpi \approx 485\,\rm pc$); the median distances for WD+WD and WD+MS binaries are 148 and 212 parsecs. Because of their closer distances, the WD+WD and WD+MS binaries are distributed roughly uniformly on the sky. The MS+MS sample, which extends to larger distances, bears clear imprint of the stratification of the Galactic disk.  Most of the binaries in the sample are part of the kinematic ``disk'' population, with a median tangential velocity $v_{\perp} = 4.74\,\rm km\,s^{-1}\times (\mu_{\rm tot}/{\rm mas\,yr^{-1}})(\varpi / \rm mas )^{-1} \approx 35\,\rm km\,s^{-1}$. There is also evidence of a kinematic ``halo'' population with $v_{\perp} \gtrsim 200\,\rm km\,s^{-1}$ that contains a few thousand binaries. 

The WD+MS and WD+WD binaries are shown separately on the CMD in Figure~\ref{fig:wd_cmds}. On top of the WD+WD binaries, we plot WD cooling tracks for carbon-oxygen cores with hydrogen atmospheres \citep{Holberg_2006, Kowalski_2006, Tremblay_2011, Bergeron_2011, Bedard2020}. WDs cool as they age, moving from the upper left to the lower right of the CMD. Cooling ages are indicated with triangular symbols along the tracks, which mark intervals of 1 Gyr. The faintest WDs in the catalog have implied cooling ages of about 10 Gyr. 31 WD+WD binaries have one component that falls below the $1.2M_{\odot}$ cooling track; i.e., with a photometrically-implied mass $M>1.2 M_{\odot}$. For WD+MS binaries, 75 WDs fall below the $1.2M_{\odot}$ cooling track. Hydrogen-atmosphere cooling tracks are not appropriate for WDs with non-DA spectral types, so spectroscopic classification must be obtained before masses and ages of individual WDs can be inferred with high fidelity.  

The catalog also contains about 10,000 high-confidence binaries in which the primary is a giant (about half these giants are in the red clump), including about 130 giant-giant binaries. These are all quite bright (both components have $G < 11$). They, along with the WD+MS binaries and the $\approx 13,000$ binaries in which one component is a subgiant, can serve as useful calibrators for stellar ages. 

Massive stars are not  well-represented in the catalog; there are 351 high-confidence primaries with $G_{\rm BP}-G_{\rm RP} < 1$ and $M_{G} < 0$. This cut corresponds roughly to $M \gtrsim 3 M_{\odot}$, though extinction complicates the mapping between $M_G$ and mass. 75\% of the high-confidence MS+MS binaries have primaries with $M_G$ between 9.5 and 3.8, corresponding approximately to $0.4 < M_1/M_{\odot} < 1.3$. For secondaries, the 75\% range is $M_G = 11.3-6.4$, corresponding to $0.2 < M_2/M_{\odot} < 0.8$. The catalog also contains about 80 binaries in which one component is likely a brown dwarf. We identify these on the CMD as objects with $M_G > 16.5$ and $G-G_{\rm RP} > 1.3$; they are all within 80 pc of the Sun and therefore are also found in the binary catalog produced by \citet{Smart2020}.

A small but noticeable fraction of sources, particularly secondaries, are scattered below the main sequence in the CMD, between the WDs and MS stars (bottom right panel of Figure~\ref{fig:basic_properties}). The majority of these sources are separated from a brighter companion by only a few arcsec; the most likely explanation for their anomalous CMD position is thus that they are MS stars with contaminated BP/RP photometry. The majority of such sources can be filtered out using cuts on \texttt{bp\_rp\_excess\_factor} and \texttt{phot\_bp/rp\_n\_blended\_transits} \citep[e.g.][]{Riello2020}, but we refrain from employing such cuts since they also remove a significant fraction of sources with acceptable photometry and astrometry. Some sources below the main sequence may also be stars with spurious parallaxes or biased colors, and a few are likely real astrophysical sources, primarily cataclysmic variables and detached but unresolved WD+MS binaries \citep[e.g.][]{Abrahams2020, Belokurov2020}.

\begin{figure}
    \centering
    \includegraphics[width=\columnwidth]{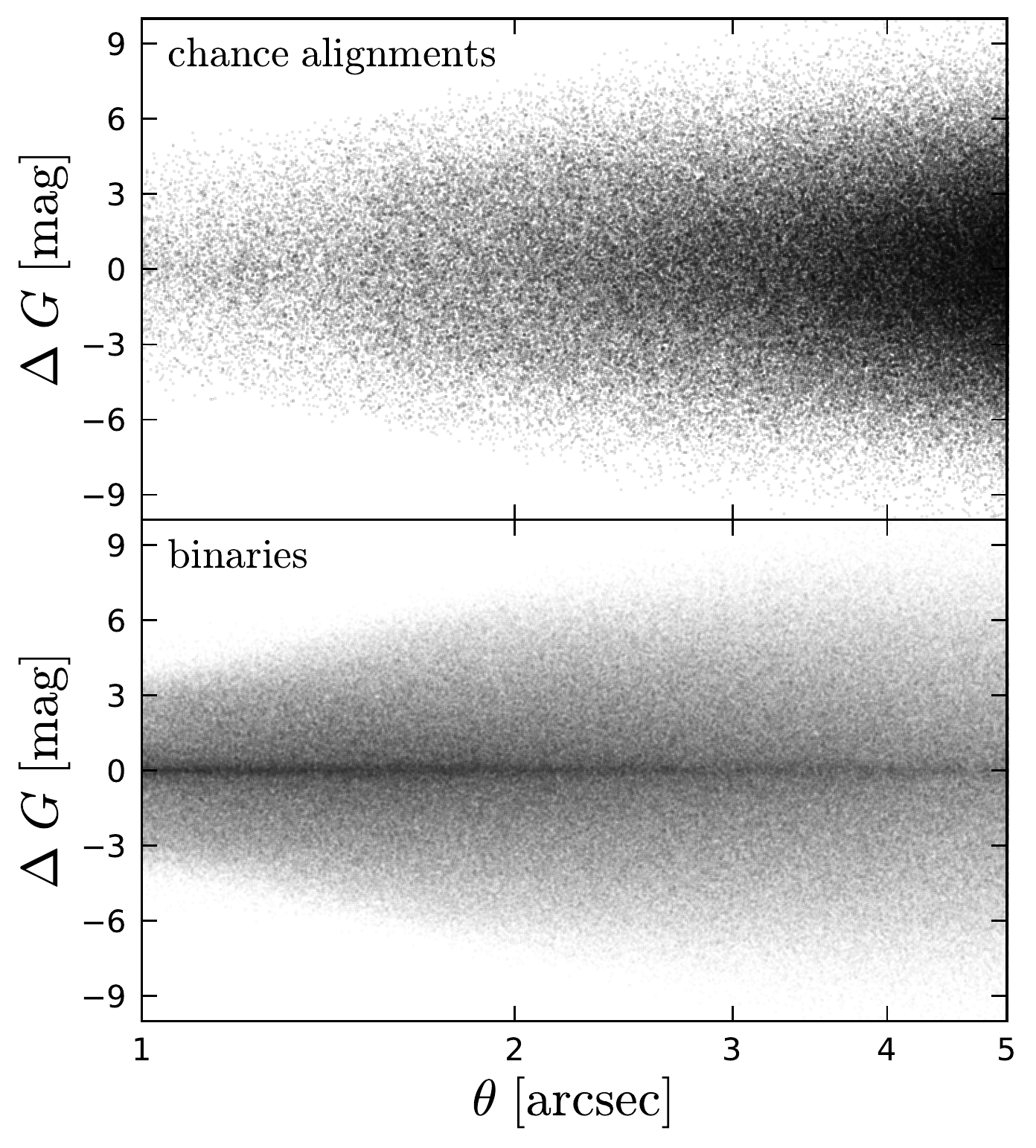}
    \caption{Magnitude difference between the two stars in chance alignments (top) and bound wide binaries (bottom) as a function of angular separation. The sign of $\Delta G$ is randomized. Chance alignments are selected following the same procedure used for real binaries, but with the requirement that the two stars' parallaxes be {\it inconsistent} rather than consistent. The chance alignments illustrate the contrast sensitivity of the {\it Gaia} eDR3} catalog: at close separations, pairs with large magnitude difference are not detected. In the bottom panel, a narrow excess population of binaries with $\Delta G\approx 0$ is visible, which is absent in the top panel. This highlights the excess population of equal-mass ``twin'' binaries.
    \label{fig:sensitivity}
\end{figure}

\subsection{Twin binaries and contrast sensitivity}
\label{sec:twins}
An excess population of equal-brightness (and presumably, equal-mass) ``twin'' binaries is also found in the catalog. Its existence is most obvious in the distribution of magnitude difference, $\Delta G$, as a function of separation, which is shown in Figure~\ref{fig:sensitivity} and compared to chance alignments. Unlike the chance alignment catalogs constructed from the mock catalog and shifted catalogs (Section~\ref{sec:chance_align}), these are selected in the same way as true binaries, but with the requirement that the parallaxes and proper motions of the two components be {\it inconsistent}. For easier visualization of the distribution of magnitude difference near $\Delta G = 0$, the sign of $\Delta G$ is randomized. The top panel illustrates the separation-dependent contrast sensitivity of the {\it Gaia} eDR3 catalog: at close angular separations, sources with significantly brighter companions are outshone. This leads to a contrast limit of $\Delta G \approx 4 \rm \,mag$ at $\theta = 1$ arcsec and  $\Delta G \approx 7 \rm \,mag$ at $\theta = 2$ arcsec. The contrast limit at a given separation is not ``sharp'', but is manifest as a smooth drop in sensitivity with increasing $\Delta G$ \citep[e.g.][]{Brandeker2019}.\footnote{At very close separations ($\theta \lesssim 0.7$ arcsec; not shown in Figure~\ref{fig:sensitivity}), the {\it Gaia} eDR3 catalog contains only equal-brightness pairs \citep[][their Figure 6]{Lindegren2020}. The contrast sensitivity is relatively smooth at $\theta > 1$ arcsec.} The contrast sensitivity is significantly improved in the binary catalog produced in this work compared to the one produced by \citetalias{Elbadry2018}: that work required both components to have relatively uncontaminated BP/RP colors and thus contained basically no binaries closer than 2 arcsec.

The bottom panel of Figure~\ref{fig:sensitivity} shows true binary candidates, in which the two components {\it do} have consistent parallaxes and proper motions. Unlike with the chance alignments, here there is an narrow excess population with $\Delta G \approx 0$. The extent and provenance of this population was studied by \citet{Elbadry2019twin}; here we simply note that it is also clearly apparent in our catalog. Because the twin excess is most prominent at close physical separations, it is somewhat more obvious in our catalog, which extends to closer angular and physical separations. 

\subsection{Comparison to other catalogs}
\label{sec:comparison}
\begin{figure*}
    \centering
    \includegraphics[width=\textwidth]{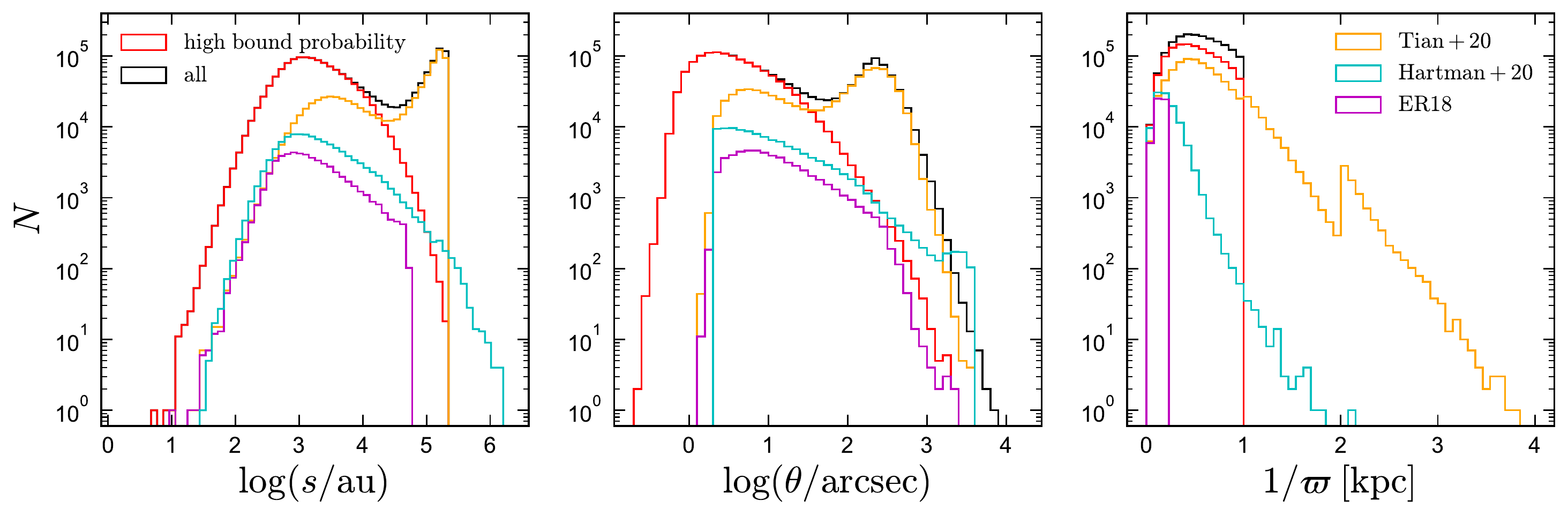}
    \caption{Comparison of the catalog produced in this work (black and red histograms) to the wide binary catalogs published by \citetalias{Elbadry2018}, \citet{Tian2020}, and \citet{Hartman2020}, all of which are based on {\it Gaia} DR2 data. This catalog expands the one produced by \citet{Tian2020} by a factor of $\sim 4$. The full catalog (black) has lower purity than the \citet{Elbadry2018} and \citet{Hartman2020} catalogs at wide separations. The subset with high bound probability (red) maintains high purity at wide separations, but does not purport to trace the intrinsic separation distribution there. }
    \label{fig:other_catalogs}
\end{figure*}

Figure~\ref{fig:other_catalogs} compares the distributions of projected physical separation, angular separation, and distance of the catalog produced here to other wide binary catalogs in the literature: 
\begin{itemize}
    \item The \citetalias{Elbadry2018} catalog (magenta) was produced from {\it Gaia} DR2 using a similar strategy to this work, but it was limited to binaries within 200 pc ($\varpi >5\,\rm mas$) and used more stringent cuts on both photometric and astrometric quality and SNR. In order to maintain high purity, it only contains binary candidates with $s < 50,000$ au. 
    \item \citet{Tian2020} expanded the \citetalias{Elbadry2018} search strategy to a larger volume (4 kpc) and used less stringent quality cuts, while still using {\it Gaia} DR2 data. Unlike this work, which uses a cut of $\varpi/\sigma_{\varpi} >5$ for both components, they required $\varpi/\sigma_{\varpi} >20$ for the primary and $\varpi/\sigma_{\varpi} >2$ for the secondary. This results in somewhat different contamination properties and completeness. They also searched out to $s=1$\,pc. In addition to to their full catalog of all candidates, \citet{Tian2020} published 3 smaller catalogs with high purity, which are not shown in Figure~\ref{fig:other_catalogs}.
    \item \citet{Hartman2020} did not use a strict distance cut, but limited their search to high-proper-motion pairs with $\mu > 40\,\rm mas\,yr^{-1}$. This preferentially selects nearby stars, since proper motion is inversely proportional to distance at fixed transverse velocity. For a ``typical'' tangential velocity of $v_{\perp}=35\,\rm km\,s^{-1}$, their proper motion cut corresponds to $d<185\,\rm pc$. However, stars on halo-like orbits with larger tangential velocities are included to larger distances; e.g., a binary with $v_{\perp} =200\,\rm km\,s^{-1}$ will have $\mu > 40\,\rm mas\,yr^{-1}$ out to a distance of 1.05 kpc. Rather than employing strict cuts on parallax and proper motion consistency, \citet{Hartman2020} used empirical estimates of the chance alignment rate as a function of position and proper motion difference from a shifted catalog (similar to our approach in Section~\ref{sec:chance_align}) to distinguish true binaries from chance alignments. Their approach has the advantage of not requiring specific cuts in parallax or proper motion difference, which are always somewhat arbitrary. A disadvantage is that it does not account for the heteroskedasticity of parallax and proper motion uncertainties -- i.e., the chance alignment probability is higher for pairs with large astrometric uncertainties, and this is not accounted for in their analysis. 
\end{itemize}

One difference between the catalog produced in this work and the other catalogs is obvious in  Figure~\ref{fig:other_catalogs}: our sample extends to smaller angular separations, and thus also physical separations. This is partly a result of the improved angular resolution of {\it Gaia} eDR3 (e.g., \citealt{Fabricius2020}) but is primarily due to a change in search strategy. \citetalias{Elbadry2018} and \citet{Tian2020} required both components of candidate binaries to have $G_{\rm BP}-G_{\rm RP}$ colors and to pass photometric quality cuts related to the \texttt{bp\_rp\_excess\_factor} reported in {\it Gaia} DR2 \citep{Evans_2018}. This set a soft resolution limit of $\sim 2$ arcsec, with a wider effective limit for pairs with large brightness contrast. We do not require colors or employ a photometric quality cut in this work; this adds an additional $\sim$400,000 binary candidates to the sample that would be excluded if we did (Table~\ref{tab:numbers}). 

Both the catalog produced in this work and the one from \citet{Tian2020} become dominated by chance alignments at $s\gtrsim 30,000$\,au. The ``high bound probability'' subset of our catalog does not, but its separation distribution falls off steeply at wide separations. This decline at wide separations is steeper than that of the intrinsic separation distribution, since a decreasing fraction of binaries at wide separations can be identified as bound with high confidence. This can be seen in comparing the red separation distribution to the magenta one from \citetalias{Elbadry2018}, which tracks the intrinsic separation distribution over $5000 < s/{\rm au} < 50,000$. The separation distribution from \citet{Hartman2020} has a similar logarithmic slope to the one from \citetalias{Elbadry2018} in this separation range and likely tracks the intrinsic separation distribution out to wider separations. At the very widest separations represented in that catalog (5+ pc, exceeding the local Jacobi radius and corresponding to an orbital period of about a Gyr), it is unlikely that pairs are actually bound. This may reflect the fact that the search strategy employed by \citet{Hartman2020} is sensitive to any pairs that are closer in phase space than chance alignments from the shifted catalog, without explicit consideration of the expected orbital velocities. That is, their search does not distinguish between bound binaries,  moving groups, or stellar streams. 

In terms of absolute numbers, the catalog represents a factor of $\sim 4$ increase in the number of high-confidence binaries over the one from \citet{Tian2020}. To our knowledge, it is the largest published  catalog of high-confidence binaries of any type. The sample could likely be expanded by a further factor of a few by loosening the distance and parallax uncertainty limits, or dropping the parallax cut entirely, while focusing on close angular separations \citep[e.g.][]{Dhital_2015}. However, the cuts we use in this paper provide a reasonable compromise between sample size, purity, and data quality.

A wide binary catalog based on {\it Gaia} eDR3 was also produced by \citet[][]{Smart2020}, which contains pairs within 100 pc. It is not shown in Figure~\ref{fig:other_catalogs}, but we find that within 100 pc, it is almost identical to ours.

\begin{figure}
    \centering
    \includegraphics[width=\columnwidth]{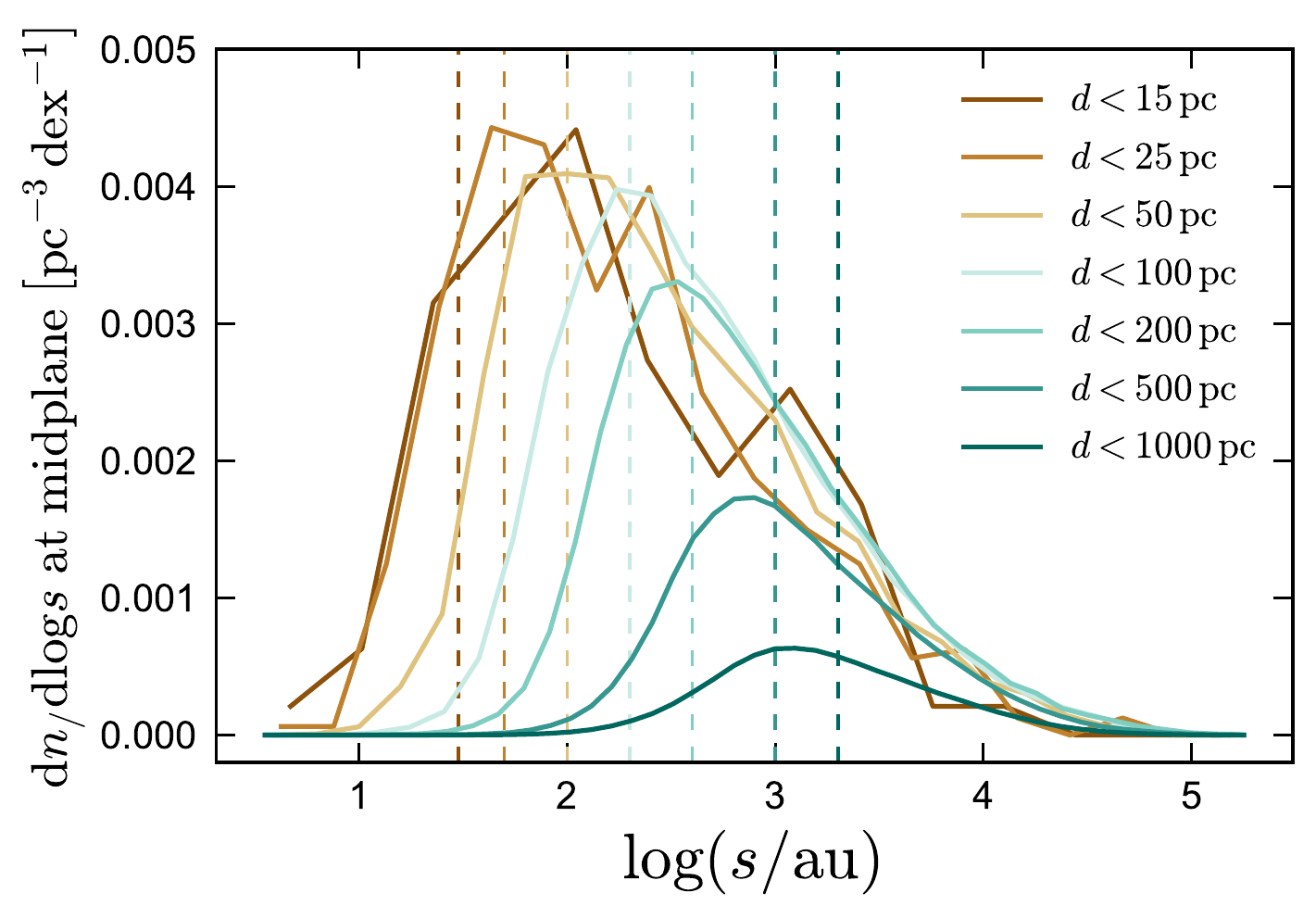}
    \caption{Midplane number density of binaries in different cumulative distance bins. Vertical lines mark $\theta=2$\,arcsec at the distance limit of each bin, roughly the angular separation at which incompleteness due to blending becomes significant. At all distances, the turnover at close separations is driven primarily by the resolution limit, but the 15 and 25 pc samples show a relatively flat separation distribution below a few hundred au. Beyond 200 pc, there is significant incompleteness due to stars being too faint to pass our fractional parallax error threshold -- i.e., the calculated number densities are lower than in the nearby samples even at wide separations, where blending should not be important.   }
    \label{fig:number_density}
\end{figure}

\subsection{Space density}
\label{sec:space_density}
Figure~\ref{fig:number_density} compares the separation distributions of binaries in different (cumulative) distance bins. We scale these by the effective stellar volume corresponding to the distance cut, i.e., by a factor proportional to the total number of stars expected in each distance sample. Were it not for the finite scale height of the Galactic disk, this factor would simply be the search volume $V=4\pi d_{\rm max}^3/3$, where $d_{\rm max}$ is the distance limit. We approximate the total stellar density within our search volume as a plane-parallel exponential distribution with the Sun at the midplane and a scale height $h_z =300\,\rm pc$ \citep{Juric2008}. We then define an effective volume $\tilde{V}$, which represents the number of stars in a sphere of radius $d_{\rm max}$ divided by the stellar density at the disk midplane:
\begin{align}
\tilde{V}	&=2\pi\int_{0}^{d_{{\rm max}}}e^{-z/h_{z}}\left(d_{{\rm max}}^{2}-z^{2}\right)dz\\
	&=2\pi\left[2h_{z}^{3}\left(e^{-d_{{\rm max}}/h_{z}}-1\right)+2h_{z}^{2}d_{{\rm max}}e^{-d_{{\rm max}}/h_{z}}+h_{z}d_{{\rm max}}^{2}\right].
	\label{eq:vtilde}
\end{align}
As expected, this expression asymptotes to $4\pi d_{\rm max}^3/3$ in the limit of $d_{\rm max} \ll h_z$. In Figure~\ref{fig:number_density}, the separation distributions for each value of $d_{\rm max}$ are divided by the appropriate value of $\tilde{V}$.
    
Dashed vertical lines in Figure~\ref{fig:number_density} mark a separation of $s_{\rm res\,\,limit}= \left(2\,{\rm arcsec}\right)\times d_{{\rm max}}$. At separations $s < s_{\rm res\,\,limit}$, incompleteness due to the {\it Gaia} eDR3 angular resolution starts to become severe. The figure shows that for $d_{\rm max} \lesssim 200\,$pc, incompleteness is due primarily to the angular resolution limit: at $s>s_{\rm res\,\,limit}$, the separation distributions in different distance bins overlap. However, for  $d_{\rm max} = 500\,$pc or 1000 pc, the catalog contains fewer binaries per effective volume than at closer distances, even at  $s>s_{\rm res\,\,limit}$. This reflects the fact that at sufficiently large distances, some binaries will have components that are too faint to pass the \texttt{parallax\_over\_error} > 5 limit, or to be detected at all. At $d=200\,\rm pc$, \texttt{parallax\_over\_error} > 5 implies $\sigma_{\varpi} < 1$\,mas. This is satisfied by most sources with $G \lesssim 20.5$ \citep{Lindegren2020}, corresponding to $M_G = 14.0$, near the bottom of the main sequence (e.g. Figure~\ref{fig:wd_cmds}). That is, the sample is expected to be almost complete for $d_{\rm max} = 200$\,pc, except for crowding/blending effects at close separations. On the other hand, at $d = 1\,\rm kpc$,  \texttt{parallax\_over\_error} > 5 implies $\sigma_{\varpi} < 0.2$\,mas, $G\lesssim 18.8$, and $M_G \lesssim 8.8$, meaning that most of the lower main sequence will be excluded. 

All of the separation distributions in Figure~\ref{fig:number_density} increase toward smaller separations at $s > s_{\rm res\,limit}$, but the distributions for $d_{\rm max} = 15$\,pc and $d_{\rm max} = 25$\,pc are similar and do appear to flatten above the resolution limit, at $s\sim 30\,\rm au$. For a typical binary in the catalog with total mass 1 $M_{\odot}$, this corresponds to a period of order 200 years, which is indeed near the peak of the approximately lognormal separation distribution for solar-type binaries \citep{Raghavan2010}. We caution that effects of astrometric acceleration also become important in this regime (Section~\ref{sec:accel}), potentially leading to spurious parallaxes and preventing pairs from being recognized as binaries by our search.

Integrating over separation, the distributions in Figure~\ref{fig:number_density} imply a total space density of $ (0.006\pm 0.001)$ wide binaries with $s>30$\,au per cubic parsec in the solar neighborhood. For context, the space density of all unresolved {\it Gaia} eDR3 sources in the solar neighborhood is $0.07\,\rm pc^{-3}$ \citep{Smart2020}, about 10 times higher. When all members of multiple systems are counted individually, the total stellar space density  in the solar neighborhood is about $0.10\,\rm pc^{-3}$ \citep[e.g.][]{Winters2020}. 

\subsection{Orbital velocities from proper motion differences}
\label{sec:triples}
Precise parallaxes and proper motions make it possible to estimate orbital velocities (projected onto the plane of the sky) from the proper motion difference between the two stars. The plane-of-the sky velocity difference $\Delta V$ can be calculated as

\begin{align}
    \Delta V=4.74\,{\rm km\,s^{-1}}\times\left(\frac{\Delta\mu}{{\rm mas\,yr^{-1}}}\right)\left(\frac{\varpi}{{\rm mas}}\right)^{-1}.
    \label{eq:deltaV}
\end{align}
Here $\Delta \mu$ is the scalar proper motion difference (Equation~\ref{eq:delta_mu}), and $\varpi$ is the parallax of the binary, for which we take the parallax of the brighter component. The corresponding uncertainty is 
\begin{align}
    \sigma_{\Delta V}=4.74\,{\rm km\,s^{-1}}\sqrt{\frac{\left(\Delta\mu\right)^{2}}{\varpi^{4}}\sigma_{\varpi}^{2}+\frac{\sigma_{\Delta\mu}^{2}}{\varpi^{2}}},
    \label{eq:sigma_deltav}
\end{align}
with $\sigma_{\Delta \mu}$ calculated from Equation~\ref{eq:sigma_delta_mu}, $\Delta \mu $ and $\sigma_{\Delta \mu}$ in $\rm mas\,yr^{-1}$, and $\varpi$ and $\sigma_{\varpi}$ in mas. We implicitly assume here that the two stars have the same parallax. Equation~\ref{eq:sigma_deltav} is almost always dominated by the first term under the radical; i.e., parallax errors dominate over proper motion errors. The median value of $\sigma_{\Delta V}$ for all high-confidence binaries in the catalog is $0.33\,\rm km\,s^{-1}$; 195,601 have $\sigma_{\Delta V} < 0.1\,\rm km\,s^{-1}$.

\begin{figure*}
    \centering
    \includegraphics[width=\textwidth]{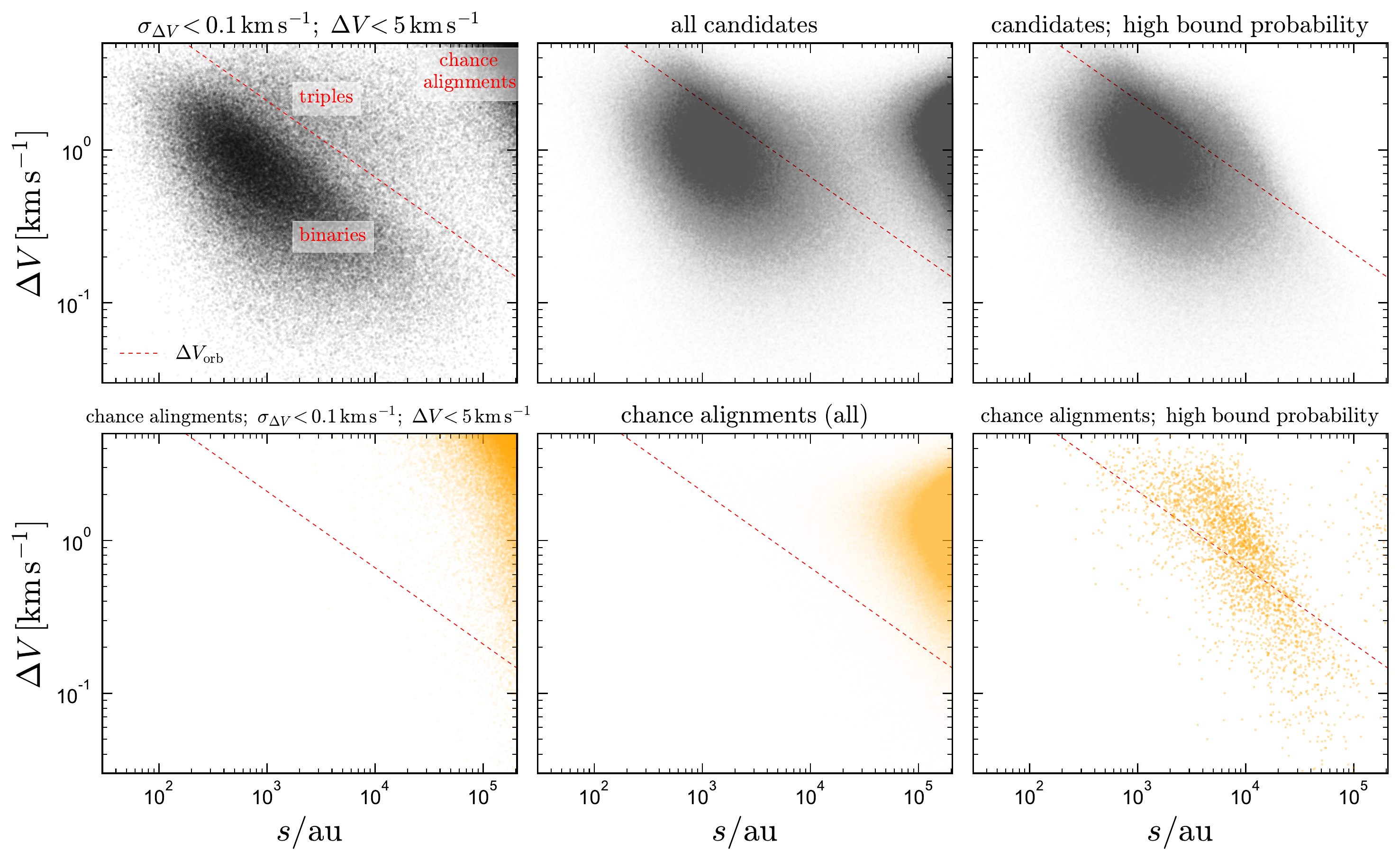}
    \caption{Upper left: plane-of-the sky velocity difference between the components of wide binary candidates selected with  $\Delta V < 5\,\rm km\,s^{-1}$ and $\sigma_{\Delta V} < 0.1\,\rm km\,s^{-1}$. Dashed red line shows the maximum expected velocity difference for a binary with total mass $5 M_{\odot}$ (Equation~\ref{eq:deltav_orb}). Binary candidates in our catalog  (top center) are required to be consistent, within 2 $\sigma$, with falling below this line, but pairs with large proper motion uncertainties can scatter well above it. The clouds of points at large separation and $\Delta V$ are chance alignments; bottom panels show chance alignments from the shifted catalogs for the same selection. Points at large separation and $\Delta V$ are excluded by the requirement of high bound probability ($\mathcal{R} < 0.1$; upper right). In the upper left panel, there is a population of triples and higher-order multiples above the dashed line, with no corresponding population in the shifted catalog (lower left). These have increased $\Delta V$ due to the gravitational effects of an unresolved close companion. They are generally gravitationally bound, but are excluded from our primary catalog.  }
    \label{fig:triples}
\end{figure*}

Figure~\ref{fig:triples} explores the $\Delta V$ values of binary candidates in the catalog, and the effects of the $\Delta V$ cuts we employ on its purity and completeness. The upper left panel shows all pairs from our initial query that have $s < 1\,\rm pc$, consistent parallaxes, $\Delta V <5\,\rm km\,s^{-1}$, and $\sigma_{\Delta V} < 0.1$. Note that the cut of $\Delta V <5\,\rm km\,s^{-1}$ is generally less strict than the one we adopt in constructing our primary catalog (Equation~\ref{eq:deltamu}), which is equivalent to $\Delta V<\Delta V_{{\rm orb}}+2\sigma_{\Delta V}$, with $\Delta V_{\rm orb}$ given by Equation~\ref{eq:deltav_orb}. 

\begin{figure}
    \centering
    \includegraphics[width=\columnwidth]{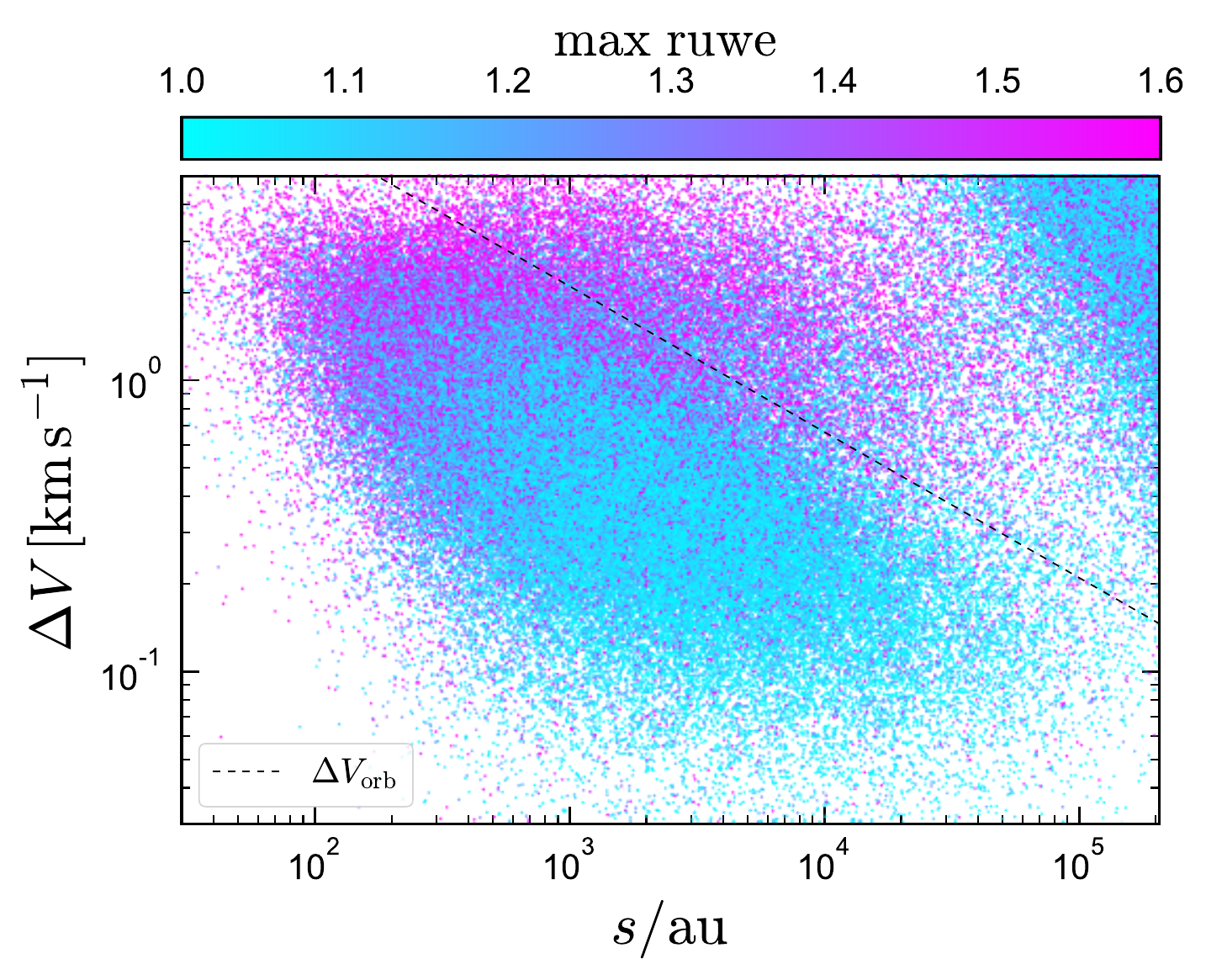}
    \caption{Binary candidates from the upper left panel of Figure~\ref{fig:triples}, now colored by the maximum \texttt{ruwe} of the two components. Most pairs with $\Delta V > \Delta V_{\rm orb}$ and $s \ll  10^5$\,au (those labeled ``triples'' in Figure~\ref{fig:triples}) have $\texttt{ruwe} \gtrsim 1.4$ for at least one component, in most cases because that component is an unresolved binary. At $s \lesssim 500$\,au, most pairs have large \texttt{ruwe} due to centroiding errors at close angular separations (see also Figure~\ref{fig:diagnostics}). }
    \label{fig:deltav_ruwe}
\end{figure}

In the upper left panel, there is a clear ridgeline of binaries with $\Delta V\sim s^{-1/2}$, as expected from Kepler's laws. This population largely falls below $\Delta V_{\rm orb}$ (dashed red line) because most of the binaries in the catalog have total masses less than $5 M_{\odot}$. There is not, however, a sharp drop-off at $\Delta V > \Delta V_{\rm orb}$. The population with $\Delta V > \Delta V_{\rm orb}$  likely consists primarily of triples and higher-order multiples, in which the plane-of-the-sky velocity of one component is affected by a close, unresolved companion \citep[e.g.][]{Clarke2020, Belokurov2020}. It is also possible that some of these pairs are moving groups that are not actually bound but remain close in phase space \citep[e.g.][]{Pittordis2019, Coronado2020}; however, we find that the population exists even among binaries on halo-like orbits, favoring multiplicity as the primary explanation for it. We also find that most pairs with $\Delta V > \Delta V_{\rm orb}$ have unusually large \texttt{ruwe}\footnote{\texttt{ruwe}, the re-normalised unit-weight error, is a measure of astrometric goodness-of-fit that corrects for global trends in the other reported goodness-of-fit indicators with magnitude and color. Values above about 1.4 indicate potential problems. } for at least one component (Figure~\ref{fig:deltav_ruwe}), suggesting that these components are unresolved binaries \citep[e.g.][]{Belokurov2020}.

A large fraction of these unresolved triples and higher-order multiples are excluded from our actual catalog by the requirement of $\Delta V<\Delta V_{{\rm orb}}+2\sigma_{\Delta V}$ rather than e.g., $\Delta V < 5\,\rm km\,s^{-1}$; this should be kept in mind when using the catalog for applications involving higher-order multiplicity. We do not use a constant $\Delta V$ cut (e.g. $\Delta V < 5\,\rm km\,s^{-1}$) in constructing the full catalog because this would result in a much higher contamination rate from chance alignments. For $5,000< s/{\rm au} < 10,000$, 22\% of pairs with $\Delta V < 5\,\rm km\,s^{-1}$ and $\sigma_{\Delta V} < 0.1\,\rm km\,s^{-1}$ are excluded from the catalog by Equation~\ref{eq:deltamu}. If we assume that large $\Delta V$ values are due mainly to subsystems, this implies that about 12\% of wide binary components in this separation range have an unresolved subsystem that imparts a large enough photocenter perturbation to significantly increase $\Delta V$. This corresponds to a subsystem separation range of $\sim 1-100$\,au, since photocenter wobbles will average out over the 34-month {\it Gaia} eDR3 baseline at closer separations, and the subsystem-induced $\Delta V$ will be small at wider separations. About 25\% of wide binary components have a subsystem in this separation range \citep{Tokovinin2002, Tokovinin2010}, so it is quite plausible that the large-$\Delta V$ pairs are mostly unresolved triples and higher-order multiples. 

The clouds of points at large separation and $\Delta V$ in Figure~\ref{fig:triples} are chance alignments. These are primarily pairs with large $\sigma_{\Delta V}$; otherwise they would be excluded by the requirement of $\Delta V<\Delta V_{{\rm orb}}+2\sigma_{\Delta V}$. The bottom panels show the distribution of chance alignments from the shifted catalog (Section~\ref{sec:chance_align}). These are distributed similarly to the large-separation cloud among binary candidates. No chance-alignment cloud is visible in the binary candidates with high bound probability (upper right), but a few chance alignments do scatter into the bound binary cloud, mostly at $\Delta V> \Delta V_{\rm orb}$ (lower right). 

We note that the interpretation of $\Delta V$ calculated from Equation~\ref{eq:deltaV} as a physical velocity difference between the two components of a binary breaks down at large angular separations, where projection effects become important \citep[e.g.][]{elbadry2019gravity}. Indeed, two stars in an ultra-wide binary can have {\it identical} space velocities but substantially different plane-of-the-sky proper motions. The magnitude of the apparent proper motion difference depends primarily on angular separation and is therefore largest for nearby binaries, which are also the binaries with the smallest $\sigma_{\Delta V}$. For the sample shown in the upper left panel of Figure~\ref{fig:triples}, projection effects become important beyond about 20,000 au, which is -- perhaps not coincidentally -- the separation beyond which the trend of $\Delta V \sim s^{-1/2}$ appears to flatten. These projections effects can be corrected if the RVs of at least one component are known \citep{elbadry2019gravity}.

\subsection{Cross-match with LAMOST}
\label{sec:lamost}

\begin{figure*}
    \centering
    \includegraphics[width=\textwidth]{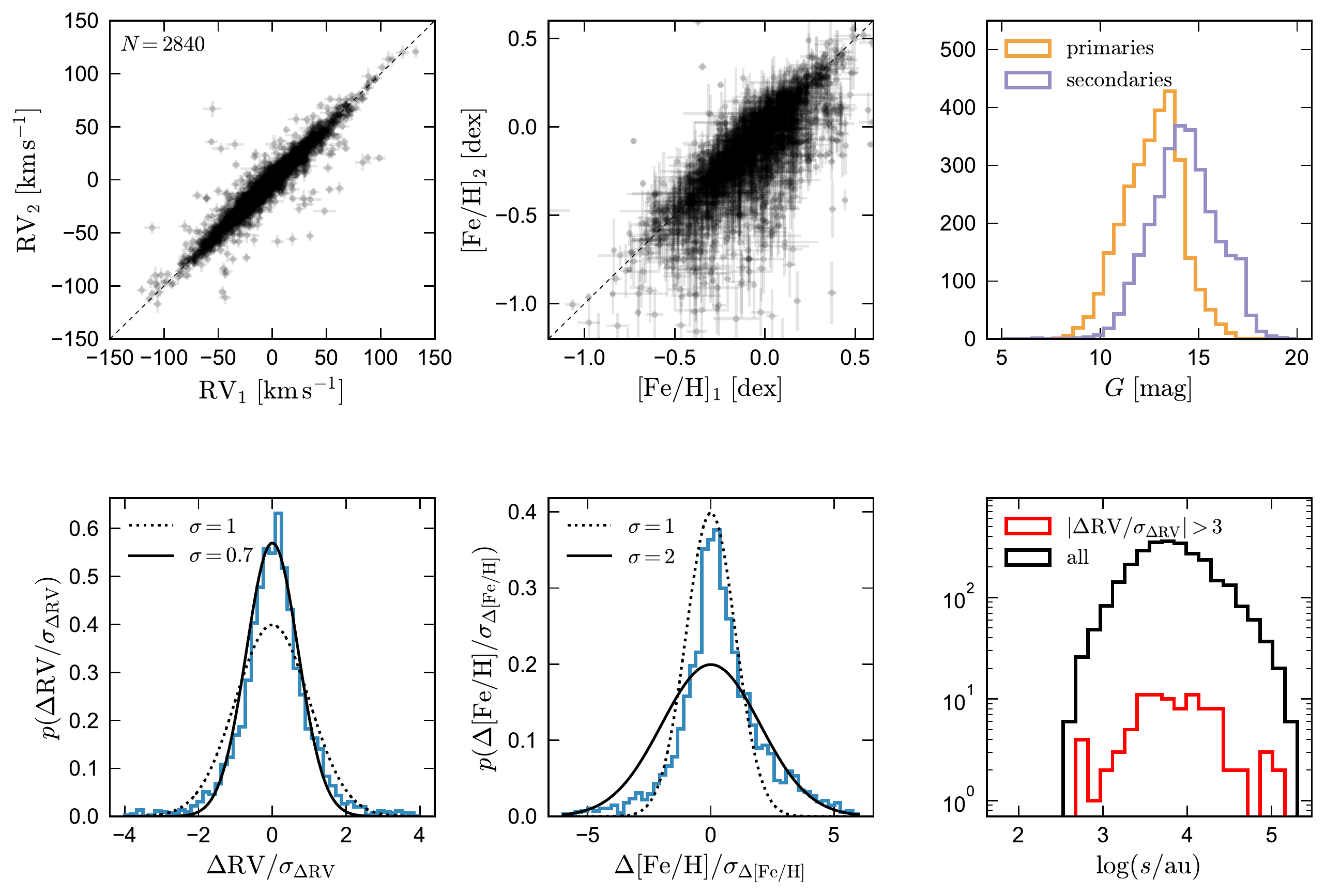}
    \caption{High-confidence binaries in which both components were observed by the LAMOST survey. We only include pairs with angular separation $\theta > 3$ arcsec to avoid blending. Top left and center: comparison of the RVs and metallicities ([\rm Fe/H]) of the primary ($x$-axis) and secondary ($y$-axis). In most binaries, the two components have very similar RV  and [Fe/H], as expected. There is evidence of a systematic bias toward lower [Fe/H] for fainter targets. Upper right: $G$ magnitudes of both components. This sample includes fainter stars than the {\it Gaia} RVS sample. Bottom left and center: distributions of uncertainty-normalized RV and [Fe/H] difference, compared to Gaussians with the listed $\sigma$. These suggest that the LAMOST RV uncertainties are overestimated by $\sim 30\%$ on average, while [Fe/H] uncertainties are underestimated and subject to temperature systematics. Bottom right: separation distributions. Pairs in which RVs are not consistent within $3\sigma$ are shown in red. If these pairs were chance-alignments, one would expect them to be concentrated at the widest separations. They are not, suggesting that they are either higher-order multiples with the RV of one component biased by an unresolved close companion, or pairs in which one component has a catastrophically wrong RV.  }
    \label{fig:lamost}
\end{figure*}

We cross-matched the binary catalog with the LAMOST survey \citep[][DR6 v2]{Cui2012}, the currently most extensive spectroscopic survey providing stellar parameters and abundances. We began with the LAMOST low-resolution  ``A, F, G and K Star'' catalog, which contains atmospheric parameters and metallicities for 5,773,552 spectra (including some duplicate observations). We cross-matched the catalog with {\it Gaia} eDR3 using the CDS Xmatch service,\footnote{http://cdsxmatch.u-strasbg.fr/} which uses {\it Gaia} proper motions to propagate source positions to epoch J2000. We matched each LAMOST observation to the nearest {\it Gaia} source within 1 arcsec. For sources with more than one LAMOST observation, we retained only the observation with the highest $g-$band SNR. This left us with LAMOST data for 4,306,131 sources, which we then matched to our catalog using {\it Gaia} source ids.  

This yielded 91,477 binaries in which at least one component has a LAMOST spectrum. This sample will be useful for a variety of applications, such as studying the dependence of the binary fraction on metallicity \citep[e.g.][]{Elbadry2019feh,  Hwang2020}. Here, we focus on a subset of the cross-match: those binaries in which {\it both} components have a LAMOST spectrum, and the angular separation is at least 3 arcseconds. The latter cut is to avoid cases where both stars fall inside a single fiber, leading to potentially biased stellar parameters and abundances \citep[e.g.][]{elbadry2018a}. The RVs of these binaries can be used to verify whether most binary candidates are bound. This test is similar to that shown in Figure~\ref{fig:rvs} with {\it Gaia} RVs, but LAMOST spectra extend to fainter magnitudes than RVs from {\it Gaia}, which are currently only available at $G\lesssim 13$.

Figure~\ref{fig:lamost} compares the LAMOSTs RVs and metallicities (i.e., [Fe/H]) of the components of these binaries. These are generally expected to be consistent for genuine wide binaries. The RVs for most binaries do indeed fall close to the one-to-one line, but there are some outliers: 91 of the 2840 binaries in the sample have RVs that are more than 3-sigma discrepant. A potential worry is that these pairs are not binaries at all, but chance alignments. To assess whether this is likely to be the case, we plot the separation distribution of the discrepant pairs in the bottom right panel of Figure~\ref{fig:lamost}. This shows that the separation distribution of pairs with large RV differences is similar to that of all pairs. Chance alignments are much more common at wide separations (Figure~\ref{fig:subsets}), so if chance alignments were the root of the discrepant RVs, one would expect these pairs to be clustered at large separations. The primary reason for the discrepant RVs is likely again higher-order multiplicity. This is likely to affect the LAMOST RVs more than it does those from {\it Gaia} (Figure~\ref{fig:rvs}), because the LAMOST RVs and their uncertainties are based on a single epoch. Unresolved short-period binaries are usually filtered out of the {\it Gaia} RV sample we consider, because $\sigma_{\rm RV}$ for that sample is calculated from the epoch-to-epoch RV dispersion, and we required $\sigma_{\rm RV} < 10\,\rm km\,s^{-1}$ for both components.

The lower left panel of Figure~\ref{fig:lamost} shows the distribution of uncertainty-normalized RV difference between the two components of binaries; i.e., $\Delta{\rm RV}/\sigma_{\Delta{\rm RV}}=\left({\rm RV_{1}-RV_{2}}\right)/\sqrt{\sigma_{{\rm RV,1}}^{2}+\sigma_{{\rm RV,2}}^{2}}$. The median $\sigma_{\Delta \rm RV}$ for this sample is $7\,\rm km\,s^{-1}$, which is larger than the typical $\sim 1\,\rm km\,s^{-1}$ orbital velocity for these binaries, so we expect the width of the main distribution to be dominated by measurement uncertainties, with the tails dominated by higher-order multiples. If the reported $\sigma_{\rm RV}$ values are accurate, $\Delta{\rm RV}/\sigma_{\Delta{\rm RV}}$ should be distributed as a Gaussian with $\sigma \approx 1$, with some outliers at higher velocity difference. The distribution is indeed approximately Gaussian, but it is narrower than $\sigma=1$; the bulk of the distribution is better approximated by $\sigma\approx 0.7$. This suggests that the LAMOST RV uncertainties are typically {\it overestimated} by $\sim 30\%$. 

The distribution of uncertainty-normalized [Fe/H] differences tells a different story. It is not well-described by a single Gaussian, but has a narrow component with $\sigma \lesssim 1$, and broad, asymmetric tails. This suggests that $\sigma_{\rm [Fe/H]}$ values are considerably underestimated for a significant fraction of the catalog. We find (not shown in the figure) that the distribution becomes narrower and more Gaussian when we only consider binaries in which the two components have similar magnitude and effective temperature. This suggests that the larger-than-expected metallicity differences are due primarily to temperature systematics in the abundance pipeline. This is particularly evident in the upper middle panel of Figure~\ref{fig:lamost}, which shows that the secondaries (which have lower $T_{\rm eff}$) systematically have lower reported [Fe/H]. It is of course possible that some binaries really have inconsistent surface abundances, but work with higher-quality spectra than those which underlie the LAMOST metallicities suggests such abundance anomalies are rare \citep{Hawkins2020}.

Many of the binaries in our catalog were also observed by other spectroscopic surveys. We defer analysis of these data to future work, but comment that the type of analysis shown in Figure~\ref{fig:lamost} will be useful in calibrating the abundances derived by surveys (and their uncertainties). For example, an earlier version of our binary catalog constructed from {\it Gaia} DR2 was recently fruitfully used by \citet{Buder2020} to assess the reliability of abundances derived by the GALAH survey.

\section{Calibrating {\it Gaia} DR3 parallax uncertainties}
\label{sec:parallax_uncert}

\begin{figure*}
    \centering
    \includegraphics[width=\textwidth]{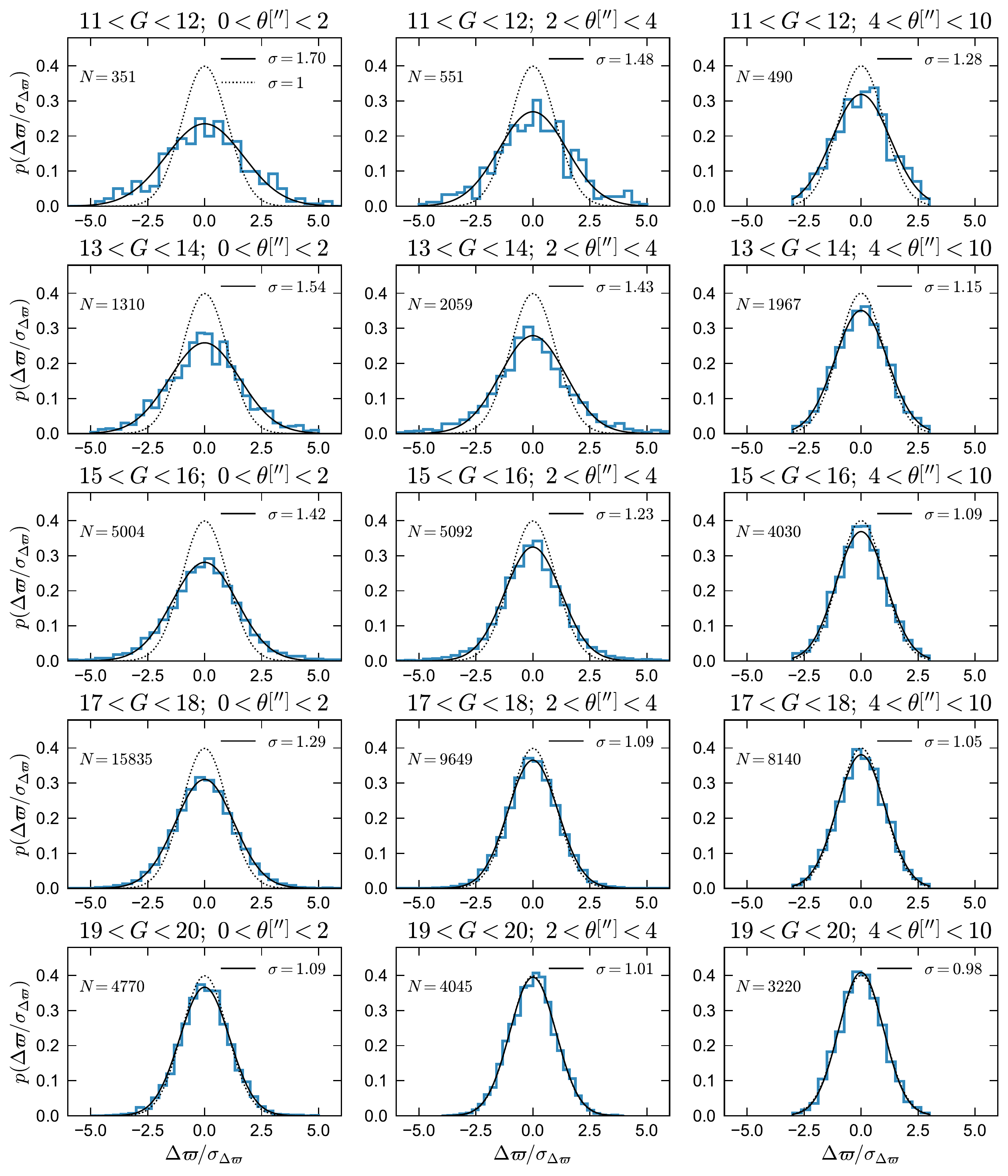}
    \caption{Distributions of uncertainty-normalized parallax difference between the two components of binaries, $\Delta \varpi/\sigma_{\Delta \varpi} = (\varpi_1 -\varpi_2)/\sqrt{\sigma_{\varpi,1}^2 + \sigma_{\varpi,2}^2}$. Each panel shows a different bin of angular separation and magnitude, with both components falling in the quoted magnitude range.  Because the two components of a binary have essentially the same distance, this quantity would be expected to follow a Gaussian distribution with $\sigma =1$ (dotted lines) if the formal parallax uncertainties were accurate. $\sigma > 1$ points toward underestimated parallax uncertainties. Solid black lines show Gaussian fits, with $\sigma$ noted in the legends. At all magnitudes, the best-fit $\sigma$ is larger at close separations, implying that $\sigma_{\varpi}$ is more severely underestimated for sources with nearby companions. At fixed separations, the fractional underestimate of $\sigma_{\varpi}$ is larger for bright stars. }
    \label{fig:histograms}
\end{figure*}

Because the two stars in a wide binary have very nearly the same distance, our catalog provides a straightforward method of validating the {\it Gaia} eDR3 parallax uncertainties. We do this by calculating the uncertainty-normalized parallax difference between the two components of each binary, $\Delta\varpi/\sigma_{\Delta\varpi}=\left(\varpi_{1}-\varpi_{2}\right)/\sqrt{\sigma_{\varpi,1}^{2}+\sigma_{\varpi,2}^{2}}$, just as we did for the LAMOST RVs. In the limit of accurate parallax uncertainties (and  small differences in the true distance to the two components), this quantity should be distributed as a Gaussian with $\sigma=1$. If the reported parallax uncertainties are underestimated, one expects a wider distribution, and possibly deviations from Gaussianity. 

Figure~\ref{fig:histograms} shows distributions of $\Delta\varpi/\sigma_{\Delta\varpi}$ for binaries at a range of angular separations and magnitudes. To isolate underestimated random errors (as opposed to systematic shifts due to e.g. variations in the parallax zeropoint), we only show binaries in which both components have magnitudes in the quoted range (corresponding to a magnitude difference of $\Delta G < 1$). This figure only shows high-confidence binaries ($\mathcal{R} < 0.1$) in which both components have \texttt{ruwe} < 1.4, indicating an apparently well-behaved astrometric solution. In this and all validation of the parallax errors, we exclude binaries that are wide and nearby enough that their physical size might measurably contribute to the parallax difference. Under the ansatz that the projected physical separation $s$ is comparable to the line-of-sight distance difference and is much smaller than the distance, we expect the true parallax difference between the two stars to scale as $\Delta\varpi_{{\rm true}}\sim1/d-1/\left(s+d\right)\approx s/d^{2}$, or 
\begin{equation}
    \Delta\varpi_{{\rm true}}=\frac{1}{206265}\,{\rm mas}\times\left(\frac{\theta}{{\rm arcsec}}\right)\left(\frac{\varpi}{{\rm mas}}\right).
    \label{eq:deltavarpi_true}
\end{equation}
This quantity is negligible compared to $\sigma_{\Delta \varpi}$ for most binaries in the catalog, but not for the nearest and widest binaries. We therefore exclude all binaries in which $\Delta\varpi_{{\rm true}}/\sigma_{\Delta \varpi} > 0.05$.

\begin{figure*}
    \centering
    \includegraphics[width=\textwidth]{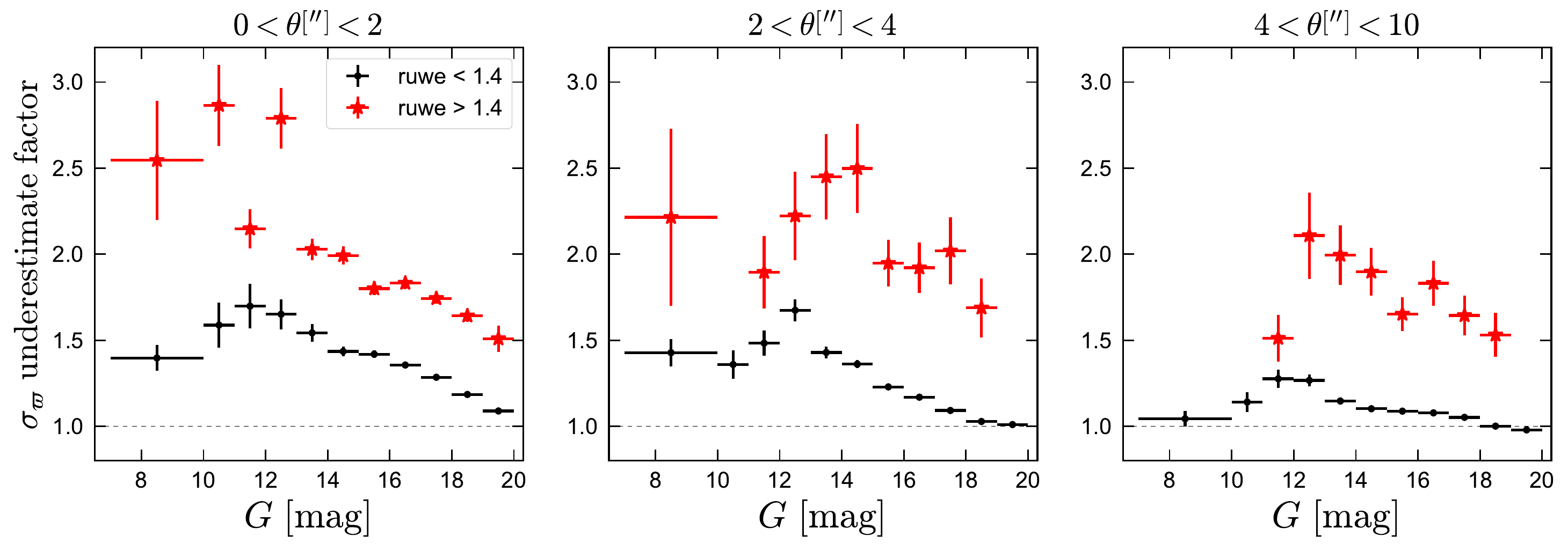}
    \caption{Fractional parallax uncertainty underestimate (i.e., the best-fit $\sigma$ in Figure~\ref{fig:histograms}) as a function of magnitude and separation. Black symbols show binaries in which both components have \texttt{ruwe} < 1.4; red symbols show those in which at least on component has \texttt{ruwe} > 1.4, indicating a potentially problematic astrometric solution. From left to right, panels show increasing angular separation. As expected, sources with \texttt{ruwe} > 1.4 have larger $\sigma_{\varpi}$ underestimates at all magnitudes and separations. Sources with \texttt{ruwe} < 1.4 have more severe underestimates of $\sigma_{\varpi}$ at close separations and for bright stars, particularly at $11 < G < 13$.}
    \label{fig:underestimate_ruwe}
\end{figure*}

\begin{figure*}
    \centering
    \includegraphics[width=\textwidth]{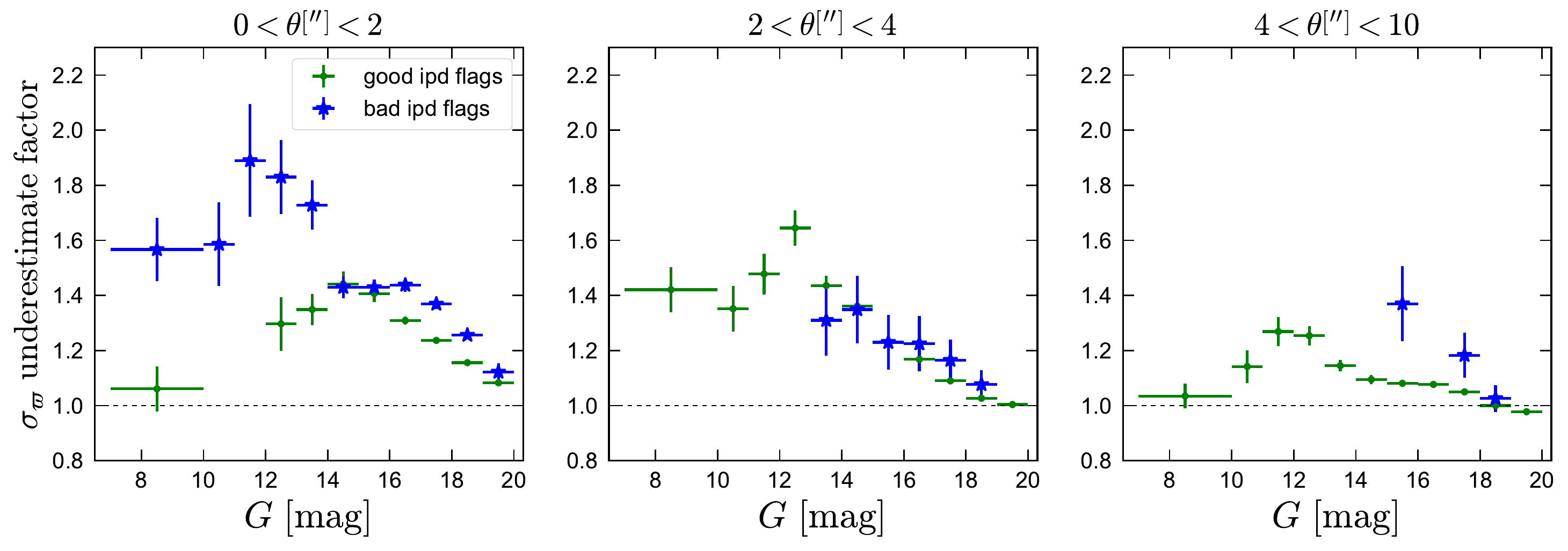}
    \caption{Fractional parallax uncertainty underestimate (i.e., the best-fit $\sigma$ reported in Figure~\ref{fig:histograms}) as a function of magnitude. All binaries considered have \texttt{ruwe} < 1.4 for both components. Green points additionally have \texttt{ipd\_gof\_harmonic\_amplitude} < 0.1 and \texttt{ipd\_frac\_multi\_peak} < 10 for both components; blue points have at least one component that does not pass these cuts. Sources that do not pass the IPD cuts have more strongly underestimated uncertainties at all separations and magnitudes. }
    \label{fig:underestimate_ipd}
\end{figure*}

Blue histograms in Figure~\ref{fig:histograms} show the observed distributions. We show 5 bins of $G$ magnitude, each 1 mag wide, and 3 bins of angular separation. At separations closer than 4 arcsec, the catalog contains binaries with $|\Delta\varpi|/\sigma_{\Delta\varpi} < 6$; at wider separations,  $|\Delta\varpi|/\sigma_{\Delta\varpi} < 3$ (Section~\ref{sec:selection}).
Also shown in Figure~\ref{fig:histograms} are Gaussian fits to the data. Because pairs with $|\Delta\varpi|/\sigma_{\Delta\varpi} < 3$ (or 6) do not enter the catalog, it is necessary to account for this truncation of the distribution. Particularly when parallax uncertainties are underestimated, simply calculating the sample standard deviation would underestimate the best-fit $\sigma$. We assume the observed values of $\Delta\varpi/\sigma_{\Delta\varpi} $ are drawn from a distribution defined as 
\begin{align}
p\left(\Delta\varpi/\sigma_{\Delta\varpi}\right)=\begin{cases}
A\exp\left[-\frac{\left(\Delta\varpi/\sigma_{\Delta\varpi}\right)^{2}}{2\sigma^{2}}\right], & \left|\Delta\varpi/\sigma_{\Delta\varpi}\right|<b\\
0, & \left|\Delta\varpi/\sigma_{\Delta\varpi}\right|>b
\end{cases}
    \label{eq:p_of_x}
\end{align}
where $A=\frac{1}{\sqrt{2\pi}\sigma\times{\rm erf}\left(b/\sqrt{2}\sigma\right)}$, erf is the error function, and $b=3$ or 6 is the sigma-threshold above which binaries are rejected. The log-likelihood for a set of uncertainty-normalized parallax differences is then 
\begin{align}
    \ln L=\sum_{i}\ln p\left(\left(\Delta\varpi/\sigma_{\Delta\varpi}\right)_{i}\right),
    \label{eq:pi}
\end{align}
where the sum is calculated over all binaries in the set. For each panel of Figure~\ref{fig:histograms}, we maximize Equation~\ref{eq:pi} to find the value of $\sigma$ that best describes the truncated Gaussian. This is plotted with a solid line, and the value of $\sigma$ is shown in the legend. For comparison, we also plot a Gaussian with $\sigma=1$, the distribution expected in the limit of accurate parallax uncertainties. To avoid having a few outliers with strongly underestimated parallax uncertainties bias the fits, we exclude binaries with $|\Delta\varpi|/\sigma_{\Delta\varpi} > 3$ and set $b=3$ in all separation bins. The figure shows that the distributions of $\Delta\varpi/\sigma_{\Delta\varpi}$ are indeed approximately Gaussian, and the fits are reasonably good representations of the data.

\begin{figure}
    \centering
    \includegraphics[width=\columnwidth]{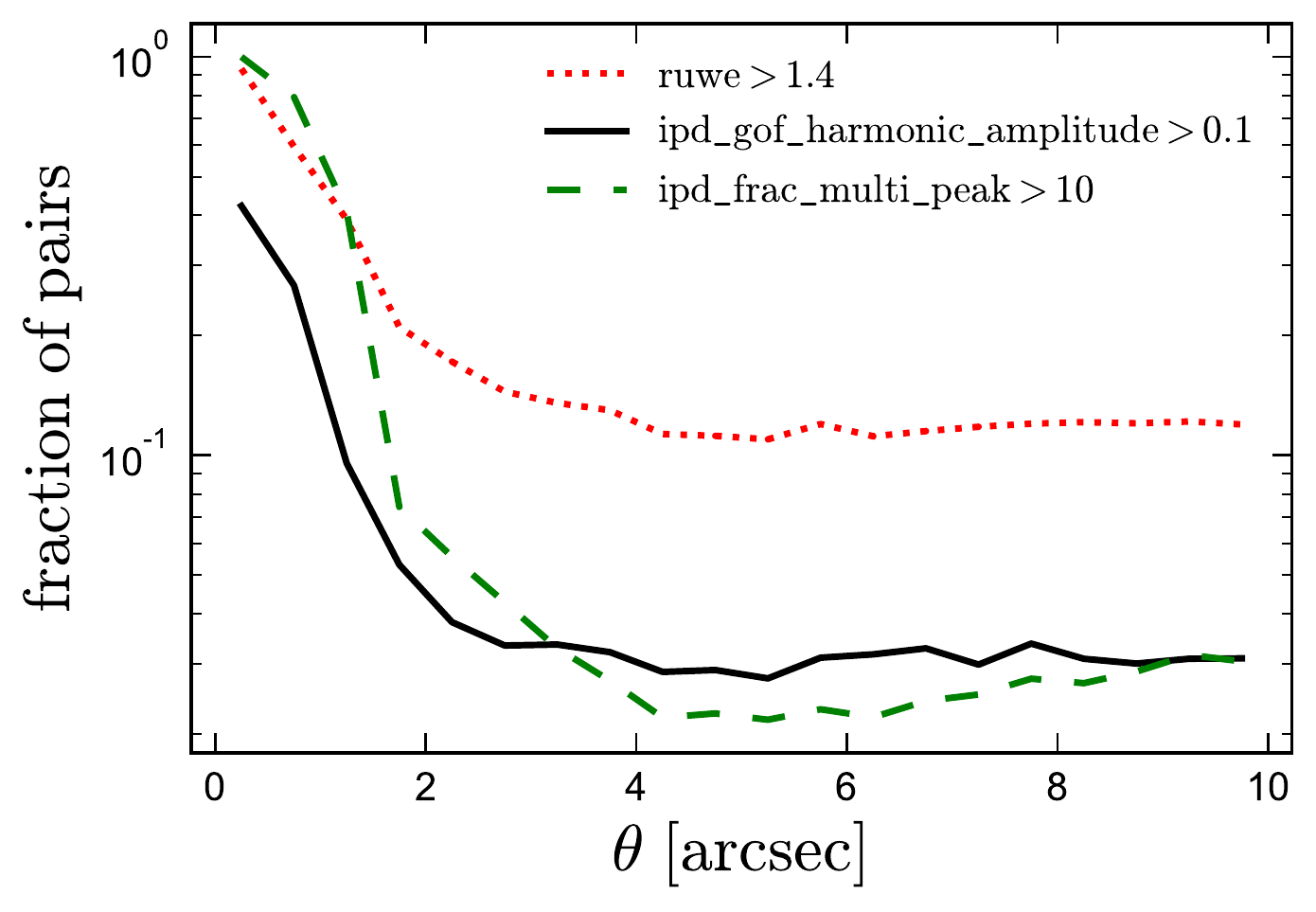}
    \caption{Fraction of pairs in which at least one component fails the indicated quality cut (larger values are indicative of a potentially problematic astrometric solution). All three indicators are significantly enhanced within $\theta \lesssim 2$ arcsec.}
    \label{fig:diagnostics}
\end{figure}

Two trends are clear in Figure~\ref{fig:histograms}: (a) at fixed apparent magnitude, the width of the observed distributions increases at closer separations, and (b) at fixed separation, their width increases toward brighter magnitudes. That is, parallaxes are more strongly underestimated for bright sources with close companions. This is illustrated more clearly in Figure~\ref{fig:underestimate_ruwe}, in which black points with error bars show the best-fit Gaussian $\sigma$ values for each bin of magnitude and angular separation, considering only sources with \texttt{ruwe} < 1.4 for both components. At all separations, parallaxes are most severely underestimated at $G\approx 13$. This is likely related to the fact that the window class (i.e., the pixel sampling scheme around detected sources; see \citealt{Rowell2020}) changes at $G\approx 13$. The largest $\sigma_{\varpi}$ underestimate factors are also accompanied by abrupt changes in the zeropoint \citep{Lindegren2020zpt}. For sources with $G<13$, a 2D window is used and is fitted with a point spread function. For $G>13$, only
the collapsed 1D scan, which is fitted with a line spread function, is available. Sources near $G=13$ have a mix of 2D and 1D windows and are likely more affected by any calibration issues.

Figure~\ref{fig:underestimate_ruwe} also shows results for binaries in which at least one component has \texttt{ruwe} > 1.4, indicative of a potentially problematic astrometric solution. As expected, the best-fit $\sigma$ values are significantly larger for these binaries, at all separations and magnitudes. 

A natural question is whether the broadened distributions of uncertainty-normalized parallax difference could be a result of contamination from chance alignments that are not actually bound, rather than underestimated parallax uncertainties. This hypothesis can be ruled out for two reasons. First, we can empirically estimate the chance-alignment rate for different subsets of the catalog (e.g. Figures~\ref{fig:chance_align_estimate} and \ref{fig:subsets}), and we find it to be extremely low for the samples we use for parallax error validation. Second, our analysis suggests that $\sigma_{\varpi}$ is most severely underestimated for bright binaries at close separations, and this is precisely the region of parameter space where the chance alignment rate is lowest (e.g. Figure~\ref{fig:subsets}).

Besides \texttt{ruwe}, {\it Gaia} eDR3 contains other diagnostics of potentially problematic astrometric fits. In particular, the parameter \texttt{ipd\_gof\_harmonic\_amplitude} quantifies how much the image parameter determination goodness-of-fit varies with scan angle; a large value is likely indicative of a marginally-resolved binary. The related \texttt{ipd\_frac\_multi\_peak} diagnostic quantifies in what fraction of scans multiple peaks are detected. Figure~\ref{fig:underestimate_ipd} separately plots the inferred $\sigma_{\varpi}$ underestimate for sources in which both components have \texttt{ipd\_gof\_harmonic\_amplitude} < 0.1 and \texttt{ipd\_frac\_multi\_peak} < 10 (green), and those in which at least one component fails one of these cuts (blue). These thresholds are motivated by the experiments performed in \citet{Smart2020}. At all separations, the implied $\sigma_{\varpi}$ underestimate is larger for sources that do not pass one of the IPD cuts; the difference is largest for bright pairs at close separations.

Figure~\ref{fig:diagnostics} shows how the prevalence of problematic sources according to the \texttt{ruwe} and IPD diagnostics depends on separation. Considering all binaries in the catalog, we plot the fraction of pairs at a given separation in which at least one component does not pass the cut listed in the legend. At separations larger than a few arcseconds, this fraction is $\sim 10\%$ for the \texttt{ruwe} cut, and $\sim 3\%$ for both the IPD cuts. However, all three diagnostics of problematic solutions increase steeply at $\theta \lesssim 2$ arcsec, with the fraction approaching unity for the \texttt{ruwe} and \texttt{ipd\_frac\_multi\_peak} cuts, and $\sim 40\%$ for the \texttt{ipd\_gof\_harmonic\_amplitude} cut. This is not unexpected: for binaries at close separations, there will necessarily be two peaks in the image. This will unavoidably lead to biases in the image parameter determination, particularly for sources with $G>13$, where the images are collapsed to 1D. There is also a danger of misattributing some scans to the wrong component in close pairs, leading to problems in the astrometric solution. In some cases, poor astrometric fits may also be due to astrometric acceleration (Section~\ref{sec:accel}). Figures~\ref{fig:underestimate_ruwe} and~\ref{fig:underestimate_ipd} show that the \texttt{ruwe} and IPD cuts are indeed useful for identifying sources with potentially problematic astrometric solutions. However, they likely do not catch all problematic sources: parallax uncertainties are underestimated somewhat even for sources that pass all cuts, and by a larger factor at close separations. 

\subsection{Comparison to Gaia DR2}
\label{sec:dr2}
\begin{figure*}

    \centering
        \includegraphics[width=\textwidth]{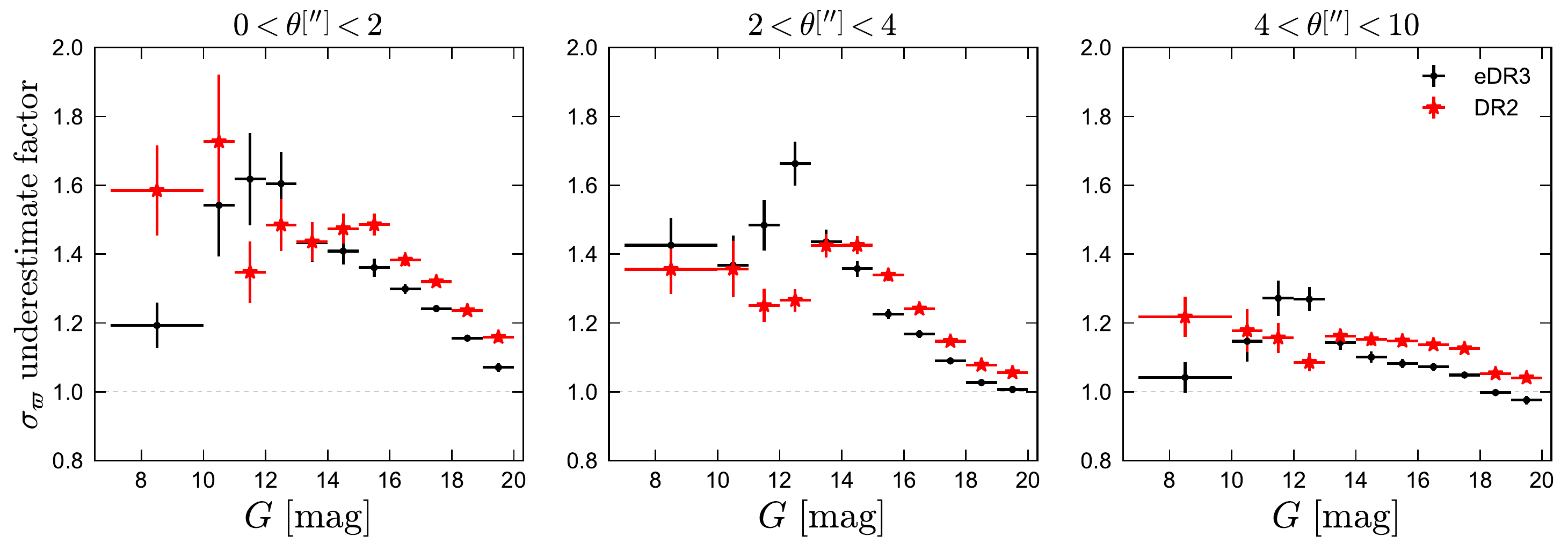}
    \caption{Same as Figures~\ref{fig:underestimate_ruwe} and~\ref{fig:underestimate_ipd}, but now comparing results obtained with {\it Gaia} DR2 vs DR3 parallaxes. All binaries considered have \texttt{ruwe} < 1.4 for both components in both DR2 and DR3. At $G\gtrsim 14$, the uncertainties are underestimated less in DR3 than in DR2. DR3 uncertainties are overestimated somewhat more severely at $11 < G< 13$.}
    \label{fig:underestimate_dr2}
\end{figure*}

Figure~\ref{fig:underestimate_dr2} compares the inferred underestimate factor of $\sigma_{\varpi}$ for DR2 and eDR3 astrometry. We attempt to match all the binary candidates in the catalog with {\it Gaia} DR2 using the \texttt{dr2\_neighbourhood} catalog in the {\it Gaia} archive. For each component of each binary, we identify the likely corresponding DR2 source as the source within 100 mas that has the smallest magnitude difference compared to DR3. There are 1,894 primaries and 15,274 secondaries in the catalog for which no corresponding DR2 source could be identified. There are also 17,514 primaries and 62,424 secondaries for which there is a corresponding source in DR2 that only has a 2-parameter solution. Still, 96\% of candidates have a corresponding DR2 source with a 5-parameter solution for both components. Figure~\ref{fig:underestimate_dr2} considers the subset of these binaries in which both components have \texttt{ruwe} < 1.4 in both DR2 and eDR3. 

Overall trends with magnitude and separation are similar in DR2 and DR3. At $G>13$, the inferred $\sigma_{\varpi}$ underestimates are smaller in DR3 at all separations, though they are still modest in DR2. Even at $G\sim 19$ and wide separations, where the inferred $\sigma_{\varpi}$ in DR3 are consistent with being accurate or very slightly overestimated, those in DR2 are underestimated by $\sim 5\%$ on average. However, at $11 < G <13$, the DR3 uncertainties are overestimated more than those in DR2. Given that the reported values of $\sigma_{\varpi}$ decreased by a factor of two on average between DR2 and DR3 in this magnitude range, the DR3 parallaxes are still ``better''  on average than those from DR2, but Figure~\ref{fig:underestimate_dr2} implies that the true gains are somewhat more modest than those reported. 

\begin{figure*}
    \centering
    \includegraphics[width=\textwidth]{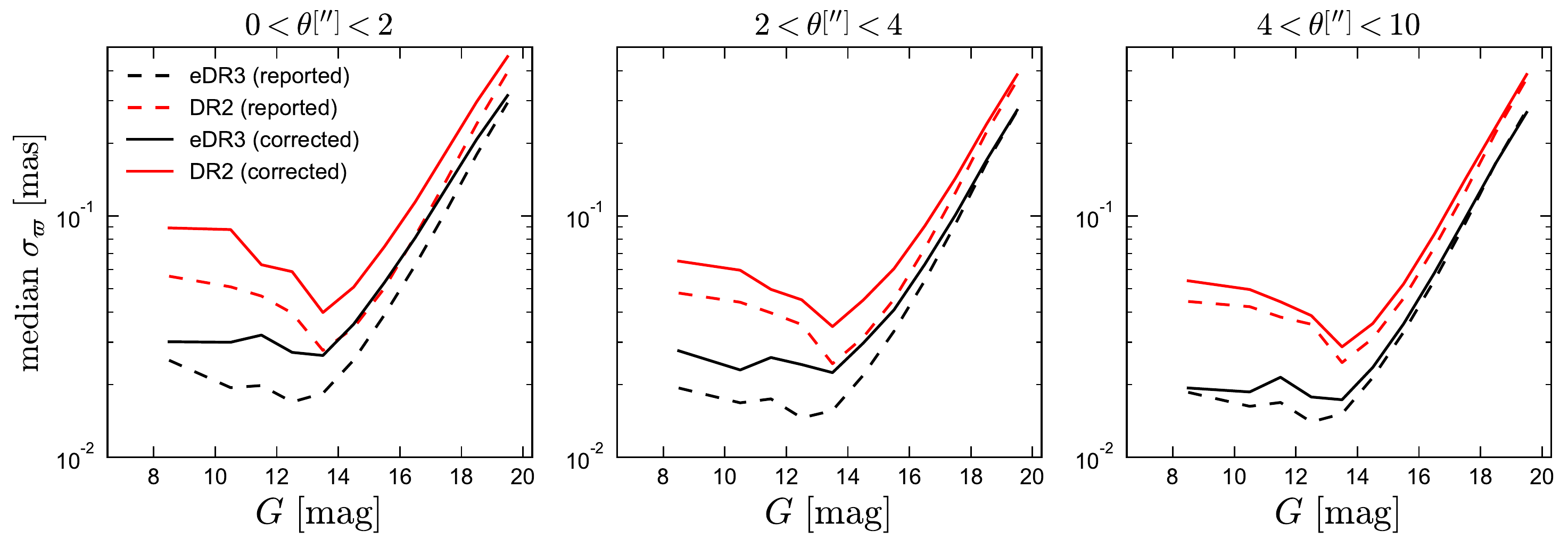}
    \caption{Median reported (dashed) and corrected (solid) parallax uncertainty as a function of $G$ magnitude, for {\it Gaia} DR3 (black) and DR2 (red). The corrected values are obtained by multiplying the reported values by the factors shown in Figure~\ref{fig:underestimate_dr2}, which are empirically determined from the reported parallax differences of the components of wide binaries. Panels show different bins of angular separation. Only stars with \texttt{ruwe} < 1.4 in both data releases are considered. The rightmost panel is appropriate for single stars outside of crowded fields.}
    \label{fig:true_sigma_varpi_dr3_dr2}
\end{figure*}

This is shown explicitly in Figure~\ref{fig:true_sigma_varpi_dr3_dr2}, which shows the reported and true (inferred) median $\sigma_{\varpi}$ for DR2 and DR3 as a function of apparent $G-$ magnitude and angular separation. Dashed lines show the median value of $\sigma_{\varpi}$ in the sample, considering both primaries and secondaries. Solid lines show the result of multiplying these values by the appropriate factors from Figure~\ref{fig:underestimate_dr2}. The corrected median $\sigma_{\varpi}$ values are at least 30\% smaller in DR3 than in DR2 at all magnitudes and separations. For bright stars ($G \lesssim 13$), the gains are generally more than a factor of 2.

\subsection{Parallax zeropoint corrections and 5-parameter vs 6-parameter solutions}
The {\it Gaia} eDR3 parallax zeropoint is known to vary with apparent magnitude, color, and ecliptic latitude. We do not attempt to account for this variation when constructing the binary catalog. Because all the sources in the catalog have $\varpi > 1\,\rm mas$, the effects of zeropoint corrections, which are typically on the order of 0.02 mas, are modest. \citet{Lindegren2020zpt} derived an empirical zeropoint for eDR3 using quasars, stars in the LMC, and binaries. Here we investigate whether ``correcting'' the parallaxes using the prescriptions they provide can reduce the inferred $\sigma_{\varpi}$ underestimate factors.

\begin{figure*}
    \centering
    \includegraphics[width=\textwidth]{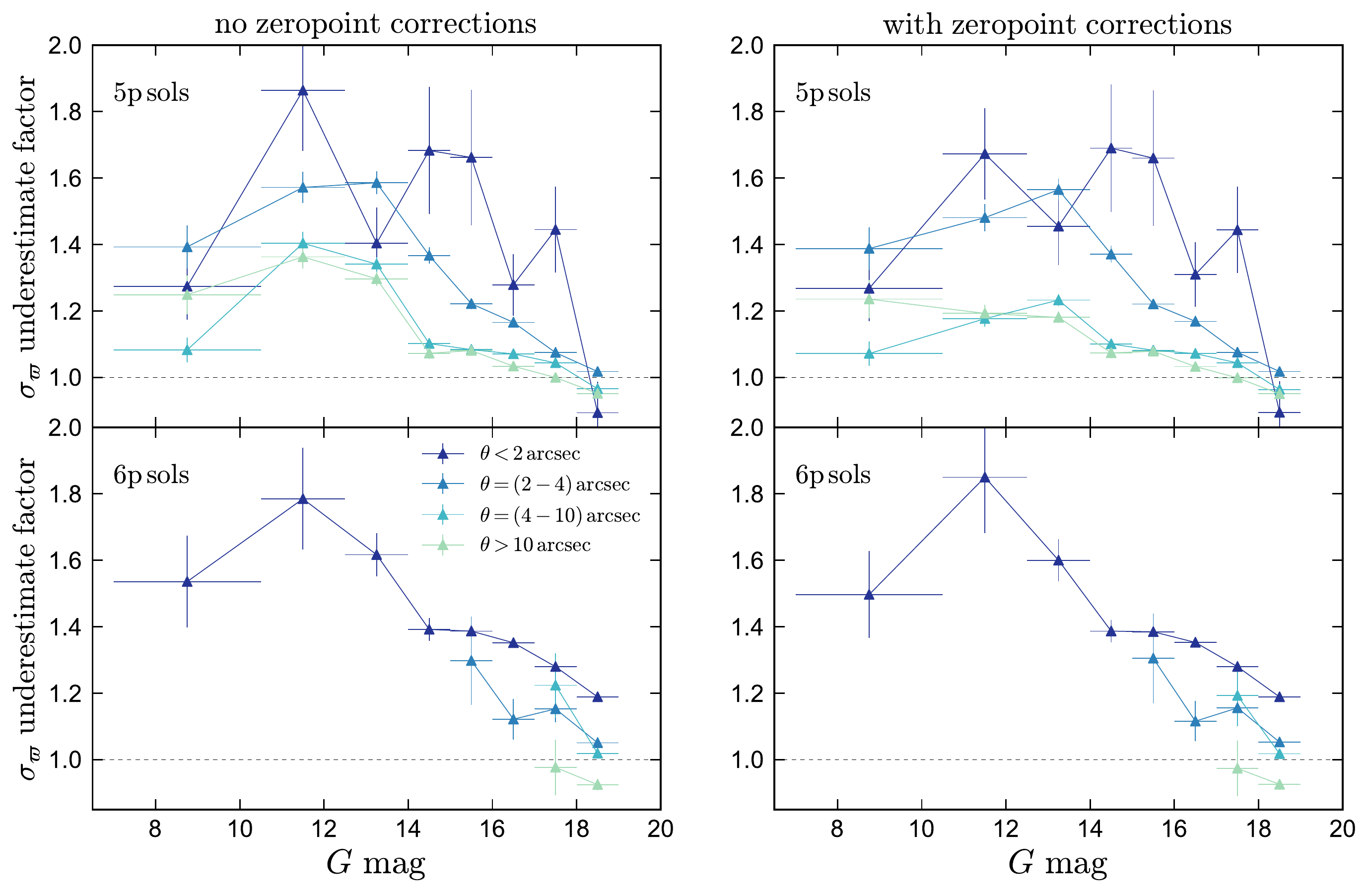}
    \caption{$\sigma_{\varpi}$ underestimate factors (similar to Figure~\ref{fig:underestimate_ruwe}) for binaries in which both components have five-parameters solutions (top) or both have six-parameter solutions (bottom). Left column uses the {\it Gaia} eDR3 parallaxes directly; right column applies the magnitude-, color-, and position-dependent parallax zeropoint correction from \citet{Lindegren2020zpt}. For bright sources, only binaries with close angular separations have six-parameter solutions. The inferred $\sigma_{\varpi}$ underestimate factors depend mainly on separation and magnitude; they are similar for 5- vs 6-parameter solutions. Applying the zeropoint correction reduces the inferred overestimate factor slightly for bright sources. }
    \label{fig:underestimate_zpt}
\end{figure*}

Figure~\ref{fig:underestimate_zpt} shows the inferred $\sigma_{\varpi}$ underestimate factors with and without the zeropoint correction. We also separately plot binaries in which both components have a 5-parameter astrometric solution and those in which both have a 6-parameter solution. Details about the differences between 5- and 6-parameters solutions are discussed in \citet{Lindegren2020}; 5-parameter solutions are generally more reliable.
Binaries with one 5- and one 6-parameter solution are excluded. At $G\lesssim 18$, most sources have 5-parameter solutions. The exception is sources with a close companion ($\theta \lesssim 2$ arcsec), which usually lack both reliable colors and 5-parameter solutions. 

At fixed separation and magnitude, the effects of applying the zeropoint correction are encouraging but modest: for widely-separated binaries with $G\approx 13$ and 5-parameter solutions, the inferred $\sigma_{\varpi}$ underestimate factor decreases from $\sim 1.30$ to 1.25. Improvements are generally smaller at closer separations and for 6-parameter solutions. The small effect is not unexpected: because the binaries we consider all have small magnitude differences, the zeropoint corrections are similar for both components, and the parallax {\it difference} does not change much when the correction is applied. At close separations, where most stars have 6-parameter solutions, the underestimate factors are similar for 5- and 6-parameter solutions. This suggest that the increased uncertainties at close separations are not primarily due to the transition from 5- to 6-parameter solutions.

\subsection{Color dependence}
\label{sec:colors}

\begin{figure*}
    \centering
    \includegraphics[width=\textwidth]{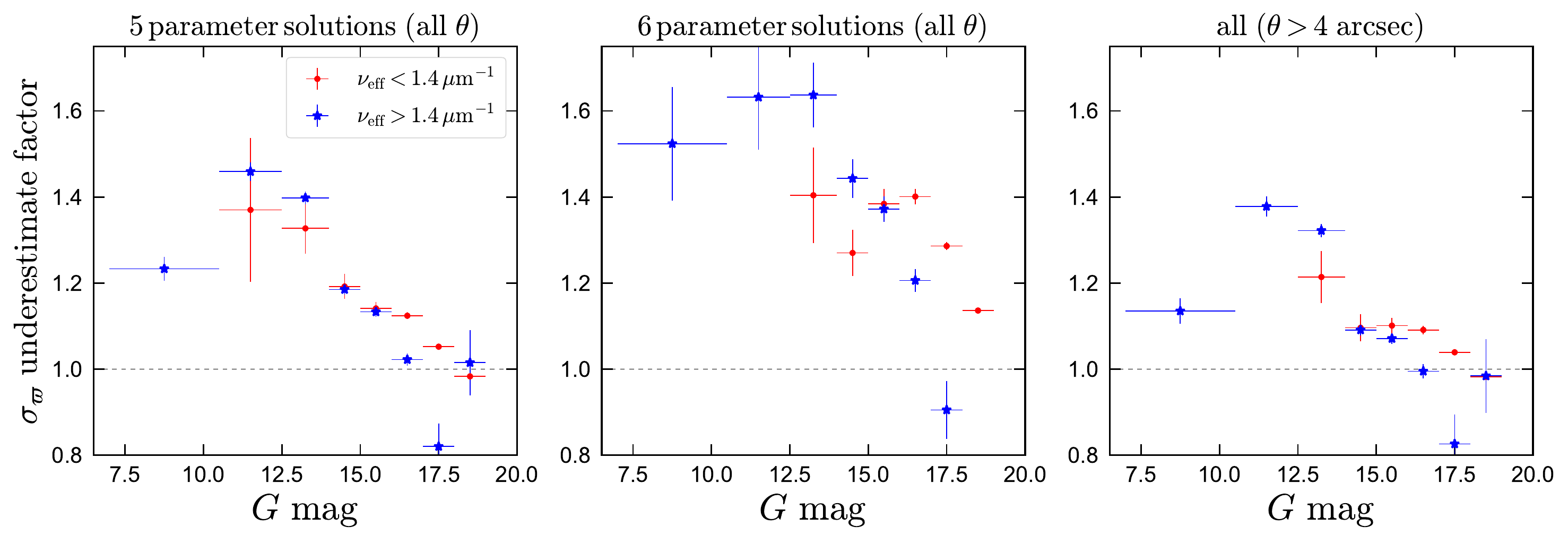}
    \caption{Similar to Figure~\ref{fig:underestimate_ruwe}, but with binaries divided by color. Only binaries in which both components have \texttt{ruwe} < 1.4 and the two components fall in the same color bin are included. $\nu_{\rm eff}$ refers to the parameters \texttt{nu\_eff\_used\_in\_astrometry} and \texttt{pseudocolour}, respectively, for 5- and 6-parameter solutions. The boundary of $\nu_{\rm eff}=1.4\,\mu \rm m^{-1}$ corresponds approximately to $G_{\rm BP}-G_{\rm RP} =1.6$. Most quasars, white dwarfs, and MS stars with $M\gtrsim 0.6 M_{\odot}$ are bluer than this; lower-mass MS stars are redder. Left and middle panels show all separations; right panel shows wide separations.  At the faint end, $\sigma_{\varpi}$ overestimates are somewhat larger for red sources. }
    \label{fig:pseudoclor}
\end{figure*}

The parallax zeropoint is also known to depend somewhat on color, likely due to the color-dependence of the PSF \citep{Lindegren2020zpt, Lindegren2020}. We therefore investigate how the $\sigma_{\varpi}$ underestimate factors depend on color in Figure~\ref{fig:pseudoclor}, where we separately consider binaries in which both components are blue and those in which both components are red. The color boundary we use, $\nu_{\rm eff}=1.4\,\mu \rm m^{-1}$, corresponds to $G_{\rm BP}-G_{\rm RP}\approx 1.6$, which is approximately the red limit of the {\it Gaia} quasar sample. Figure~\ref{fig:pseudoclor} shows that for sources with $G>16$, the inferred $\sigma_{\varpi}$ underestimates are somewhat larger for red sources. We verified that there are not significant angular separation differences between the red and blue pairs at fixed magnitude, so color is the most likely driving variable. 

\subsection{Fitting function to inflate $\sigma_{\varpi}$}
\label{sec:fitting}

We fit a function to our inferred $\sigma_{\varpi}$ inflation factors as a function of $G$ magnitude, which can be used to empirically correct $\sigma_{\varpi}$ values reported for isolated sources. To derive a correction appropriate for single sources with well-behaved astrometry, we consider binaries with $\theta > 5\,\rm arcsec$ and \texttt{ruwe} < 1.4 for both components, and we apply the zeropoint correction from \citet{Lindegren2020zpt} to both components' parallaxes. Because we observe a general decline in the inflation factor with increasing $G$ and a peak at $G\approx 13$ (Figure~\ref{fig:fittingfunction}), we fit a polynomial plus a Gaussian bump: 

\begin{equation}
    \label{eq:fitting_function}
    f\left(G\right)=A\exp\left[-\frac{\left(G-G_{0}\right)^{2}}{b^{2}}\right] + p_{0} + p_{1}G + p_{2}G^{2}
\end{equation}
We find $A=0.21$, $G_0 = 12.65$, $b = 0.90$,  $p_0 = 1.141$, $p_1 = 0.0040$, and $p_2 = -0.00062$. This is also plotted in Figure~\ref{fig:fittingfunction}. Multiplying by $f$ will -- on average -- correct reported $\sigma_{\varpi}$ values for {\it Gaia} eDR3 and single-source solutions in DR3. The correction is appropriate for sources with $7 < G < 21$ that have \texttt{ruwe} < 1.4, have no comparably bright sources within a few arcsec, and have already had their parallaxes corrected by the zeropoint from \citet[][]{Lindegren2020zpt}. The effects of having a close companion on $\sigma_{\varpi}$ likely depend on the brightness contrast and a variety of other factors; a rough estimate of the magnitude of the inflation can be obtained from Figure~\ref{fig:underestimate_ruwe}-\ref{fig:underestimate_ipd}. The correction can be reasonably applied to both 5- and 6-parameter astrometric solutions. We do not fit separate corrections for red and blue sources, but we note that at the faint end, the inflation factors are generally somewhat smaller for blue sources. 

\subsection{Angular correlations in parallaxes}
The $\sigma_{\varpi}$ inflation factors inferred in this work and predicted by Equation~\ref{eq:fitting_function} should be interpreted as lower limits. {\it Gaia} eDR3 parallaxes are subject to systematic trends on degree scales (and larger) due to the scanning law \citep[e.g.][]{Fabricius2020, Lindegren2020}. The angular separations of most of the binaries in the catalog are significantly smaller than this (Figure~\ref{fig:basic_properties}), so the ``local'' parallax zeropoint for the two stars is usually very similar. These local positional variations in the zeropoint are not accounted for in the correction from \citet{Lindegren2020zpt} and also inflate the effective parallax uncertainties. With a typical scale of about $10\,\mu \rm as$, these may contribute significantly to the uncertainties at $G\lesssim 13$. 

A contemporaneous study by \citet{Zinn2021} validated {\it Gaia} eDR3 parallaxes and their uncertainties using bright giants in the Kepler field ($9\lesssim G \lesssim 13$) with independent distance estimates from asteroseismology. They tested our parallax uncertainty inflation function (Equation~\ref{eq:fitting_function}) and found it to perform well; i.e., no further uncertainty inflation was required after it was applied. The angular size of the Kepler field ($\sim$10 deg) is larger than the scale on which the strongest angular correlations in {\it Gaia} eDR3 are manifest \citep[e.g.][their Figure 14]{Lindegren2020}. This suggests that any additional uncertainty inflation required due to angular correlations in the zeropoiont is modest. 

\subsection{Comparison to other work}
The reliability of parallax uncertainties reported in {\it Gaia} eDR3 has been investigated by several other works. \citet[][their Figure 19]{Fabricius2020} used the dispersion in parallaxes reported for distant objects -- quasars, stars in the LMC, and stars in dwarf spheroidal (dSph) galaxies, which should all have negligible true parallaxes -- to estimate  $\sigma_{\varpi}$ inflation factors as a function of $G$ magnitude. Their results from quasars and dSph stars, which are only available at the faint end $(G \gtrsim 16)$ are broadly consistent with our results. In the LMC, they infer inflation factors that are larger at fixed magnitude than our results at wide separations, or their results for quasars and dSph at the same magnitude. This discrepancy is very likely a result of crowding: the source density in the LMC is large enough that a significant fraction of sources have another source within a few arcsec, and as we have shown (e.g. Figure~\ref{fig:underestimate_dr2}), sources with companions within a few arcsec have more severely underestimated parallaxes. This likely also explains why \citet[][]{MaizApellaniz2021} found somewhat larger uncertainty inflation factors in globular clusters than we do with widely separated binaries. In the Kepler field, \citet[][]{Zinn2021} found a mean uncertainty estimate of (22$\pm$ 6)\% for sources with $9\lesssim G \lesssim 13$  (with most sources at the faint end of this range); this is consistent with our results.

\begin{figure}
    \centering
    \includegraphics[width=\columnwidth]{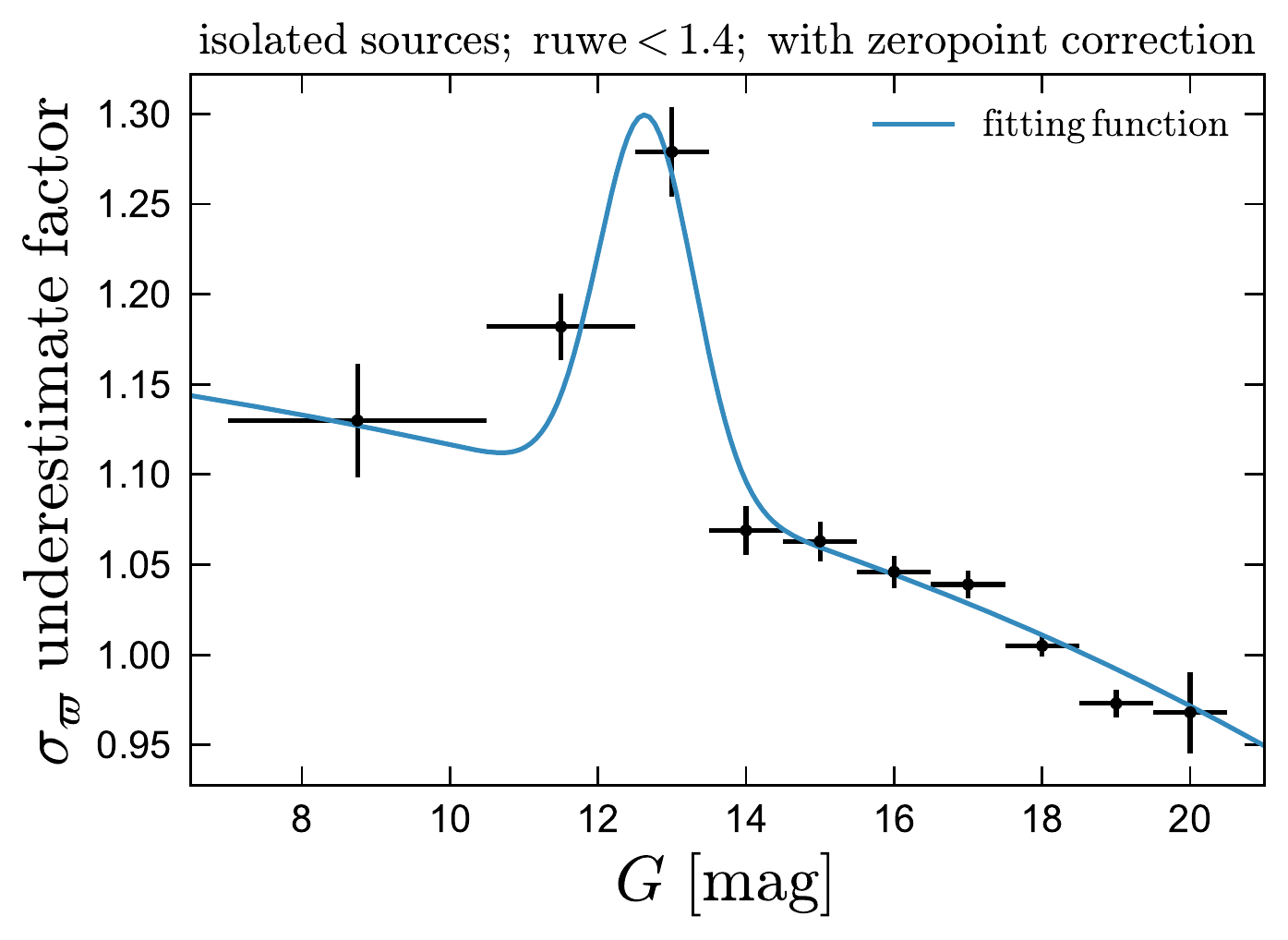}
    \caption{Parallax uncertainty inflation factors inferred from widely-separated pairs with \texttt{ruwe} < 1.4 and parallaxes corrected by the zeropoint from \citet{Lindegren2020zpt}. This is the $\sigma_{\varpi}$ underestimate factor for typical single sources with well-behaved astrometry. A fitting function (Equation~\ref{eq:fitting_function}) is provided. } 
    \label{fig:fittingfunction}
\end{figure}

\section{Summary and discussion}
\label{sec:discussion}
We have constructed a catalog of 1.2 million high-confidence, spatially resolved wide binaries using {\it Gaia} eDR3 and have used them to empirically validate the reported parallax uncertainties in {\it Gaia} eDR3. Overall, the results are very encouraging: outside of crowded regions (i.e., for stars with no comparably bright companion within a few arcseconds), parallax uncertainties for sources with well-behaved astrometric fits are underestimated by {\it at most} 30\% (at $G\approx 13$), and by considerably less at fainter magnitudes.  Our main results are as follows:

\begin{enumerate}
    \item: {\it Catalog description}: The full catalog contains 1.2 million high-confidence binaries, including 15,982 WD+MS binaries and 1362 WD+WD binaries (Figures~\ref{fig:basic_properties} and~\ref{fig:wd_cmds}), and $\sim$20,000 binaries containing giants and subgiants. The binaries span a projected separation range of a few au to 1 pc, have heliocentric distance up to 1 kpc, and include both (kinematic) disk and halo binaries. All binaries have reasonably precise astrometry, with $\varpi/\sigma_{\varpi} > 5$ for both components. The full catalog of 1.8 million binary candidates become dominated by chance alignments at $s\gtrsim 30,000$\,au, but high-quality subsets can be selected that are relatively pure out to separations as large as 1\,pc (Figure~\ref{fig:subsets}). The catalog builds on previous efforts to identify binaries using {\it Gaia} DR2, expanding the sample of known high-confidence binaries by a factor of 4 (Figure~\ref{fig:other_catalogs}). This increase in sample size owes partly to the higher astrometric precision provided by {\it Gaia} eDR3, and partly to improvements in the binary identification and vetting strategy. 
    
    \item: {\it Quantifying and controlling chance alignments}: We estimate the contamination rate from chance alignments using two approaches: a mock catalog that does not contain any true binaries, and a version of the {\it Gaia} catalog in which stars have been artificially shifted from their true positions, removing real binaries but preserving chance-alignment statistics (Figure~\ref{fig:chance_align_estimate}). Both approaches show that the full catalog has high purity at $s < 10,000$\, but becomes dominated by chance alignments at $s \gtrsim 30,000$\,au. Using the shifted chance alignment catalog, we show how one can select subsamples that have lower contamination rates, including some that are pure out to 1\,pc (Figure~\ref{fig:subsets}). We also include in the catalog an estimate of the probability that each binary candidate is a chance alignment; this is constructed empirically from the distribution of known chance alignments in a seven-dimensional space of observables (Figure~\ref{fig:hist_ratios} and Appendix~\ref{sec:kde}). We use radial velocities from {\it Gaia} (Figure~\ref{fig:rvs}) and LAMOST (Figure~\ref{fig:lamost}) to validate these probabilities.
    
    \item {\it Orbital velocities}: The high-precision of {\it Gaia} astrometry makes obvious the plane-of-the-sky velocity difference of the components of binaries due to orbital motion (Figure~\ref{fig:triples}). About 200,000 binaries in the catalog have sufficiently accurate astrometry that the plane-of-the-sky velocity difference between the components can be measured with accuracy $\sigma_{\Delta V} < 100\,\rm m\,s^{-1}$. A Keplerian decline in the velocity difference, $\Delta V \propto s^{-1/2}$, is visible out to $s\sim 20,000$\,au, where projection effects become important. The requirement of proper motions consistent with Keplerian orbits excludes a significant fraction of hierarchical triples and higher-order multiples from the catalog (Figure~\ref{fig:deltav_ruwe}).
    
    \item {\it Validation of Gaia DR3 parallax uncertainties}: We use the sample of high-confidence binaries to validate the published parallax uncertainties included in {\it Gaia} eDR3. This analysis makes use of the fact that the two stars in a binary have essentially the same distance and thus should generally have reported parallaxes that are consistent within their uncertainties (Figure~\ref{fig:histograms}). We find that the published uncertainties are accurate for faint stars ($G\gtrsim 18$) that have well-behaved astrometric solutions and do not have a companion within a few arcseconds (Figure~\ref{fig:underestimate_ruwe}). They are underestimated somewhat for brighter stars, particularly in the range of $11 < G < 13$, where the published uncertainties should be multiplied by a factor of 1.3 on average. The degree to which uncertainties are underestimated is larger for sources with large \texttt{ruwe}, \texttt{ipd\_gof\_harmonic\_amplitude}, and \texttt{ipd\_frac\_multi\_peak} (Figure~\ref{fig:underestimate_ruwe} and~\ref{fig:underestimate_ipd}), and is larger for red sources than blue sources (Figure~\ref{fig:pseudoclor}). The reported parallax uncertainties are generally more reliable in {\it Gaia} eDR3 than they were in DR2, except at $11 < G < 13$ (Figure~\ref{fig:underestimate_dr2}). In an absolute sense, the $\sigma_{\varpi}$ values improved by at least 30\% from DR2 to eDR3, at all magnitudes and separations (Figure~\ref{fig:true_sigma_varpi_dr3_dr2}). We provide an empirical fitting function to correct reported $\sigma_{\varpi}$ values  (Figure~\ref{fig:fittingfunction}).
    
    Parallax uncertainties are underestimated more for binaries with angular separations less than a few arcsecoonds. Sources with resolved close companions are more likely to have high \texttt{ruwe} and IPD diagnostics related to binarity (Figure~\ref{fig:diagnostics}), but the underestimates of $\sigma_{\varpi}$ is enhanced at close separations even for pairs in which both components have low \texttt{ruwe} and IPD flags. This is true both for sources with 5- and 6-parameter solutions (Figure~\ref{fig:underestimate_zpt}).
    
\end{enumerate}


\subsection{Scientific uses for the catalog}
\label{sec:usingcatalog}
This paper was primarily concerned with assembling the wide binary catalog. Here we note a few possible uses for the sample, which will be pursued in future work.  

\begin{itemize}
    \item {\it Calibrating stellar ages }: A useful property of wide binaries is that the two stars have basically the same total age, but can have different evolutionary states. If the age of one component can be constrained (e.g., because it is a WD, a subgiant, or a giant with asteroseismic mass constraints), that age constraint can be transferred to the companion \citep[e.g.][]{Chaname2012, Fouesneau2019, Qiu2020}. This in turn can be used to calibrate more poorly-understood age indicators for MS stars, such as gyrochronology, stellar activity, and X-ray luminosity \citep[e.g.][]{Soderblom2010}.
    \item {\it The initial-final mass relation}: The WD+WD sample will be useful for constraining the initial-final mass relation (IFMR) for WDs: the masses and cooling ages of both WDs can (often) be well-constrained from photometry, and the IFMR can be constrained by the fact that both WDs have the same total age (and presumably, follow the same IFMR; e.g.,  \citealt{Andrews2015}). The subset of the WD+MS sample in which the age of the non-WD component can be constrained independently will also be useful for constraining the IFMR \citep[e.g.][]{Catalan2008}.
    \item {\it WD masses from gravitational redshift}: Because the WD and MS star in a wide binary have essentially the same RV, the difference in their apparent RVs is due primarily to the WD's gravitational redshift (which is typically $20-100\,\rm km\,s^{-1}$). This provides a useful way of measuring WD mass that are essentially model-independent if the distance is well-constrained \citep[e.g.][]{Koester1987}. Masses from gravitational redshift will be particularly useful for measuring the mass distributions of WDs with rare spectral types and poorly-understood formation histories, such as the ``Q-branch" WDs revealed by the {\it Gaia} CMD \citep[][]{Bergeron2019, Cheng2020}.
    \item {\it Abundances for WD progenitors}: a MS companion provides a window into the chemical abundances of a WD's progenitor, which are otherwise inaccessible. Among other applications, this provides an avenue to compare the primordial and final abundances of disintegrating planets around polluted WDs, which represent a significant fraction of the WD population \citep[][]{Koester2014, Farihi2016}.
    \item {\it Wide binary spin alignment}: An open question in binary star formation is how aligned the spins vectors of binaries are, and how this varies with separation \citep[e.g.][]{Justesen2020}. Of particular interest is whether the spins of excess ``twin'' binaries, which may have formed in circumbinary disks, are more aligned than those of non-twins at the same separation \citep{Elbadry2019twin}. Spin inclinations can be measured by combining a spectroscopic measurement of $v \sin i$ with a rotation period measured from spots (which requires a light curve) and a radius from parallax and temperature. About 6,000 of the high-confidence binaries in the catalog have both components with $G<14$ and angular separations $\theta > 30$ arcseconds. These are ideal for follow-up study with TESS, because they are separated widely enough that high-quality light curves are available for both components separately. About 39,000 binaries in the catalog are in the K2 fields, and 5,000 are in the Kepler field. 
    \item {\it Calibration of spectroscopic surveys}: The surface abundances of stars in wide binaries are generally very similar \citep[e.g.][]{Hawkins2020}. This enables diagnosis of systematics in the abundances reported by surveys (e.g. Figure~\ref{fig:lamost}). Similar analyses can be done with stars in clusters, but binaries are more abundant and populate abundance space more densely than clusters.

    \item {\it Dynamical probes}: At separations wider than about 10,000 au, wide binaries are susceptible to dynamical disruption through gravitational encounters with other stars, compact objects, or molecular clouds \citep[e.g.][]{Weinberg_1987}. This makes the wide binary separation distribution a sensitive probe of the population of possible perturbers \citep[e.g.][]{Yoo_2004, Tian2020}.
\end{itemize}

\subsection{Constraining the parallax zeropoint with binaries}
A further application of the {\it Gaia} binary sample, which we have not explored in this work, is  calibration of the parallax zeropoint. We have intentionally limited our analysis to binaries in which both stars have almost the same magnitude, (and, because almost all the stars are on the main sequence, the same color). This avoids complication arising from the magnitude- and color-dependence of the parallax zeropoint, which should be nearly the same for both stars. If we were to consider binaries with substantially different primary and secondary magnitudes, magnitude-dependence of the zeropoint would be manifest as a shift in the {\it mean} signed parallax difference: that is, the distributions in Figure~\ref{fig:histograms} would no longer be centered on zero. If the absolute zeropoint at a particular magnitude and color can be pinned down from external data (e.g. quasars  at the faint/blue end), binaries then allow for determination of the zeropoint at all other magnitudes and colors. Some analysis along these lines was carried out by \citet[][]{Lindegren2020zpt} and \citet[][]{Fabricius2020}. 

A challenge to carrying out this type of calibration with our current catalog is Lutz-Kelker bias \citep{Lutz1973}: because the fainter secondaries have larger parallax errors, their parallaxes in our sample will on average be overestimated more than those of the primaries. That is, in the absence of any magnitude-dependence of the zeropoint, the mean value of $\varpi_1-\varpi_2$ would be negative. This bias must be eliminated or accounted for in order for reliable determination of the zeropoint from binaries to be feasible. The most straightforward path forward is likely to select binaries without explicit cuts on parallax.

\subsection{Astrometric acceleration}
\label{sec:accel}
We have shown that parallax uncertainties are more severely underestimated at close angular separations (e.g. Figure~\ref{fig:underestimate_ruwe}), and that a larger fraction of close binaries have \texttt{ruwe} > 1.4 for at least one component (Figure~\ref{fig:diagnostics}). Here we consider whether this is likely due to actual astrometric acceleration, or other issues.

Whether orbital acceleration of a binary is detectable depends on a variety of factors, including the eccentricity, orientation, and phase of the orbit. Here we derive a crude estimate. We consider a face-on circular orbit with  $M_1 + M_2 = 1 M_{\odot}$ and $M_1 \gg M_2$, with an angular separation $\theta$ viewed at a distance $d$. The semi-major axis is $a=1\,{\rm au}\times\left(d/{\rm pc}\right)\left(\theta/{\rm arcsec}\right)$, and the orbital period is $P=1\,{\rm yr}\times\left(d/{\rm pc}\right)^{3/2}\left(\theta/{\rm arcsec}\right)^{3/2}$. During the 34-month baseline of {\it Gaia} DR3, the azimuthal angle swept out by the secondary is 
\begin{align}
\phi&=2\pi\times\left(\left(34\,{\rm months}\right)/P\right)\\ &=0.018\,{\rm radians}\times\left(d/{\rm \left(100\,pc\right)}\right)^{-3/2}\left(\theta/{\rm arcsec}\right)^{-3/2},
    \label{eq:phi}
\end{align}
where $\phi=2\pi $ would signify a full orbit.

In the limit of small $\phi$, the total orbital motion that is perpendicular to the instantaneous proper motion vector at the first observation is $\vartheta_{\perp}\approx\frac{1}{2}\theta\phi^{2}$, from Taylor expanding $x=\theta \cos \phi $. The total perpendicular deviation from linear motion is thus 
\begin{align}
\vartheta_{\perp}&\approx0.16\,{\rm mas}\times\left(\frac{d}{100\,{\rm pc}}\right)^{-3}\left(\frac{\theta}{{\rm arcsec}}\right)^{-2}\\&\approx0.16\,{\rm mas}\times\left(\frac{d}{100\,{\rm pc}}\right)\left(\frac{s}{{\rm 100\,au}}\right)^{-2}.
    \label{eq:perpendicular_motion}
\end{align}
The deviation from the {\it best-fit} single-star orbit will likely be a factor of a few smaller than this.

To determine whether orbital acceleration is plausibly detectable, this quantity can be compared to the typical astrometric precision (e.g. Figure~\ref{fig:true_sigma_varpi_dr3_dr2}). At the typical distance of binaries in the catalog, $d\approx 500$\,pc, the predicted deviation for $\theta=1$ arcsec is on the order of 0.001 mas, well below the sensitivity of {\it Gaia} eDR3. Astrometric acceleration due to orbital motion is thus not expected to be detectable for the large majority of binaries in our catalog, and it is therefore likely that the more strongly underestimated $\sigma_{\varpi}$ at close separations is primarily due to other issues, such as centroiding errors or some scans being attributed to the wrong component.

Astrometric acceleration should, in principle, be non-negligible for the nearest and closest binaries in the catalog. Considering only binaries with $s < 100\,\rm au$, the median deviation predicted by Equation~\ref{eq:perpendicular_motion} is 0.3 mas, which is larger than $\sigma_{\varpi}$ for the majority of that sample. To investigate whether there is evidence of acceleration in our sample at close separations, we compared the DR2 and eDR3 proper motions of both components, under the assumption that acceleration should manifest as a change in mean proper motion from epoch 2015.5 to 2016.0 \citep[e.g.][]{Kervella2019}. It is important to note that the coordinate systems of DR2 and eDR3 are not identical. An ad-hoc correction was applied to the eDR3 coordinate frame to remove a $\sim$0.1\,$\rm mas\,yr^{-1}$ rotation that was present in the coordinate system for bright stars in DR2 \citep[][]{Lindegren2020}. Properly aligning the coordinate systems between the two releases is nontrivial \citep[e.g.][]{Brandt2018}. For this reason, and because proper motion uncertainties, like parallax uncertainties, are likely underestimated somewhat, it is beyond the scope of our investigation to determine which proper motion differences are significant. Instead, we simply consider how the fraction of sources with proper motion differences above an particular threshold depends on separation. We find that the fraction of binaries with inconsistent proper motions for one or both components is strongly enhanced at close angular separations, as would be expected in the presence of accelerations. For the full binary catalog, the fraction of pairs that have at least one component with DR2 and eDR3 proper motions inconsistent within $3\sigma$ is 18\% at $\theta > 4$ arcsec, but 50\% at $\theta < 1$ arcsec, and 85\% at $\theta < 0.5$ arcsec. 

However, the fraction of sources with inconsistent proper motions is significantly enhanced even at $d > 500$\,pc, where Equation~\ref{eq:perpendicular_motion} suggests that any perpendicular acceleration should be negligible. Indeed, at fixed magnitude, the fraction of sources with inconsistent proper motions depends primarily on angular, not physical, separation.  This suggests spurious astrometry (due to bias in the image parameter determination or source misidentification) for sources with close companions is the primary cause for the apparent acceleration. Although Equation~\ref{eq:perpendicular_motion} suggests that acceleration should often be detectable in our sample at $s < 50$\,au, the expected sensitivity is not yet realized there due to problematic astrometry for barely-resolved sources. The detectability of accelerations with {\it Gaia} was also investigated by \citet{Belokurov2020}. They found that while \texttt{ruwe} is often enhanced in close binaries, the enhancement can be reliably tied to orbital motion only in the regime where a significant (order unity) fraction of the orbit is covered by the {\it Gaia} time baseline.

\section{Catalog description}
\label{sec:cat_description}
The full binary catalog will be hosted at CDS. It can also be accessed at \href{https://zenodo.org/record/4435257}{https://zenodo.org/record/4435257}. All columns in the \texttt{gaiaedr3.gaia\_source} catalog are copied over for both components. We also include the columns \texttt{source\_id}, \texttt{parallax}, \texttt{parallax\_error}, \texttt{pmra}, \texttt{pmdec}, \texttt{pmra\_error},  \texttt{pmdec\_error}, and \texttt{ruwe} from {\it Gaia} DR2 for both components; these have the prefix \texttt{dr2\_}.

Columns ending in ``1'' and ``2'' refer to the primary and secondary component, respectively. The primary is always the component with the brighter $G$ magnitude. We also include columns \texttt{pairdistance} (angular separation $\theta$, in degrees), \texttt{sep\_AU} (projected separation $s$, in au), \texttt{R\_chance\_align} ($\mathcal{R}$; Equation~\ref{eq:ratio}), and \texttt{binary\_type} (e.g. MSMS, WDMS, etc.; see Table~\ref{tab:numbers}). The ordering in \texttt{binary\_type} does not account for primary/secondary designations; i.e., all binaries containing a WD and a MS star are designated WDMS, irrespective of whether it is the WD or MS component that is brighter.

The shifted chance alignment catalog is also available. It contains the same columns as the binary candidate catalog, except the {\it Gaia} DR2 columns. Because one component of each pair has been shifted from its true position in the \texttt{gaia\_source} catalog, the \texttt{ra} and \texttt{dec} columns in it do not match those reported in the  \texttt{gaia\_source} catalog.

\section*{Acknowledgements}
We thank the anonymous referee for a constructive report, Jackie Faherty for help creating visualizations of the catalog, and Eliot Quataert, Dan Weisz, Jan Rybizki, and Anthony Brown for helpful comments. We acknowledge earlier discussions with Tim Brandt that proved seminal for this paper. We are grateful to Geoff Tabin and In-Hei Hahn for their continued hospitality during the writing of this manuscript. KE was supported by an NSF graduate research fellowship and a Hellman fellowship from UC Berkeley. T.M.H. acknowledges support from the National Science Foundation under Grant No. AST-1908119. 

This project was developed in part during the 2020 virtual eDR3 Unboxing Gaia Sprint. This work has made use of data from the European Space Agency (ESA) mission {\it Gaia} (\url{https://www.cosmos.esa.int/gaia}), processed by the {\it Gaia} Data Processing and Analysis Consortium (DPAC, \url{https://www.cosmos.esa.int/web/gaia/dpac/consortium}). Funding for the DPAC has been provided by national institutions, in particular the institutions
participating in the {\it Gaia} Multilateral Agreement. Guoshoujing Telescope (the Large Sky Area Multi-Object Fiber Spectroscopic Telescope LAMOST) is a National Major Scientific Project built by the Chinese Academy of Sciences. Funding for the project has been provided by the National Development and Reform Commission. LAMOST is operated and managed by the National Astronomical Observatories, Chinese Academy of Sciences. This research made use of Astropy,\footnote{http://www.astropy.org} a community-developed core Python package for Astronomy \citep{astropy:2013, astropy:2018}. This research made use of the cross-match service provided by CDS, Strasbourg.

\section*{Data Availability}
All the data used in this paper is publicly available. The {\it Gaia} data can be retrieved through the {\it Gaia} archive (https://gea.esac.esa.int/archive), and the LAMOST data are available at http://dr6.lamost.org. The binary catalog and code to produce it can be found at \href{https://zenodo.org/record/4435257}{https://zenodo.org/record/4435257}. 


\bibliographystyle{mnras}


\appendix

\section{Chance alignment probabilities}
\label{sec:kde}
We estimate the local density in parameter space of binary candidates and known chance alignments from the shifted catalog using a Gaussian kernel density estimate (KDE). The parameters (``features'') we use are listed in Table~\ref{tab:parameters}. Most of them are described in Section~\ref{sec:selection}. 

We also add a measure of the local sky density, $\Sigma_{18}$. This represents the number of sources per square degree that (a) pass the cuts of our initial query (Section~\ref{sec:selection}) and (b) are brighter than $G=18$. We calculate the value of $\Sigma_{18}$ around every binary candidate, counting the number of sources within 1 degree of the primary and dividing by $\pi$. Values of $\Sigma_{18}$ range from  280 toward the Galactic poles to 8700 toward the Galactic center. A significant fraction of the sources toward the Galactic center are likely background stars that are not actually within the 1 kpc search volume. 

We rescale the features so that they have similar dynamic range. This is accomplished by applying a few simple functions, which are listed in the ``scaled parameter'' column of Table~\ref{tab:parameters}. The rescaled parameters all have a dynamic range of about 4. We then calculate a 7-dimensional Gaussian KDE using a bandwidth $\sigma= 0.2$. 

The distance metric in this parameter space is somewhat ill-defined due to the different units and distributions of the features. We nevertheless proceed boldly, making no claim that the set of features, rescalings, or the choice of kernel are optimal. Our choices are designed to make the kernel (a) narrow enough that it does not smooth over the sharpest features in the data, and (b) wide enough to prevent density peaks around individual, discrete binaries (overfitting).

When calculating the KDE for the binary candidates, we use a leave-10\% out method wherein the density at the positions of 10\% of the binary candidates is evaluated using a KDE constructed from the other 90\%. To minimize discreteness noise in the chance-alignment KDE, we produce 30 different realizations of the shifted chance alignment catalog, shifting the declination of each star by a random variable $\mathcal{U}(-0.5,0.5)$ degrees for each realization. We combine the realizations when calculating the KDE, and then divide the calculated density by 30 to reflect the number of pairs in a single realization.

\begin{figure*}
    \centering
    \includegraphics[width=\textwidth]{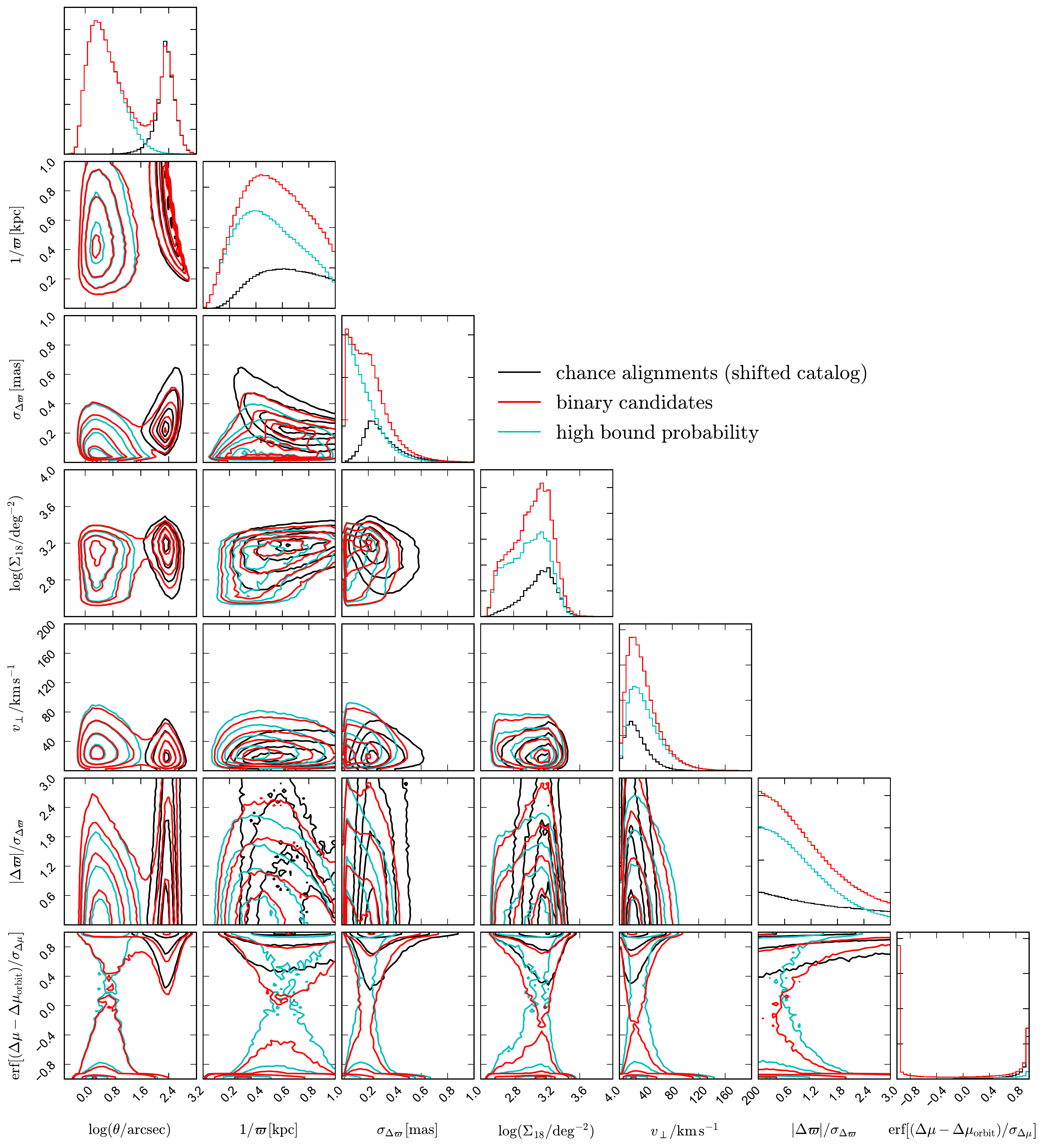}
    \caption{Parameter distribution of chance alignments from the shifted catalog (black), binary candidates (red), and high-confidence binary candidates with $\mathcal{R} < 0.1$ (section~\ref{sec:chance_alignment_prob}; cyan). Compared to chance alignments, the high-probability binaries have smaller angular separations, closer distances, larger parallax uncertainties, higher local source densities, larger tangential velocities, and more consistent parallaxes and proper motions. }
    \label{fig:big_corner_plot}
\end{figure*}

Figure~\ref{fig:big_corner_plot} compares the distributions of shifted chance alignments, all binary candidates, and candidates with $\mathcal{R} < 0.1$, in the space of features used for the KDE. It is clear that there are two modes in the binary candidate distribution, only one of which has a corresponding population in the chance alignment catalog. The clearest divisions between chance-alignments and binaries are in the dimensions of angular separation and proper motion difference (see also Figure~\ref{fig:triples}), but their distributions also differ in other features. For example, a binary candidate is more likely to be genuine if the parallax errors are small (low $\sigma_{\Delta \varpi}$), or if it is found in a region of low stellar density (low $\Sigma_{18}$). 

\begin{figure}
    \centering
    \includegraphics[width=\columnwidth]{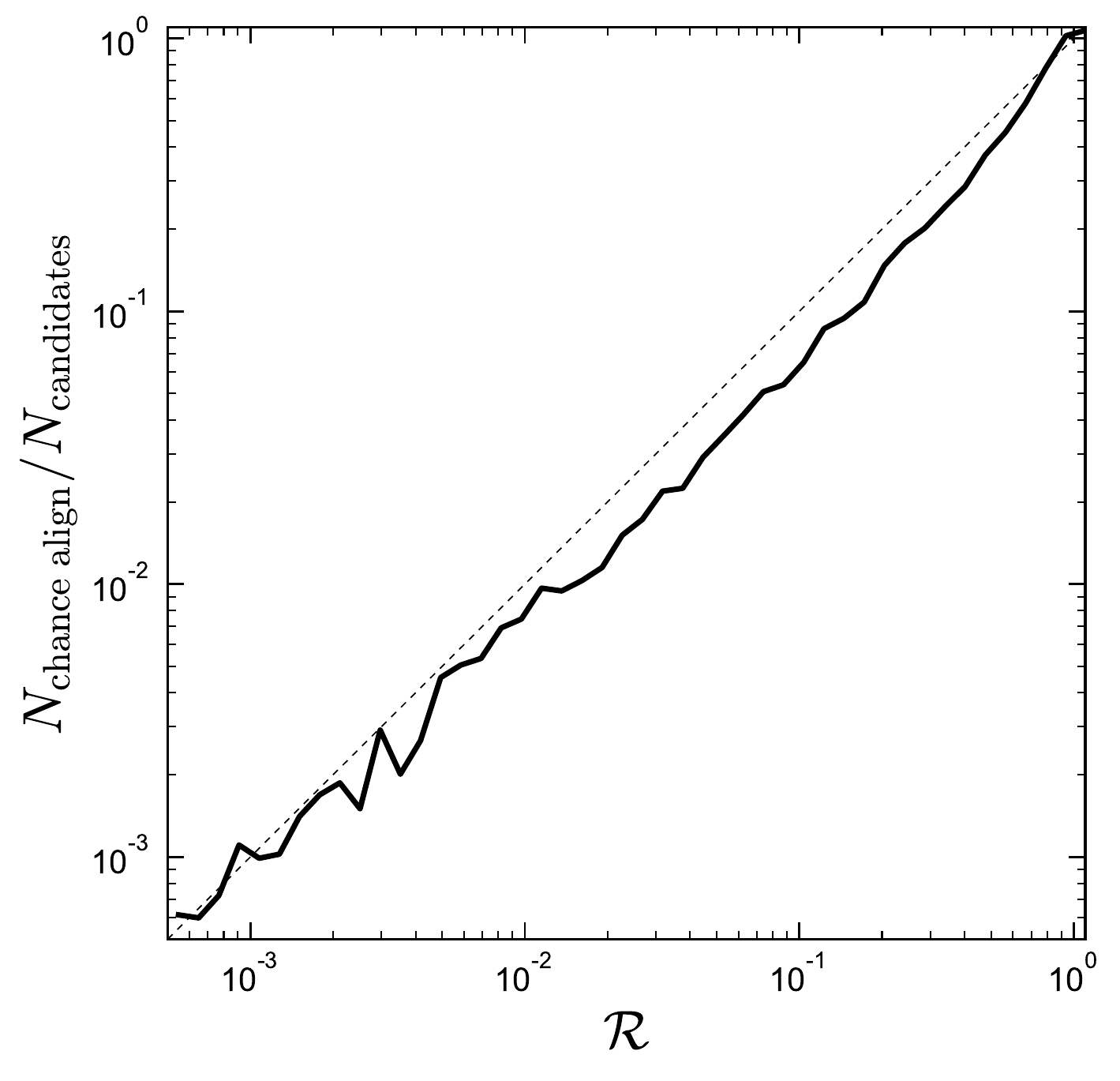}
    \caption{$\mathcal{R}$ is the ratio of the local ``density'' of chance-alignments from the shifted catalog to that of binary candidates (Equation~\ref{eq:ratio}). We compute $\mathcal{R}$ for binary candidates and for chance alignments from a realization of the shifted catalog. We then plot the ratio of the number of pairs in the chance alignment catalog that have a given $\mathcal{R}$ value to pairs in the binary candidate catalog with the same $\mathcal{R}$ value. Dashed line shows a one-to-one relation for comparison. }
    \label{fig:ratio_diagnostic}
\end{figure}

Figure~\ref{fig:ratio_diagnostic} shows the ratio of the number of chance alignments (from the shifted catalog) that have a given $\mathcal{R}$ value to the number of binary candidates with similar $\mathcal{R}$. If $\mathcal{R}$ is interpreted as the probability that a candidate is a chance alignment, one would expect this ratio to follow the one-to-one line. It does indeed fall close to the one-to-one line (dashed) but with some deviations, likely due to over-smoothing of the KDE. This figure suggests, for example, that about 6\% of binary candidates with $\mathcal{R}\sim 0.1$ are chance alignments, implying that chance alignment probabilities inferred when $\mathcal{R}$ is interpreted as a probability are conservative.

A small fraction of binary candidates in the catalog have extremely small $\mathcal{R}$ values; i.e., 0.5\% have $\log(\mathcal{R}) <-10$, and 0.09\% have $\log(\mathcal{R}) <-20$. These are primarily at close separations, where the chance alignment probability is indeed very low, but in this regime the $\mathcal{R}$ values should not be interpreted as probabilities due to the finite size of the shifted chance alignment catalog. 

\begin{table*}
\begin{tabular}{llllll}
Parameter & units & scaled parameter & (1,99)\% range & scaled (1,99)\% range & Description \\
\hline
$\theta$                  &   arcsec     &   $\log \theta$  &  (0.7, 607) & (-0.16, 2.78)  &  angular separation       \\
$\varpi_1$                  &   mas    &    $4/\varpi_1$   &  (1.02, 10.7) & (0.37, 3.93)   &   parallax (primary)     \\
$\sigma_{\Delta \varpi}$  &  mas   &     $4 \sigma_{\Delta \varpi}$   & (0.02, 0.74) &  (0.08, 2.96)  &   parallax difference error      \\
$\Sigma_{18}$                       &    deg$^{-2}$    &   $4\log\left(\Sigma_{18}\right)$    &  (345, 2790) &    (10.15,  13.78)  &   $G< 18$ local source density     \\
$v_{\perp,1}$               & km\,s$^{-1}$      &    $v_{\perp,1}/50$        &  (3.4, 121) &    (0.07,  2.43)  &     tangential velocity (primary) \\
$\Delta \varpi /\sigma_{\Delta \varpi}$       &     --   &   $|\Delta \varpi |/\sigma_{\Delta \varpi}$     & (-3.24, 3.30)  & (0.02, 3.95)  &    normalized parallax difference    \\
$\left(\Delta\mu-\Delta\mu_{{\rm orbit}}\right)/\sigma_{\Delta\mu}$         &   --     &   $2{\rm erf}\left[\left(\Delta\mu-\Delta\mu_{{\rm orbit}}\right)/\sigma_{\Delta\mu}\right]$    & (-88, 1.97) &  (-2, 1.99)  &  scaled proper motion difference       \\

\hline
\end{tabular}
\caption{\label{tab:parameters} Features used by our Gaussian KDE in computing the local density of binaries and chance alignments. The feature vector is the ``scaled parameter'' column, in which all variables have been rescaled to have comparable dynamic range. We list the middle-98\% ranges of both the raw and scaled parameters.}
\end{table*}

\subsection{Sources with spurious astrometry}
\label{sec:spurious}
A non-negligible fraction of sources in {\it Gaia} eDR3 have spurious astrometric solutions, meaning that they have large reported parallaxes and small reported uncertainties, but the parallaxes are significantly in error. The types of problems that can cause spurious solutions -- typically crowding and marginally resolved sources -- are generally equally likely to produce positive and negative parallaxes. The impact of spurious sources on our sample can thus be assessed by considering sources with significant negative parallaxes. 

To this end, we repeat our initial ADQL query (Section~\ref{sec:selection}) but require \texttt{parallax} < -1 and \texttt{parallax\_over\_error} < -5. This yields 2,877,625 sources, implying that about 4.5\% of the sources returned by the initial query have spurious solutions. We add the sources to our initial sample, treating their parallaxes as if they were positive. We then repeat the neighbor-counting procedure described in Section~\ref{sec:clusters} for these sources, again removing objects with more than 30 neighbors. Of the 2,877,625 known spurious sources, only 380,379 (13\%) survive this cut. That is, spurious sources are overwhelmingly found in regions of high source density, and a majority of them are removed by the first pass of cleaning.

We then carried out the full catalog construction procedure, now operating on an input sample that includes the initially selected sources as well as the known spurious sources with negative parallaxes, where the sign of the parallax is inverted for the known spurious sources. This yielded 15,852 candidate pairs in which one component is from the spurious sample. As expected, these pairs are concentrated at large separations; only 187 (593) have projected separations $s < 10,000$ au ($s < 30,000$ au). Finally, we repeat the calculation of $\mathcal{R}$ on the candidates in which at least on component is known to be spurious, yielding 133 pairs with $\mathcal{R} < 0.1$. This implies that about 1 in 10,000 binary candidates with $\mathcal{R} < 0.1$ contains a source with a spurious parallax as defined here.  

\bsp	
\label{lastpage}
\end{document}